\documentclass[twocolumn,showpacs,preprintnumbers,amsmath,amssymb]{revtex4}

\usepackage{graphicx}
\usepackage{dcolumn}
\usepackage{bm}
\usepackage{color}
\usepackage{graphicx}
\usepackage{dcolumn}
\usepackage{bm}
\usepackage{ragged2e}
\usepackage{chngcntr}
\usepackage{epsfig}
\usepackage{amssymb}
\usepackage{lineno}
\usepackage{blindtext}
\usepackage{mathtools}
\usepackage{hyperref}
\usepackage{cleveref}

\begin{document}

\title{Notes on the two main routes of period doubling and $\frac{1}{2}$ order subharmonic oscillations in a bubble oscillator}

\author{AJ Sojahrood$^{1,2}$}
 \email{amin.jafarisojahrood@ryerson.ca}
\author{R.E. Earl$^{3}$}%
\author{M. C. Kolios$^{1,2}$}%
\author{R. Karshafian$^{1,2}$}%

\affiliation{$^{1}$ Department of Physics, Ryerson University, Toronto, Canada\\
	$^{2}$ Institute for Biomedical Engineering, Science and Technology (IBEST) a partnership between Ryerson University and St. Mike's Hospital, Toronto, Ontario, Canada
	\\
}%
\affiliation{$^{3}$ Department of Mechanical Engineering, Mc Gill University, Montreal, Canada
}%
\date{\today}

\begin{abstract}
 Period doubling route to chaos is one of the well-known characteristics of nonlinear oscillators. A bubble is a highly nonlinear oscillator that exists in various phenomena and applications ranging from material science and underwater acoustics to medical ultrasound. The occurrence of period doubling (PD) in the oscillations of the bubbles is concomitant with the generation of $\frac{1}{2}$ order subharmonics (SHs). SH oscillations are used to monitor the bubble activity and are employed as a measure of stable cavitation. As bubbles are intravascular contrast agents, SH oscillations are used in diagnostic ultrasound to increase the contrast of the blood vessels to the tissue. Despite the importance of PD in a bubble oscillator, the dynamical behavior of the bubble during stable SH and ultra-harmonic (UH) oscillations have not thoroughly studied yet. In this work, through applying a comprehensive bifurcation method, we study the nonlinear radial oscillations of the bubble oscillator as a function of ultrasound driving pressure. The frequency of the driving force is chosen as the linear resonance frequency ($f_r$) and linear SH resonance frequency ($f_{sh}=2f_r$) of the bubble. Results show that, when the bubble is sonicated with $2f_r$, PD doubling occurrence is more likely to result in non-destructive oscillations. The evolution of the bubble stable P2 dynamics forms the shape of a bow-tie for bubbles with an initial diameter of 740nm and above. When $f=2f_r$, the phase portrait of the P2 attractor is distinctly different from a P2 attractor when $f=f_r$, and SH component of the backscattered pressure is maximum. When sonicated by $2f_r$, due to lower oscillation amplitude and gentler bubble collapse, the bubble can sustain stable P2 oscillations for a longer duration and over a broader range of applied acoustic pressure.
\end{abstract}
\maketitle
\section{Introduction}
An acoustically excited bubble is an example of a highly nonlinear and complex oscillator [1-27]. Bubbles exist in several phenomena in nature; they are involved in underwater acoustics and oceanography studies [12,27,28]; they have a key role in enhanced chemical reaction in sonochemistry [29-32]; they are the building block of sonoluminescence [30,31]  and they have several advantageous applications in medical ultrasound [33-40](e.g. contrast-enhanced imaging [33-36], drug delivery [34,35], blood-brain barrier (BBB) opening [37], enhanced heating in high-intensity ultrasound treatments [38], shock wave lithotripsy [39], histotripsy [40] and sonothrombolysis [41,42]).\\
The complex dynamics of the bubbles have been the subject of numerous numerical [1-27] and experimental studies [4,43-51]. The pioneering work of [1] has extensively studied the bifurcation structure of the bubble oscillator and revealed the nonlinear nature of the system and period doubling route to chaos.  The chaotic dynamics of the bubble oscillator has recently been extensively studied using the methods of nonlinear dynamics [15-26].  The existence of period 2,3, 4 and higher periods have been confirmed in several numerical and recent experimental studies of single bubble dynamics [1-27].\\
One of the main characteristics of nonlinear oscillators is the period doubling route to chaos [1,4]. The occurrence of period doubling in the oscillations of the bubbles is concomitant with the generation of $\frac{1}{2}$ order subharmonics (SHs) and ultraharmonics (UHs) in the backscattered pressure signal from the bubbles.  $\frac{1}{2}$ order SHs has been used as an indicator for stable cavitation to monitor treatments [52,53,54]; in bubble sizing [55], in contrast-enhanced ultrasound to detect the signal from blood [54-58], for non-invasive measurement of the pressure inside vessels [59-61] and as an indicator  for the pressure threshold of BBB opening [62,63] among several other applications.\\
It is known that as the acoustic pressure is increased, the nonlinear response will become chaotic and bubble radius grows beyond a limit that may result in bubble destruction. When chaos occurs, the SH amplitude experiences fluctuations or may even disappear.  Because of the spread of the signal energy over a wider frequency range, chaotic oscillations won’t be useful in imaging methods or monitoring treatments as they may not be distinguishable from broadband noise due to bubble destruction. Thus, in this paper, SH oscillations are of main interest for acoustic pressures between two limit values: the threshold for the onset of SH oscillations and the critical pressure at which the nonlinear response becomes chaotic or results in bubble destruction. Knowledge of these limits is essential for the optimization of applications that depend on the SH oscillations of the bubbles.\\
Pioneering theoretical work of Eller [64] and Prosperetti [65,66,67,68] investigated the pressure threshold for the generation of subharmonics for uncoated free bubbles. Later, studies have theoretically and experimentally investigated the pressure threshold for the generation of SHs in the encapsulated bubbles [69,70,71,72]. The focus of these studies was on the determination of the conditions required to achieve the lowest pressure threshold that can produce  $\frac{1}{2}$ order SHs. In the theoretical works [64,65,66,67,68,69,70], the equations for the bubble radial oscillations were linearized and used to determine the lowest pressure threshold.\\ 
Pioneering theoretical work [64,65,66,67,68,69] has shown that sonication of uncoated bubbles with twice their linear resonance frequency ($f_r$) will result in the generation of SHs at the lowest pressure threshold.  This frequency can be referred to as the linear $\frac{1}{2}$ order SH resonance frequency ($f_{sh}=2f_r$). Recent numerical works []71,72] investigated the SH threshold in uncoated and encapsulated bubbles.  Their method was based on calculating the SH component of the backscattered pressure from different bubble sizes. They found that at low pressures, there is no SH component that is distinguishable from the noise level; however, by increasing the acoustic pressure, the SH component appears and grows quickly. This is followed by a gradual saturation of the SH component and eventual disappearance. In their work, the excitation pressure just above which a distinct subharmonic peak appears was selected as the SH threshold. They found that for small bubbles (less than 1 micron), increased damping weakens the bubble response at twice the resonance frequency, leading to a shift of the minimum SH pressure threshold from twice the resonance frequency toward the resonance frequency.\\ Recent theoretical work of Prosperetti [67] investigated the SH threshold of coated bubbles and showed that the subharmonic threshold can be considerably lowered with respect to that of an uncoated free bubble if the mechanical response of the coating varies rapidly in the neighborhood of certain specific values of the bubble radius (e.g. changes in shell parameters due to buckling of the shell). [73] numerically investigated the ambient pressure dependence of the SH generation from contrast bubbles.\\
Despite the studies investigating the SH threshold of the bubbles [64-72], the bifurcation structure of the bubble oscillator in the regime of  $\frac{1}{2}$ order SHs (especially when sonicated with $f_{sh}$) has not been investigated in detail. Additionally, the evolution of the nonlinear bubble dynamics at higher amplitudes of P2 oscillations and the exposure conditions to generate sustainable non-destructive high amplitude P2 bubble oscillations are not understood in detail.\\
Detailed investigation of the possible routes to PD and their dynamical properties will help in better understanding of the conditions for the generation, maximization and sustainability of  $\frac{1}{2}$ order SHs. To achieve this, in this work, we have studied the dynamics of the bubble oscillators by paying closer attention to the dynamics of the bubbles undergoing period doubling (PD).  The bifurcation structure of the bubble radial oscillations has been investigated as a function of pressure under sonication with $f_r$ and $2f_r$. To gain more detailed information on the oscillatory behavior of the bubble, bifurcation structure of the bubble is constructed using two different methods [74,75] and the results are plotted alongside each other. The fist method is the conventional method of constructing the bifurcation diagrams, which is based on plotting the oscillation amplitude after every period of the acoustic wave [1] and the second method, extracts the maxima of the radial oscillations of the bubble. We have shown in [74,75] that using these two methods in tandem, we can reveal more detailed information about the oscillatory behavior of the bubble (e.g. UH and superharmonic oscillations).\\
Our results show that the bubble oscillator undergoes a new bifurcation route to chaos when it is driven with $2f_r$. The bifurcation is in fact a period doubling, however it evolves in a shape of  bow tie for bubbles above 740 nm in diameter; below this size the bubble exhibits a simple PD route to chaos.   The dynamical properties of the two different period-2 oscillations are investigated in detail using the time series of the radial oscillations, the phase portraits of the two P2 attractors as well as the frequency spectra of the (re-radiated pressure) backscattered waves by bubbles. When sonicated with $2f_r$, the P2 oscillations are more likely to result in non-destructive oscillations with gentler bubble collapses and a stronger SH component in the backscatter signal. Stable SH oscillations are less likely to be possible when the bubble is sonicated with its resonance frequency.\\
This study provides fundamental insight over the nature of different period 2 oscillations. This can help to better optimize the applications regarding the bubble dynamics and will aid a better understanding of the PD phenomenon in case of more complex systems like encapsulated microbubbles with nonlinear shell behaviour [76] or bubbles entrapped in tissues [77,78,79]. 
\section{Methods}
Since the purpose of this study is the detailed investigation of the nature of P2 oscillations and their fundamental characteristics in a bubble system, we have chosen the uncoated bubble as the oscillator of interest. Addition of the encapsulating shells will add more complexity to the dynamics and will be the subject of future studies. The fundamental information on the bubble dynamics in the absence of the shell will help provide a better understanding of the more complex features that will appear in case of the coated bubbles and will help in separating the shell effects from the abstract bubble system.
\subsection{The Bubble model}
The dynamics of the bubble model including the compressibility effects to the first order of Mach number can be modelled using Keller-Miksis equation [80]:

\begin{equation}
\rho[(1-\frac{\dot{R}}{c})R\ddot{R}+\frac{3}{2}\dot{R}^2(1-\frac{\dot{R}}{3c})]=(1+\frac{\dot{R}}{c})(G)+\frac{R}{c}\frac{d}{dt}(G)
\end{equation}
where $G=P_g-\frac{4\mu_L\dot{R}}{R}-\frac{2\sigma}{R}-P_0-P_A sin(2 \pi f t)$. $P_g$ is the gas pressure in the bubble and is given by $P_g=(P_0+\frac{2\sigma}{R})*(\frac{R_0}{R})^{3\gamma}$\\
In this equation, R is radius at time t, $R_0$ is the initial bubble radius, $\dot{R}$ is the wall velocity of the bubble and $\ddot{R}$ is the wall acceleration	$\rho{}$ is the liquid density (998 $\frac{kg}{m^3}$), c is the sound speed (1481 m/s), $P_g$ is the gas pressure, $\sigma{}$ is the surface tension (0.0725 $\frac{N}{m}$), $\mu{}$ is the liquid viscosity (0.001 Pa.s), $P_A$ and \textit{f} are the amplitude and frequency of the applied acoustic pressure. The values in the parentheses are for water at 293$^0$K. In this paper the gas inside the bubble is Air ($\gamma$=1.4) and water is the host media.\\
\subsection{Backscattered pressure}
Oscillations of a bubble generate a backscattered pressure ($P_Sc$) which can be calculated by [81]:
\begin{equation}
P_{sc}=\rho\frac{R}{d}(R\ddot{R}+2\dot{R}^2)
\end{equation}
where $d$ is the distance from the center of the bubble (and for simplicity is considered as 1m in this paper) [15].
Equation 1 is solved using the 4th order Runge-Kutta technique using the ode45 function in Matlab (this function also has a 5th order estimation); the control parameters of interest are $R_0$, $f$ and $P_A$.  The resulting radial bubble oscillations are visualized using bifurcations diagrams. Bifurcation diagrams of the normalized bubble oscillations $\frac{R}{R_0}$ are presented as a function of driving pressure for a given frequency. Detailed analysis is presented for  selected control parameters using a) the radius versus time curves, b) phase portrait analysis and c) the frequency spectrum of the backscattered pressure.
\subsection{Bifurcation diagrams}
Bifurcation diagrams are valuable tools to analyze the dynamics of nonlinear systems where the qualitative and quantitative changes of the dynamics of the system can be investigated effectively over a wide range of the control parameters. In this paper, we employ a more comprehensive bifurcation analysis method introduced in [74,75].\\
\textbf{2.3.a)	Conventional bifurcation analysis}\\
When dealing with systems responding to a driving force, to generate the points in the bifurcation diagrams vs. the control parameter, one option is to sample the R(t) curves using a specific point in each driving period. The approach can be summarized in:

\begin{equation}
P \equiv (R(\Theta))\{(R(t),  \dot{R}(t) ):\Theta= \frac{n}{f} \}\hspace{0.1cm} where \hspace{0.1cm} n=400,401...440
\end{equation}
Where $P$ denotes the points in the bifurcation diagram, $R$ and $\dot{R}$
are the time dependentradius and wall velocity of the bubble at a given
set of control parameters of ($R_{0}$, $P_{0}$, $P_{A}$, $c$, $k$, $\mu$,
$\sigma$, $f$) and $\Theta$ is given by $\frac{n}{f}$.  Points on the bifurcation diagram are constructed by plotting the solution of $R(t)$ at time points that are multiples of the driving acoustic period. The results are plotted for $n=400-440$ to ensure a steady state solution has been reached for all bubbles. Due to smaller viscous effects, bigger bubbles require longer cycles to reach steady state.\\
\textbf{2.3.b)	Method of peaks}
As a more general method, bifurcation points can be constructed by setting one of the phase space coordinates to zero:      
\begin{equation}
Q \equiv max(R)\{(R, \dot{R} ):\dot{R}= 0\}
\end{equation}
In this method, the steady state solution of the radial oscillations for each control parameter is considered. The maxima of the radial peaks ($\dot{R}=0$) are identified (determined within 400-440 cycles of the stable oscillations) and are plotted versus the given control parameter in the bifurcation diagrams. 
The bifurcation diagrams of the normalized bubble oscillations ($\frac{R}{R_0}$) are calculated using both methods a) and b). When the two results are plotted alongside each other, it is easier to uncover more important details about the SuH and UH oscillations, as well as the SH and chaotic oscillations.\\
\subsection{Analysis}
Bubbles of 400 nm-20 $\mu$ m diameter were considered. The linear damped resonance curves of the bubbles were calculated by numerically solving Eq. 1 for different frequency values and for an acoustic pressure of 1kPa. The linear damped resonance frequency ($f_r$) was determined as the frequency by which the oscillation amplitude was maximum. To avoid transient oscillations, for each simulation parameter, all analysis was performed within the last 40 cycles of a 440 cycle acoustic pulse. The process of choosing the maximum resonance frequency is similar to [82].\\
The bifurcation structure of the bubble radial oscillations were be plotted as a function of ($\frac{R}{R_0}$) with respect to the applied acoustic pressure when the driving frequencies were $f_r$ and $f_{sh}=2f_r$. $f_{sh}$ is called the linear SH resonance frequency. Results were compared for $f=f_r$ and $f=f_sh$.
The evolution of the two different Period 2 attractors was studied in more detail by examination of the time-series of the radial oscillations, maximum wall velocities, phase portraits and frequency spectra of the backscattered pressure at different stages of the dynamical evolution of the system. For each sonication frequency and pressure the maximum wall velocity and maximum non-destructive wall velocity ($\frac{R}{R_0}\leq2$ [83], for a review on the minimum threshold of bubble destruction refer to [15]) were calculated for the regimes of non-chaotic oscillations.  The results were compared for cases of $f=f_r$ and $f=f_{sh}$.
The pressure ranges which result in non-destructive bubble oscillations ($\frac{R}{R_0}\leq2$ [15,82]) and non-chaotic oscillations were determined. For these determined parameter ranges, the maximum fundamental (FU), subharmonic (SH) and ultraharmonic (UH) amplitude of the backscattered acoustic pressure wave were calculated. The results were compared for cases of $f=f_r$ and $f=f_{sh}$.
\section{Results}
\subsection{Bifurcation structure of micron size bubbles}
Figure  1 shows the bifurcation structure of the normalized radial oscillations of the three bubble sizes chosen ($R_0=0.5,1.5 \& 2.5 \mu m$) with respect to pressure; the left column represents the case where the sonication frequency is $f_r$ and right column represents the case where the sonication frequency is $f_{sh}$ ($f_{sh}=2f_r$).  The red graphs represent the structure constructed by the conventional method and blue graphs represent the structure constructed by the maxima method.\\ 
Comparison between the two columns reveals 3 important findings:\\
1-	The pressure threshold for P2 oscillations are lower when $f=f_{sh}$ and the possible pressure range of P2 oscillations are considerably larger.\\
2-	When $f=f_{sh}$, PD occurs over a process that looks like a  bow tie.\\
3-	The amplitude of P2 oscillations are considerably ($\approx 80\%$) smaller when $f=f_{sh}$.\\
Figures 1a, 1c, and 1e show that when $f=f_r$, the linear period-1 (P1) oscillations monotonically increase with pressure increase; then at a pressure threshold bubble oscillations undergo period doubling (PD).  For lower pressure values (e.g. 1kPa-10 kPa in fig 1a) the two bifurcation diagrams are on top of each other. This means that the wall velocity of bubble oscillations is in phase with the driving acoustic filed and oscillations are resonant [27]. As the pressure increases, the two curves diverge; this is because at higher pressures resonance frequency decreases [15] and oscillations become off-resonant. 
\begin{figure*}
\begin{center}
	\scalebox{0.43}{\includegraphics{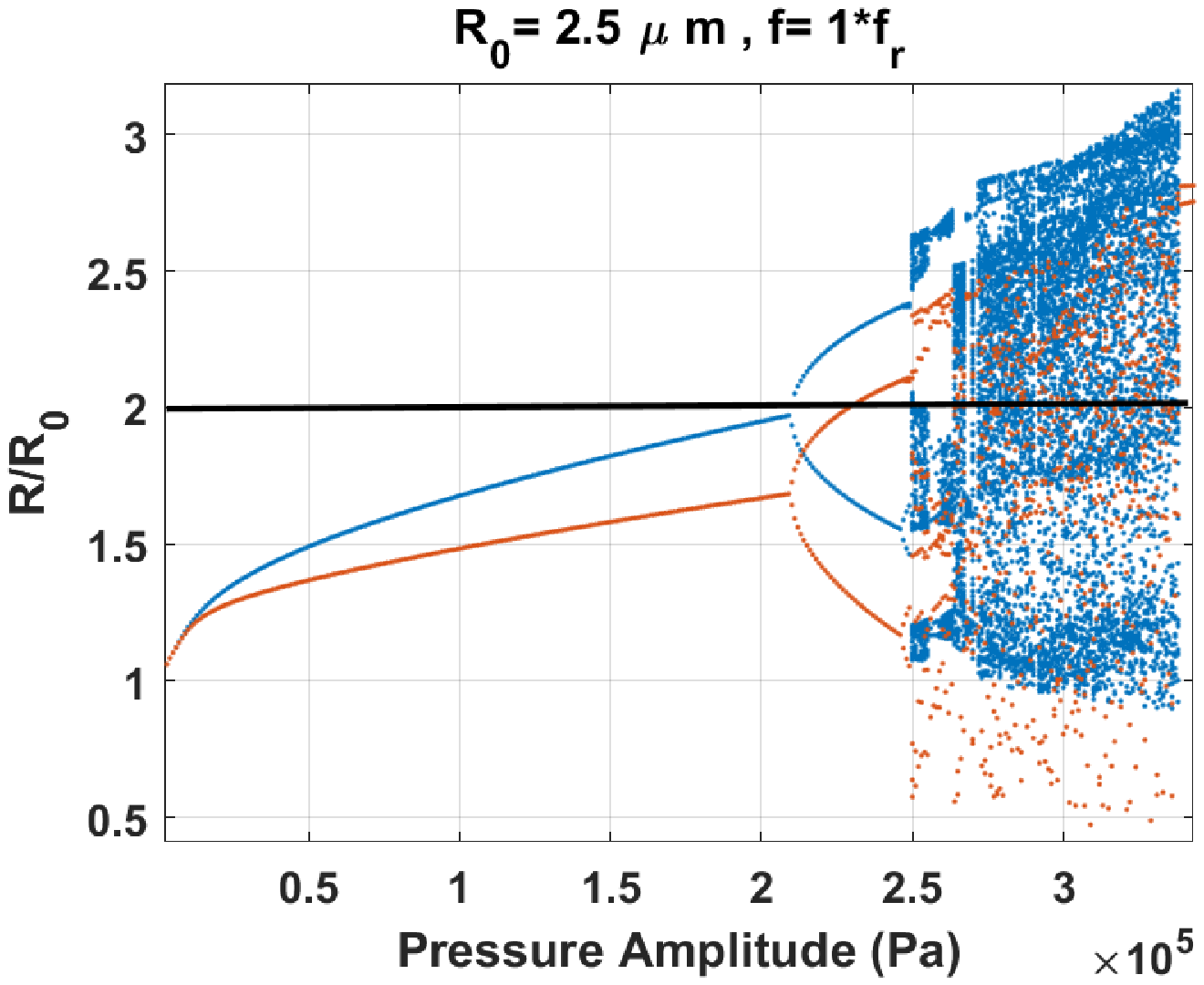}} \scalebox{0.43}{\includegraphics{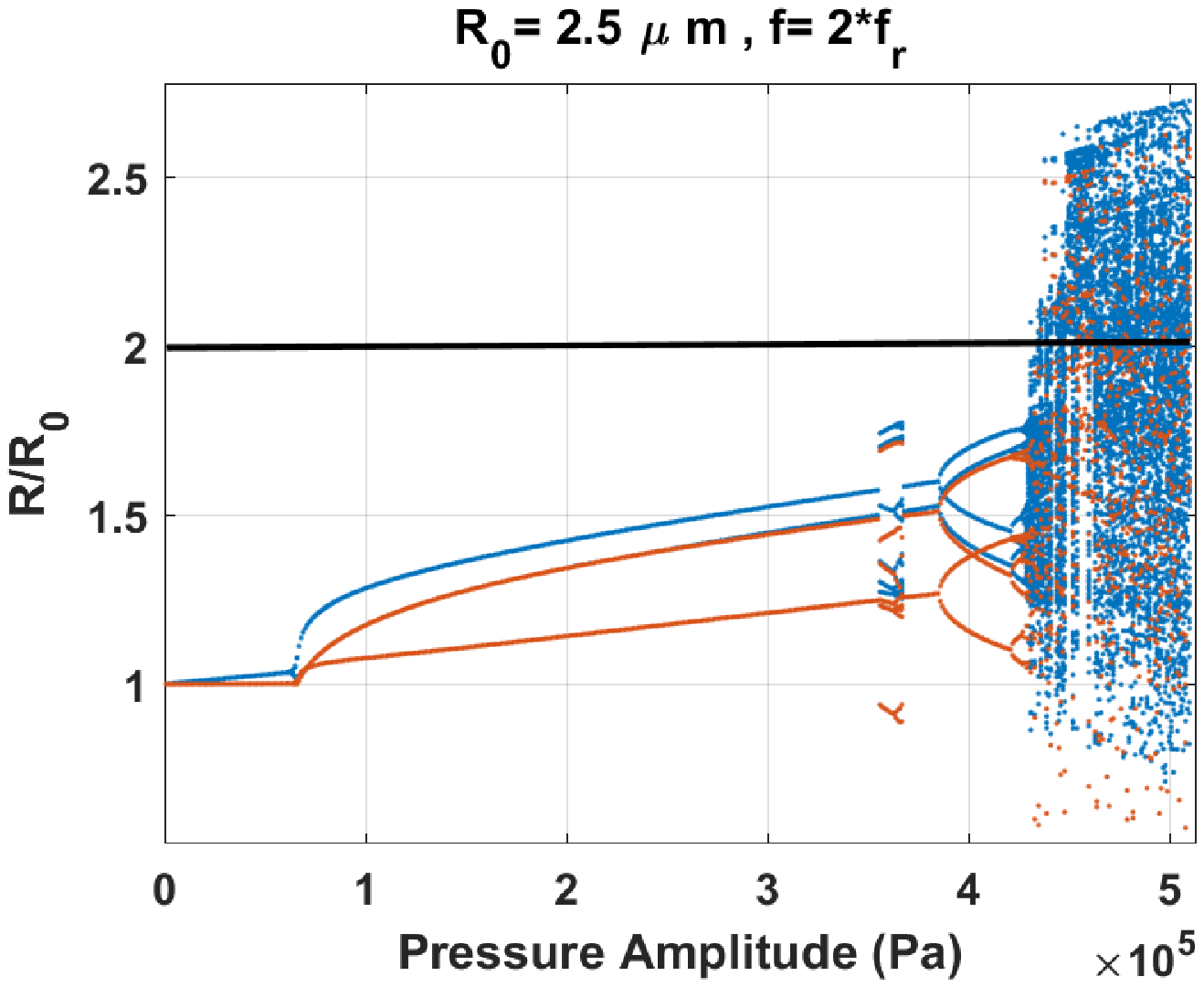}}\\
	\hspace{0.5cm} (a) \hspace{6cm} (b)\\
	\scalebox{0.43}{\includegraphics{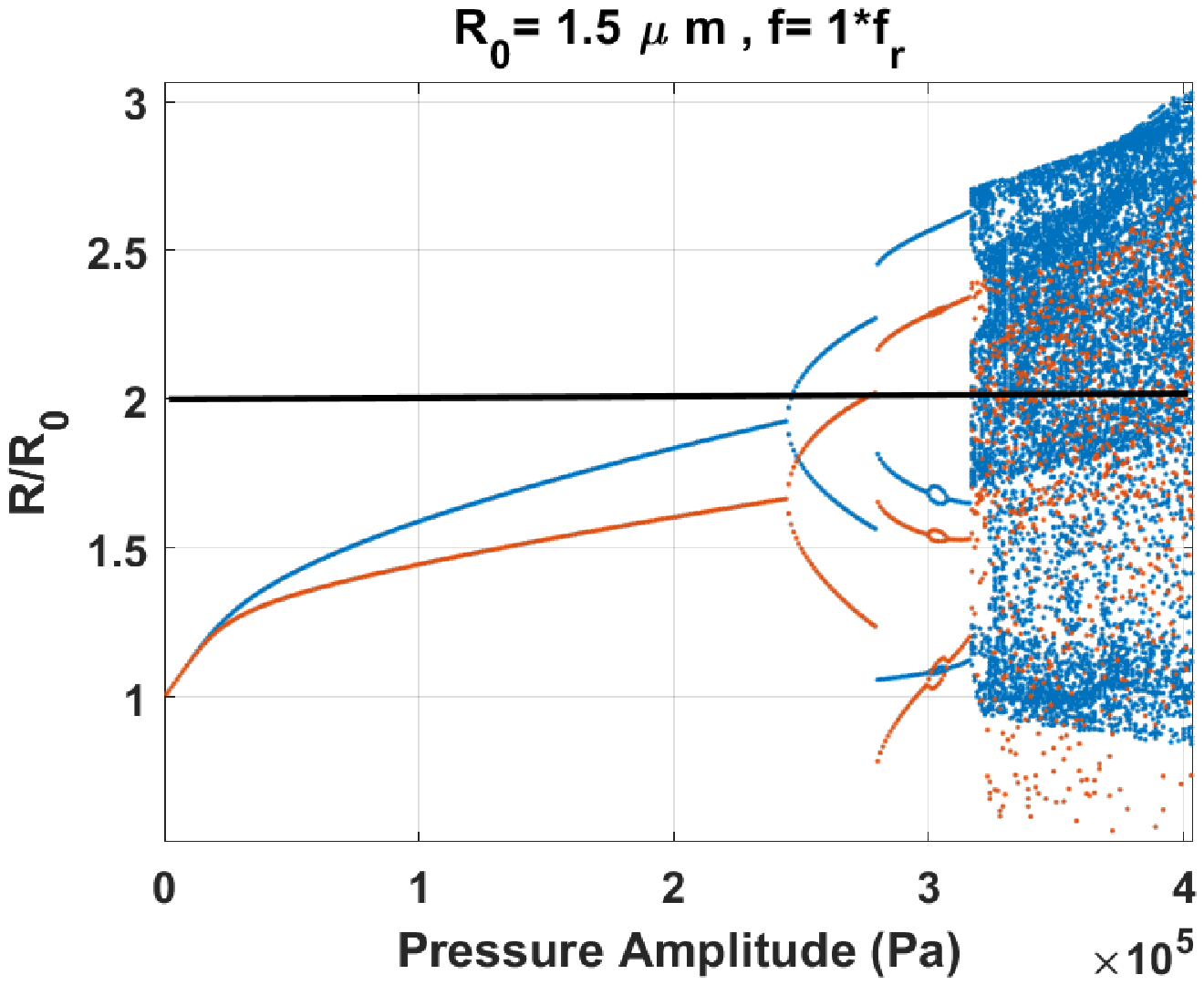}} \scalebox{0.43}{\includegraphics{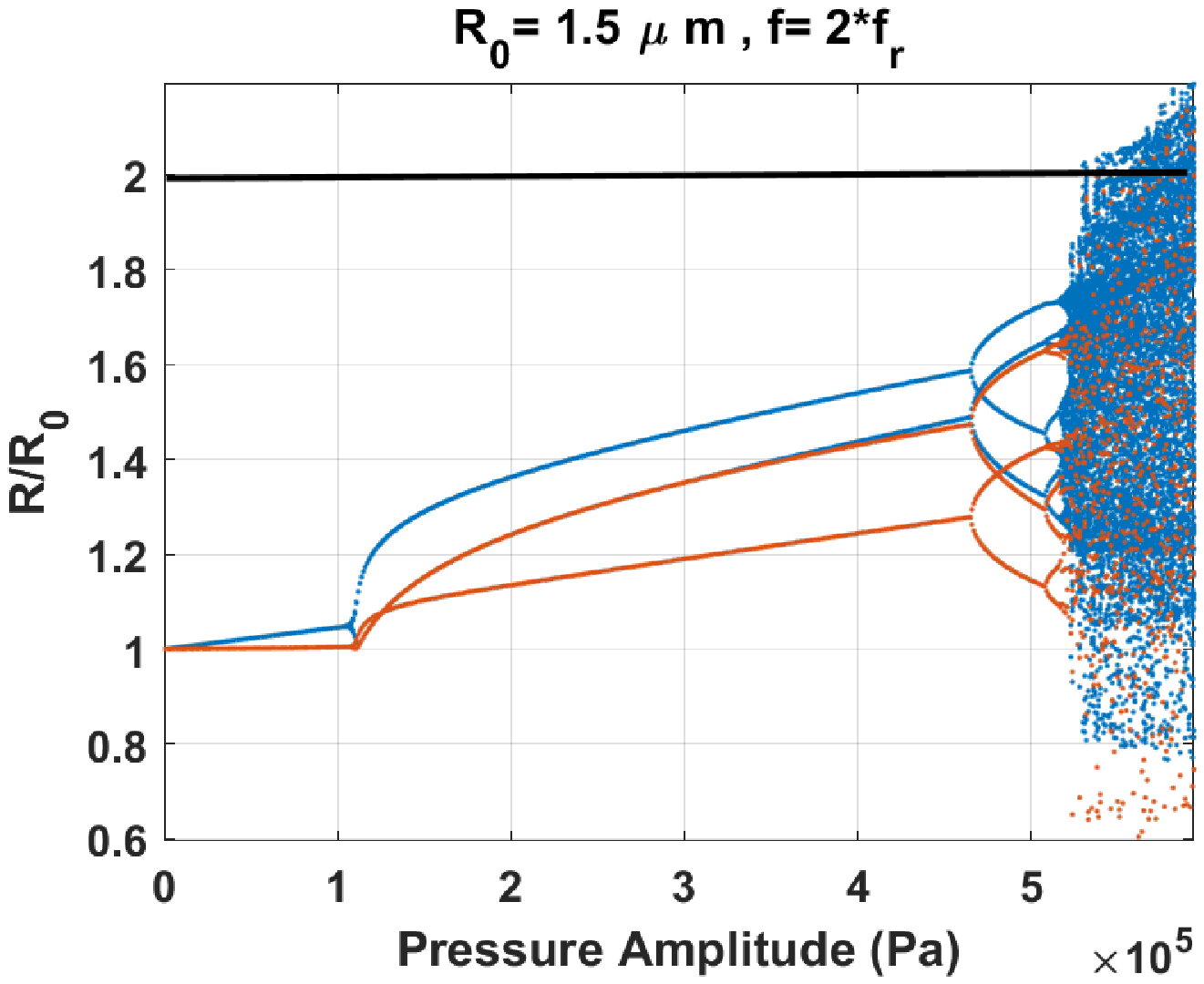}}\\
	\hspace{0.5cm} (c) \hspace{6cm} (d)\\
	\scalebox{0.43}{\includegraphics{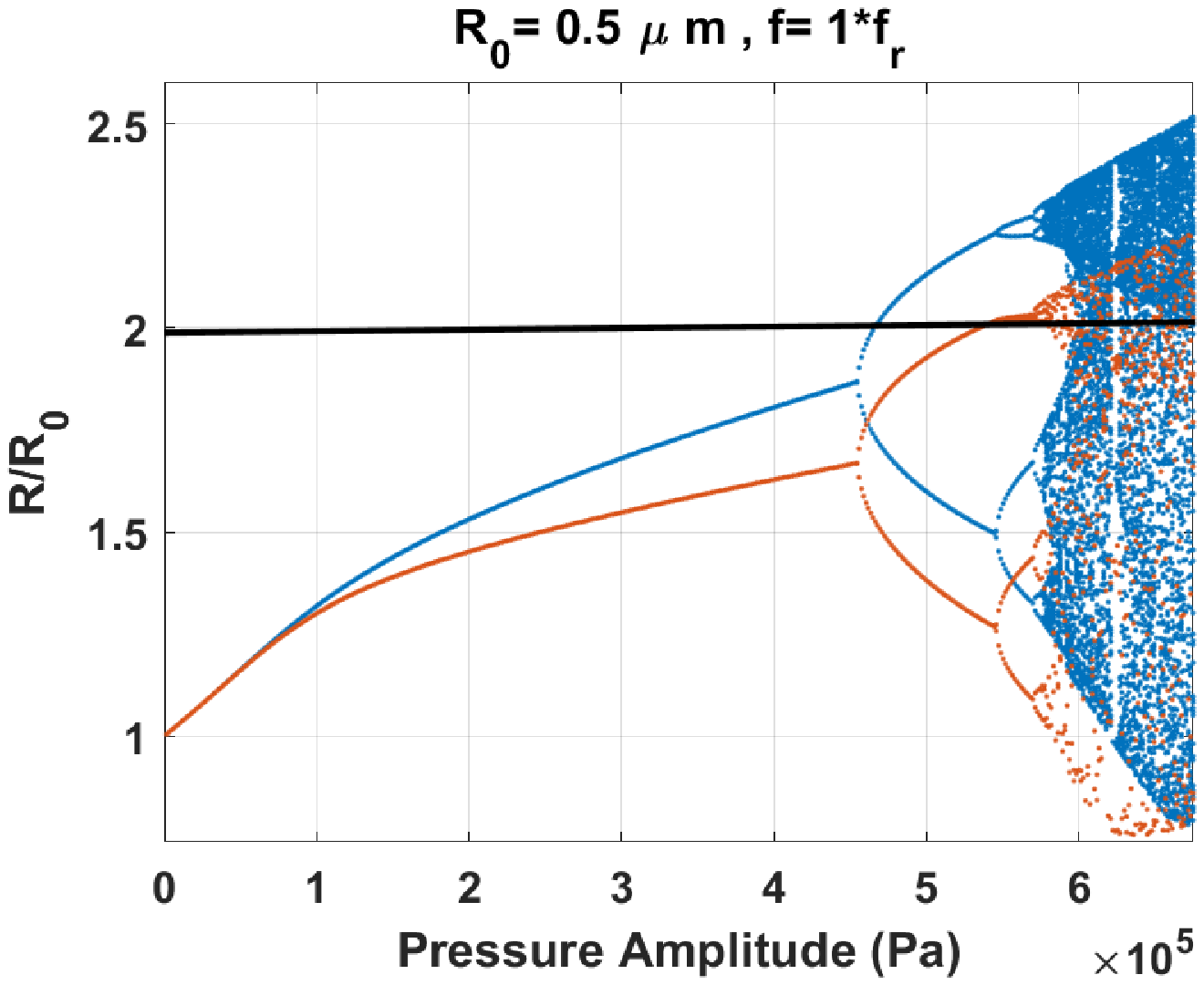}} \scalebox{0.43}{\includegraphics{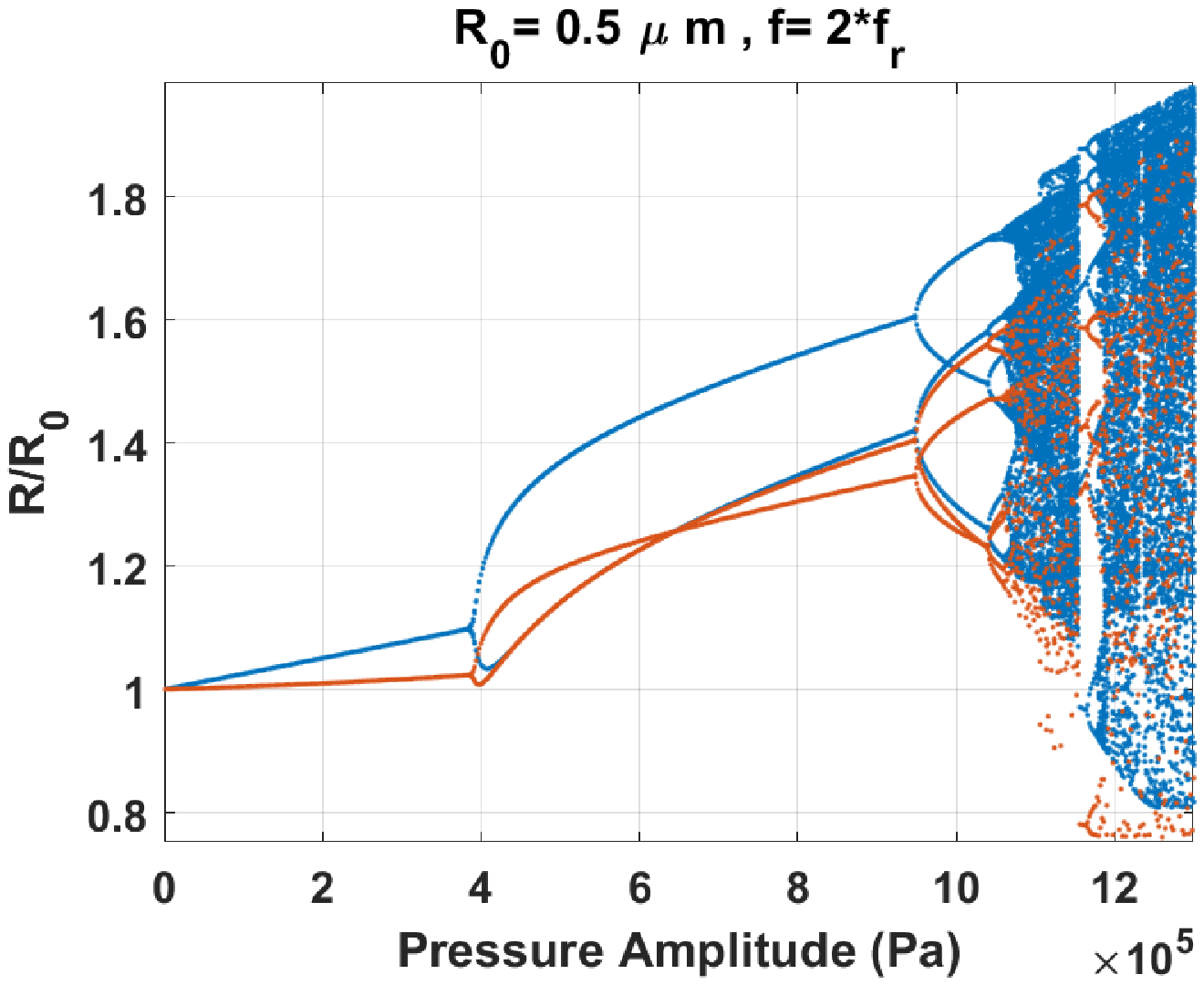}}\\
	\hspace{0.5cm} (e) \hspace{6cm} (f)\\
	\caption{Bifurcation structure (conventional method in red and method of peaks in blue) of the normalized radial oscillations ($\frac{R}{R_0}$) of the bubble as a function of pressure for a) $R_0$=2.5 $\mu m$ $\&$ $f=f_r$ b) $R_0=2.5 \mu m$ $\&$ $f=2f_r$ , c) $R_0=1.5 \mu m$ $\&$ $f=f_r$, d)$R_0=1.5 \mu m$ $\&$ $f=2f_r$, e) $R_0=0.5 \mu m$ $\&$ $f=f_r$, f) $R_0=0.5 \mu m $ $\&$ $ f=2f_r$.}
\end{center}
\end{figure*}
The pressure threshold for PD is size dependent and is higher for smaller bubbles due to the stronger viscous effects on smaller bubbles which is consistent with analytical predictions [62-66]. The bubble with $R_0=2.5 \mu m$ underwent a period doubling at ~208 kPa while the pressure threshold for the bubble with $R_0=0.5 \mu m$ was ~446 kPa. Both methods of bifurcation construction show a period doubling succeeding linear oscillations.\\ When $f=f_r$, the period doubling phenomenon however occurs when the $\frac{R}{R_0}$ amplitude is very close to 2; indicating that bubble would most likely undergo inertial collapse. 
Figures 1b, 1d and 1f show the bifurcation structure of the bubble when $f=2f_r$ for $R_0$=2.5, 1.5 and 0.5 $\mu m$ respectively. Compared to sonication with $f=f_r$, the linear oscillation amplitude and $\frac{R}{R_0}$ growth rate with pressure increase are smaller. The pressure threshold of PD is lower than sonication with $f=f_r$; the bubble with $R_0=2.5 \mu m$ and the bubble with $R_0=0.5 \mu m$ underwent PD at 58 and 310 kPa respectively. The $\frac{R}{R_0}$ oscillation amplitude of the P2 oscillations are much smaller than 2 indicating that the bubble may sustain long-lasting P2 oscillations without destruction. Additionally, the pressure range that results in P2 oscillations is broader (e.g. when $R=R_0$, for $f=f_r$ pressure range of P2 oscillations is $\approx$ 40 kPa while for $f=2f_r$ this pressure range increases to 400 kPa). Fig 1 shows that the pressure range of P2 oscillations becomes broader as R0 decreases, likely due to the stronger effects of viscous damping (e.g. when f=2fr the pressure range of P2 oscillations are $\approx$ 400kPa, $\approx$ 480 kPa and $\approx$ 600 kPa respectively for $R_0$= 2.5, 1.5 and 0.5 $\mu m$).\\
The most interesting difference between the cases of sonication with $f=f_r$ (left column) and $f=2f_r$ (right column) is on the process of the period doubling bifurcation. The period doubling for $f=2f_r$ is through a bow-tie shape like period doubling bifurcation (when the graphs obtained by the conventional method are analyzed (red curve)). The bubble starts the period doubling and as the pressure increases, the two branches of the $\frac{R}{R_0}$ on the red curve converge for a very short pressure range and then diverge; the initial lower amplitude of $\frac{R}{R_0}$ branch grows and becomes stronger than the initially higher branch. This phenomenon makes the P2 bifurcation process to look like a bow-tie when $f=2f_r$.\\
When the blue curve (representing the maxima of the $\frac{R}{R_0}$) is investigated, it is seen that the bubble undergoes a PD concomitant with the PD in the red curve (at the same pressure threshold for PD) but one branch of the $\frac{R}{R_0}$ of the P2 curve disappears quickly as the pressure increases. This indicates that the period of oscillations is 2; however, the $\frac{R}{R_0}$ oscillations vs time have only one maximum. Above a second pressure threshold, the second maxima re-emerges and its amplitude is exactly equal to the highest amplitude of the $\frac{R}{R_0}$ branch of the P2 oscillations in red curve. This indicates that the velocity of the $\frac{1}{2}$ SH oscillations of the bubble is essentially in phase with the acoustic driving force once every two acoustic cycles. Thus, full SH oscillations are developed. In case of $f=f_r$ and $R_0=1.5 \mu m$ (fig 1c) we see a short window of P3 oscillations and for the bubble with $R_0=2.5 \mu m$, we see a short pressure window of P6 oscillations for $f=2f_r$ (Fig 1b). These will be discussed in more detail in the Appendix B. 
\subsection{Bifurcation structure of nano size bubbles}
Fig 2, shows the bifurcation structure of $\frac{R}{R_0}$ of the bubbles with $R_0$ of 400, 300, and 200 nm (top to bottom respectively) versus acoustic pressure. The left column represent the case where the sonication frequency is set to the linear resonance of the bubble ($f_r$) and right column represent the case of sonication with $2f_r$. \\ 
Comparison between the two columns reveals 3 important facts:\\
1-	The pressure threshold for P2 oscillations are lower when $f=f_{sh}$ and the possible pressure range of P2 oscillations are considerably larger.\\
2-	The amplitude of P2 oscillations are considerably ($\approx 80 \%$) smaller when $f=f_{sh}$.\\
3-	When $f=2f_r$, for smaller bubbles PD initiates at a higher $\frac{R}{R_0}$ due to increases in damping.\\
\begin{figure*}
\begin{center}
	\scalebox{0.43}{\includegraphics{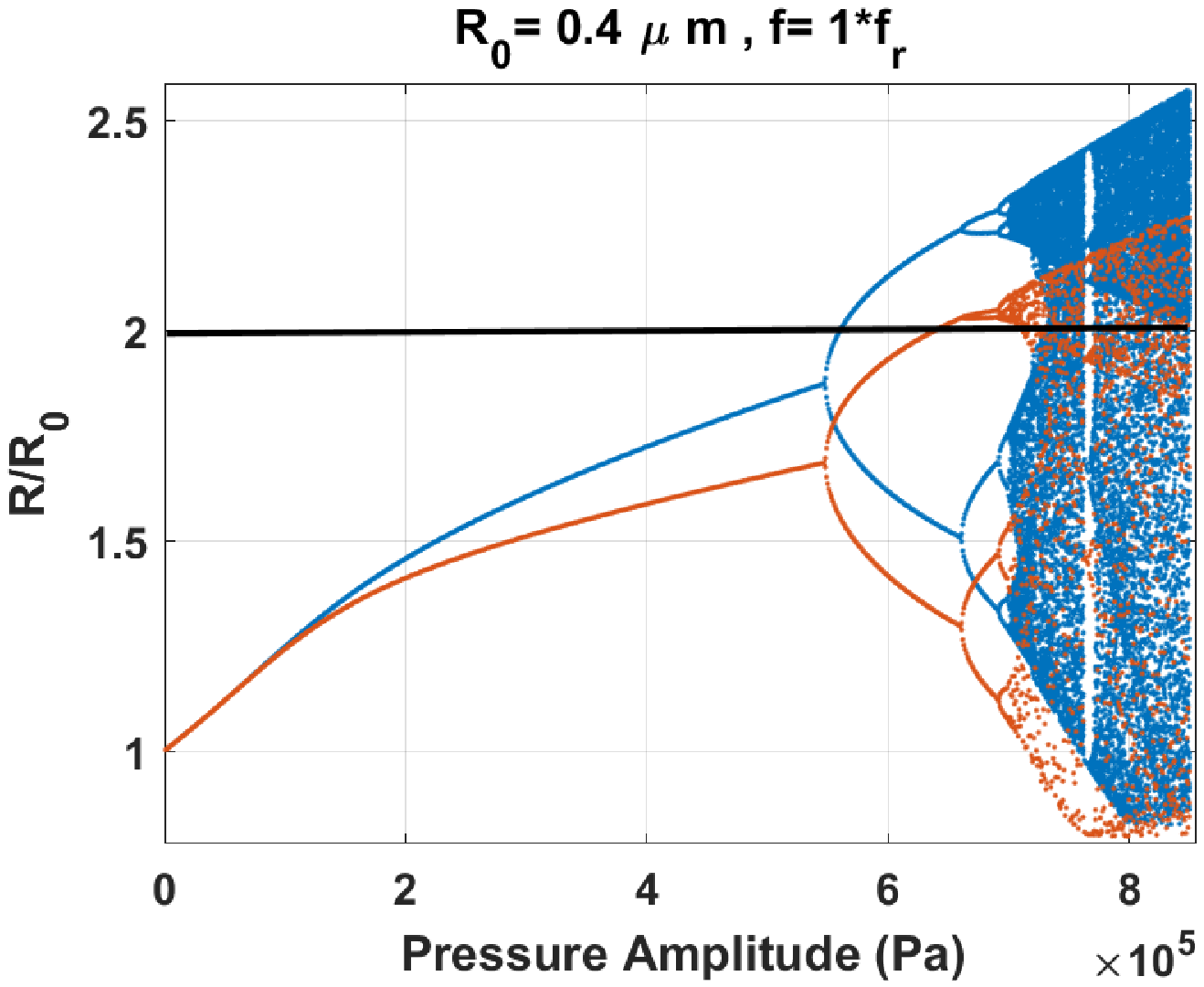}} \scalebox{0.43}{\includegraphics{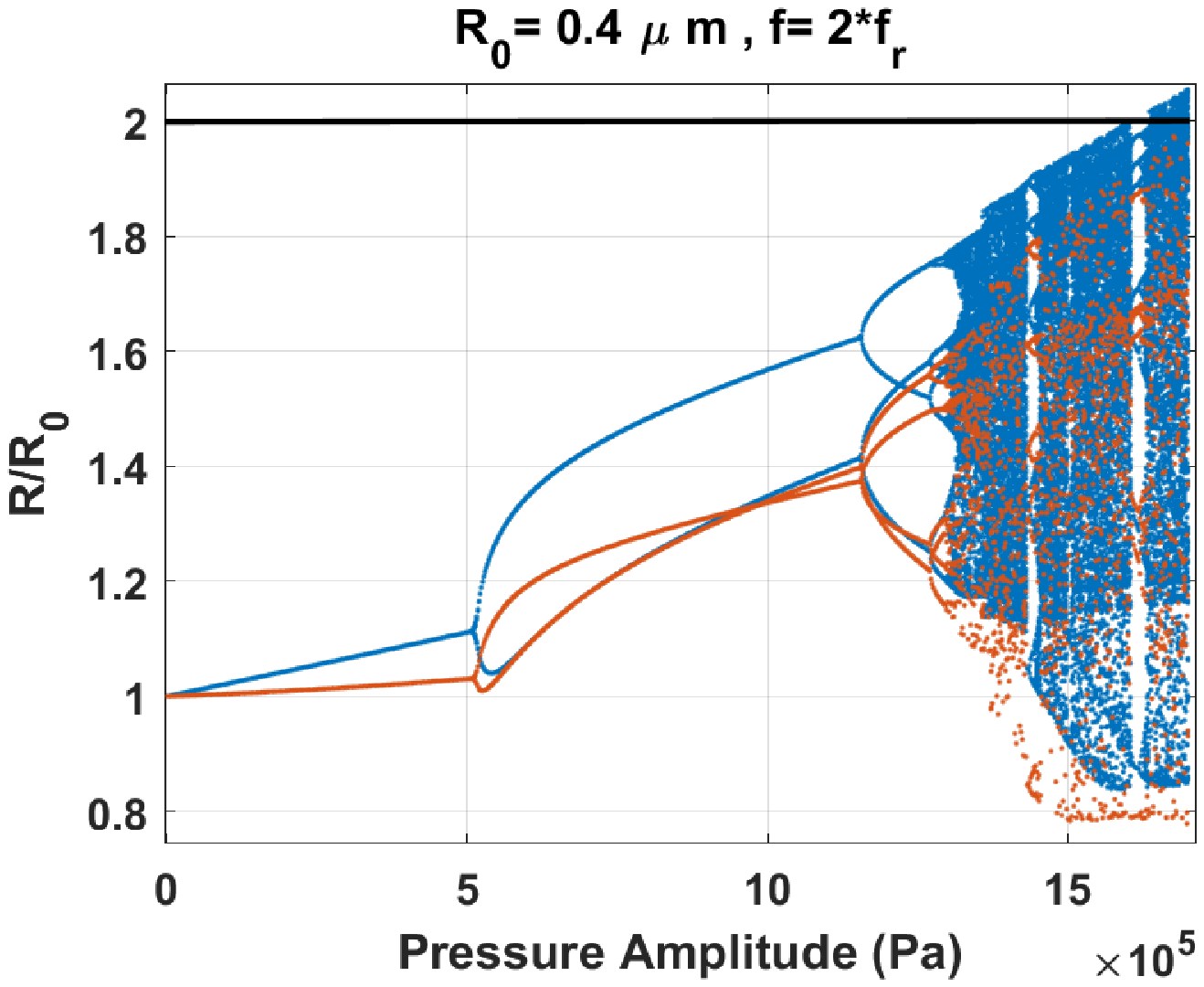}}\\
	\hspace{0.5cm} (a) \hspace{6cm} (b)\\
	\scalebox{0.43}{\includegraphics{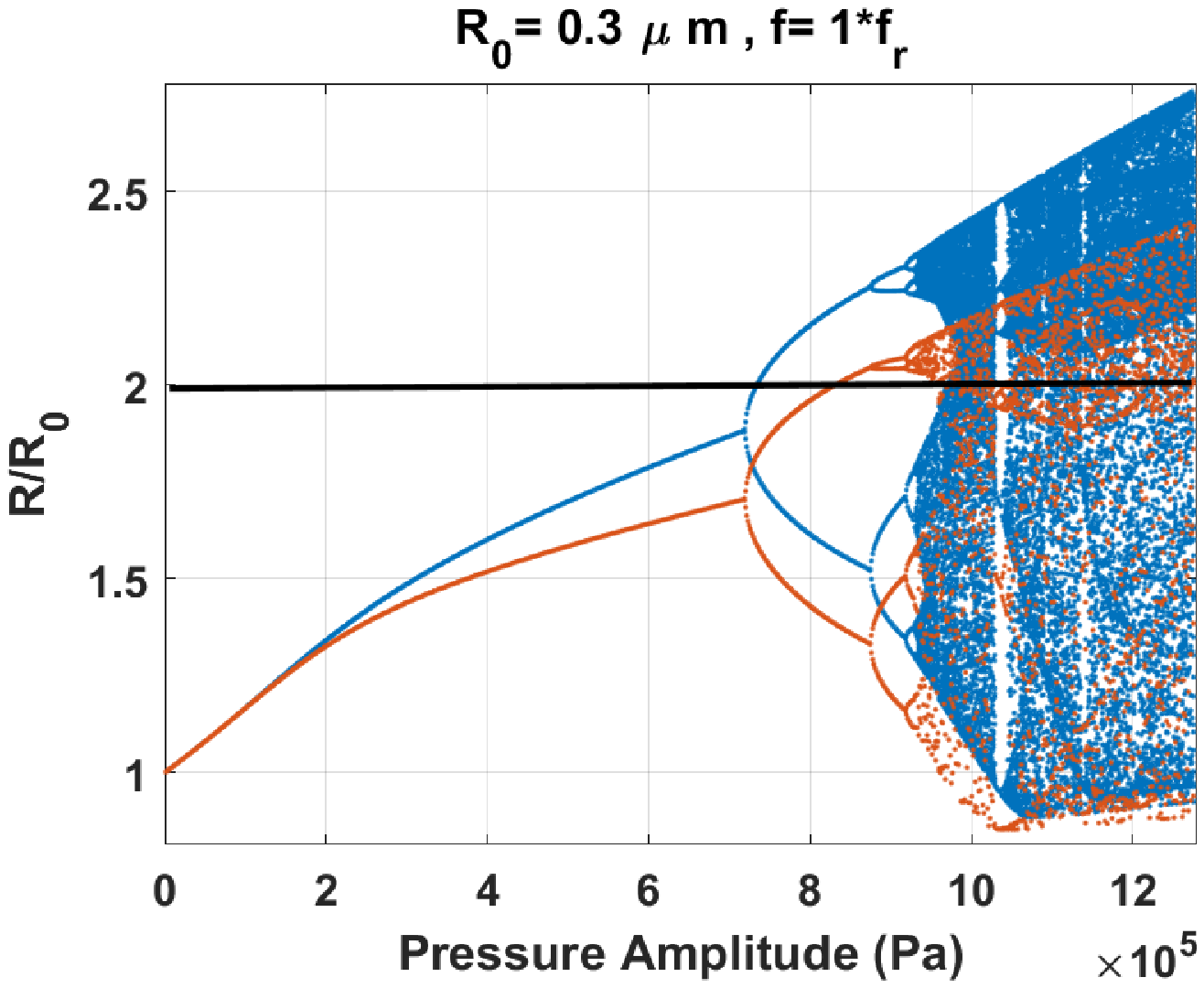}} \scalebox{0.43}{\includegraphics{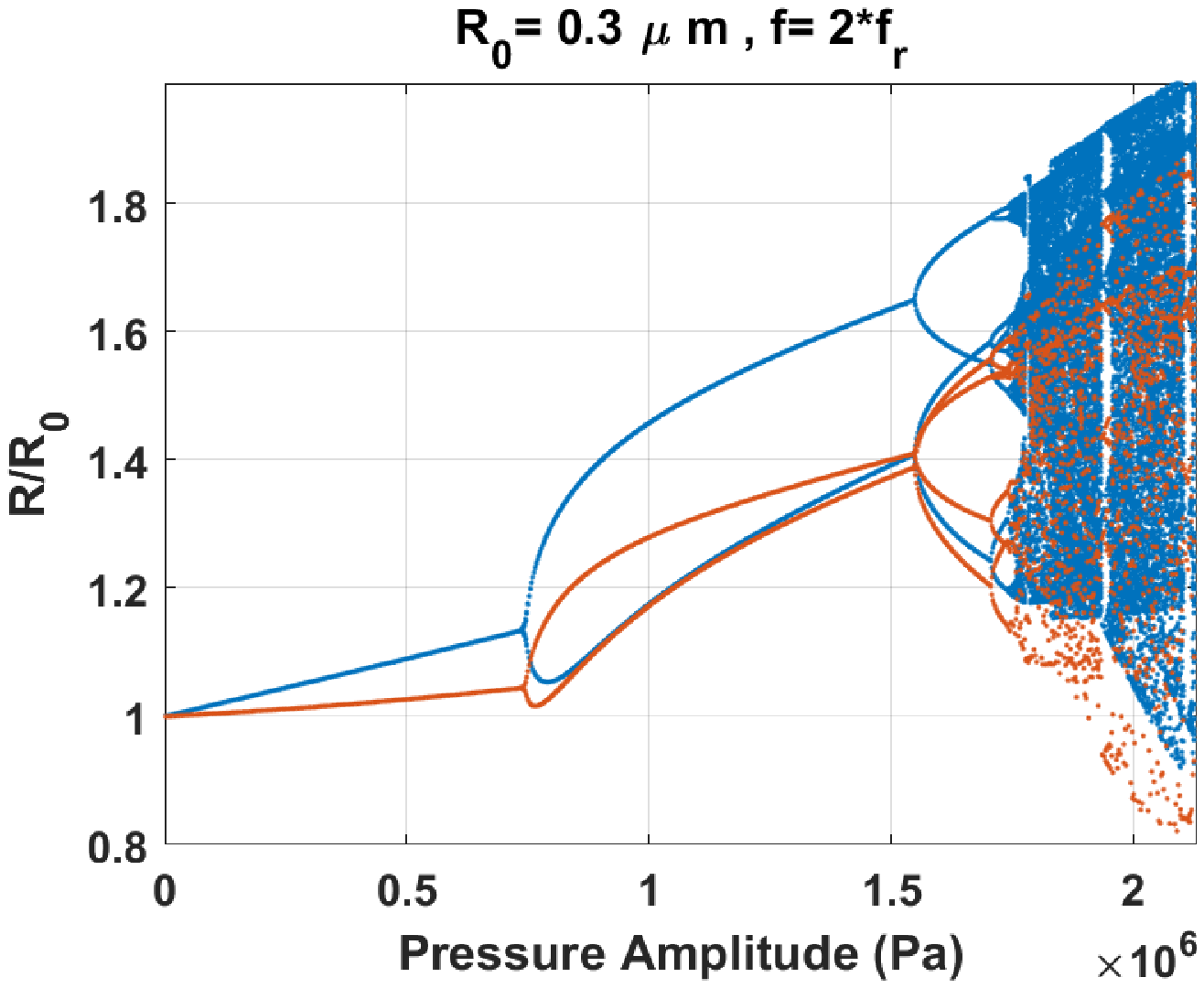}}\\
	\hspace{0.5cm} (c) \hspace{6cm} (d)\\
	\scalebox{0.43}{\includegraphics{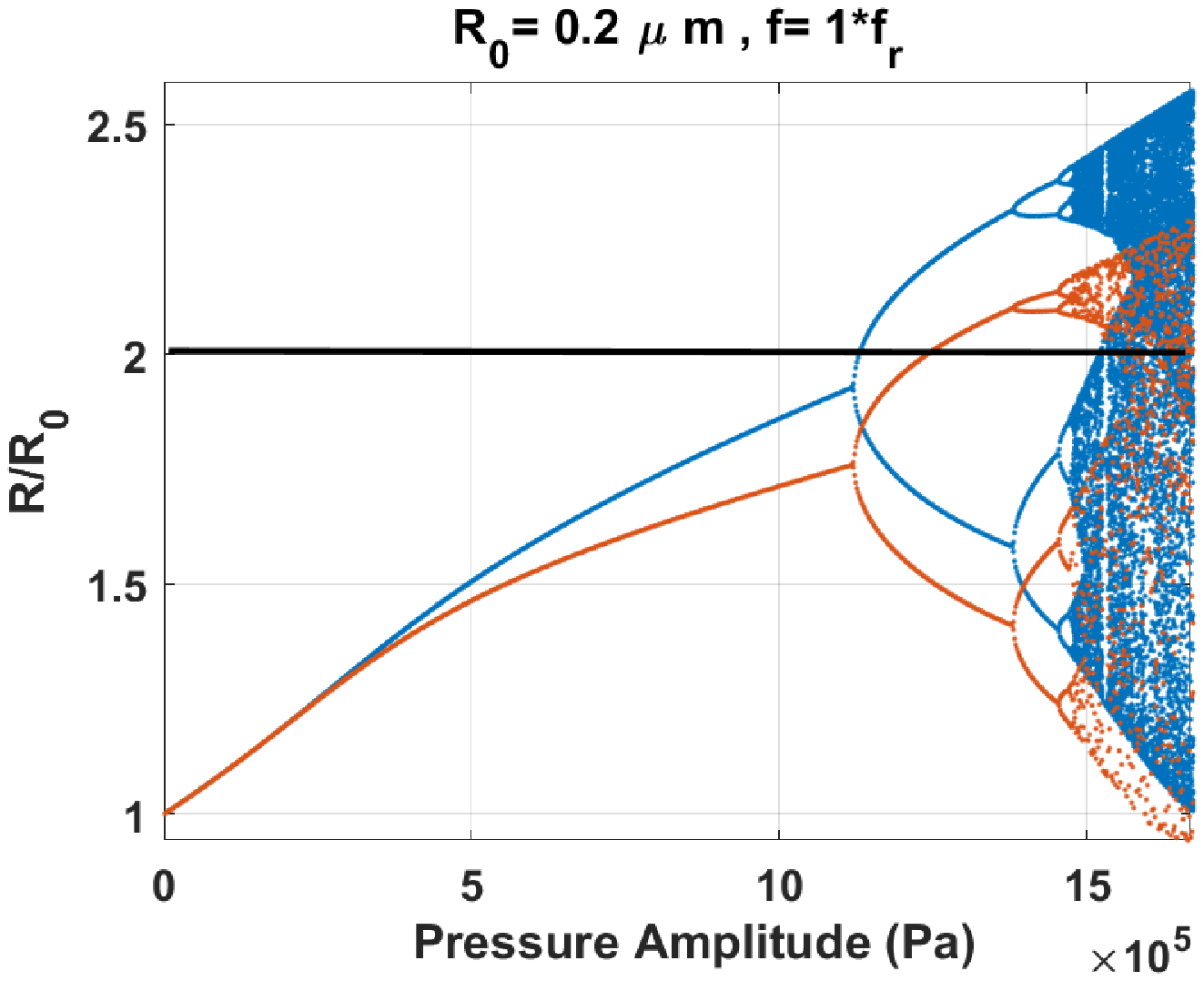}} \scalebox{0.43}{\includegraphics{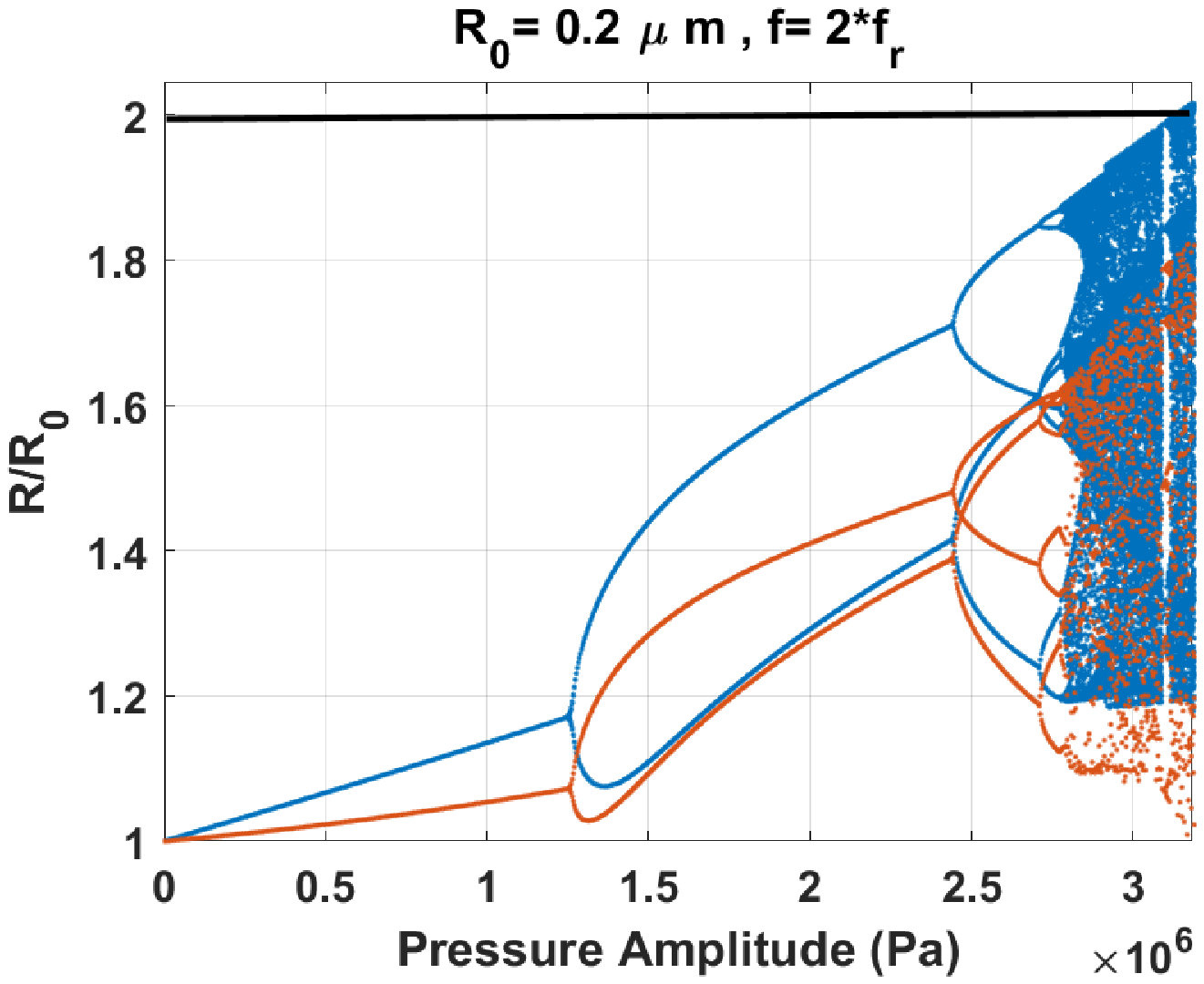}}\\
	\hspace{0.5cm} (e) \hspace{6cm} (f)\\
	\caption{Bifurcation structure (conventional method in red and method of peaks in blue) of the normalized radial oscillations ($\frac{R}{R_0}$) of the bubble as a function of pressure for a) $R_0=400 nm$ $\&$ $f=f_r$ b) $R_0=400 nm$ $\&$ $f=2f_r$ , c) $R_0=300 nm $ $\&$ $f=f_r$, d)$R_0=300 nm$ $\&$ $f=2f_r$, e) $R_0=200 nm$ $\&$ $f=f_r$, f) $R_0=200 nm $ $\&$ $ f=2f_r$.}
\end{center}
\end{figure*}
When the bubble is sonicated with $f_r$; the radial oscillations increase monotonically with pressure elevation and the bubble undergoes PD route to chaos above a pressure threshold (e.g. 700 kPa for the bubble with $R_0$=300 nm). The blue curve and red curve are initially on top of each other (wall velocities are in phase with the driving acoustic pressure) but they divert as the pressure increases. For smaller bubbles the pressure range where the red and blue curve have the same value are wider (e.g. 200 kPa for the 600 nm bubble and ~460 kPa for the 400 nm bubble).  When $f=f_r$, the $\frac{R}{R_0}$ amplitude of bubble oscillations at the time of PD is very close to 2; this indicates that bubbles may not be able to sustain non-destructive oscillations. The bifurcation diagrams generated by both the conventional and peaks method demonstrate concomitant PD; and both graphs show 2 solutions. This shows that the oscillations are of P2 with two maxima.\\
When bubbles are sonicated with $f=2f_r$; however, $\frac{R}{R_0}$ amplitude is well below the value of 2 when PD occurs. The bubble keeps P2 oscillations with an amplitude relatively below 2 (e.g. 1.15 at 0.1 MPa in fig 2f ) indicating that the bubble is more likely to sustain non-destructive P2 oscillations when sonicated with $2f_r$. Additionally the pressure range of P2 oscillations are much broader compared to the case of sonication with fr (e.g. ~1 MPa in fig 2d).  Unlike the cases of bubbles with $R_0$ larger than 500 nm; when sonicated with $f=2f_r$, the bubble oscillations exhibit two maxima for the whole range of P2 oscillations. After a pressure threshold, the lower maxima of the P2 oscillations lies on top of one of the branches of the P2 oscillations in the conventional bifurcation diagram; this indicates that the velocity of bubble oscillations is in phase with the driving acoustic period once every two acoustic cycles. Furthermore, the P2 mechanism for bubbles with $R_0$ $\leq$ 370 nm does not exhibit the bow tie shape; in other words, when sonicated with 2fr a bow tie P2 bifurcation only happens for bubbles larger than 740 nm. Comparing figs 1 and 2, when $f=f_r$, chaotic oscillations only develop when $\frac{R}{R_0}$ $>$ 2.3; thus, in practice, resonant bubbles may not sustain chaotic oscillations due to the high possibility of destruction. However, when $f=2f_r$, and for bubbles with $R_0 \lessapprox ~2.5 \mu m$, chaotic oscillations can develop when $\frac{R}{R_0}$ $\leq$ 2; thus, these bubbles may sustain chaotic oscillations when $f=2f_r$.
\subsection{Pressure threshold and range of P2 oscillations} 
\begin{figure*}
\begin{center}
	\scalebox{0.43}{\includegraphics{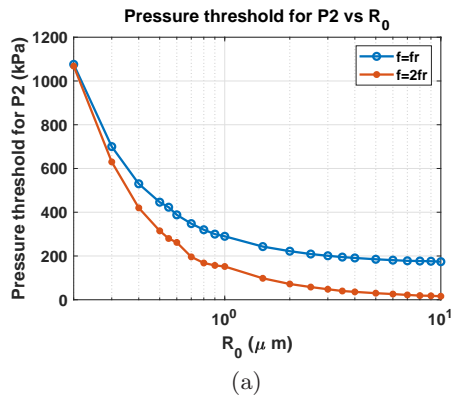}}\\
	(a) 
	\caption{Pressure threshold of P2 oscillations as a function of $R_0$ when bubble is sonicated with $f=f_r$ and $f=2f_r$.}
\end{center}
\end{figure*}
Figure 3 illustrates the pressure threshold of period doubling (PD) as a function of $R_0$ for $f=f_r$ and $f=2f_r$ and is created by analyzing the bifurcation diagrams of bubbles with sizes between 400nm-20 $\mu m$ (Appendix A: Figs A.1a and A.2a). The pressure threshold ($P_t$) of PD is lower when the bubble that is sonicated with $2f_r$; however, as $R_0$ decreases the difference between the $P_t$($f_1$) and $P_t$($f_2$) decreases. The pressure threshold of PD is lower for bigger bubbles; this is due to weaker effects of viscosity on larger bubbles. These results are in agreement with analytical predictions of Prosperetti [66].\\
Figure 4 is made by analyzing the bifurcation diagrams of bubbles of 400nm-20 $\mu m$ size (Appendix A: Figs A.1a and A.2a). Figure 4a demonstrates the range of P2 oscillations as a function of $R_0$ for $f=f_r$ and $f=2f_r$. When bubble is sonicated with 2fr, the range of acoustic pressures that can result in P2 oscillations are broader than when compared to $f=f_r$ by an order of magnitude (e.g. for $R_0=4\mu m$, the P2 pressure range is 34 and 304 kPa for $f=f_r$ and $f=2f_r$ respectively). Figure 4b demonstrates the pressure range of P2 oscillations when $\frac{R}{R_0}$ $\leq$ 2. When $f=f_r$, bubbles with $R_0$$>$2.5 $\mu m$ undergo PD when $\frac{R}{R_0}$$>$2, thus they may not exhibit non-destructive SH oscillations; however, when $f=2f_r$, all the studied bubble sizes (200nm$\leq$$R_0$$\leq$10 $\mu m$) exhibit non-destructive P2 oscillations over a broad range of acoustic pressures. Thus, if non-destructive SH oscillations are desired in an application, the sonication frequency should be set as twice the resonance frequency of the bubbles.\\
\begin{figure*}
\begin{center}
	\scalebox{0.33}{\includegraphics{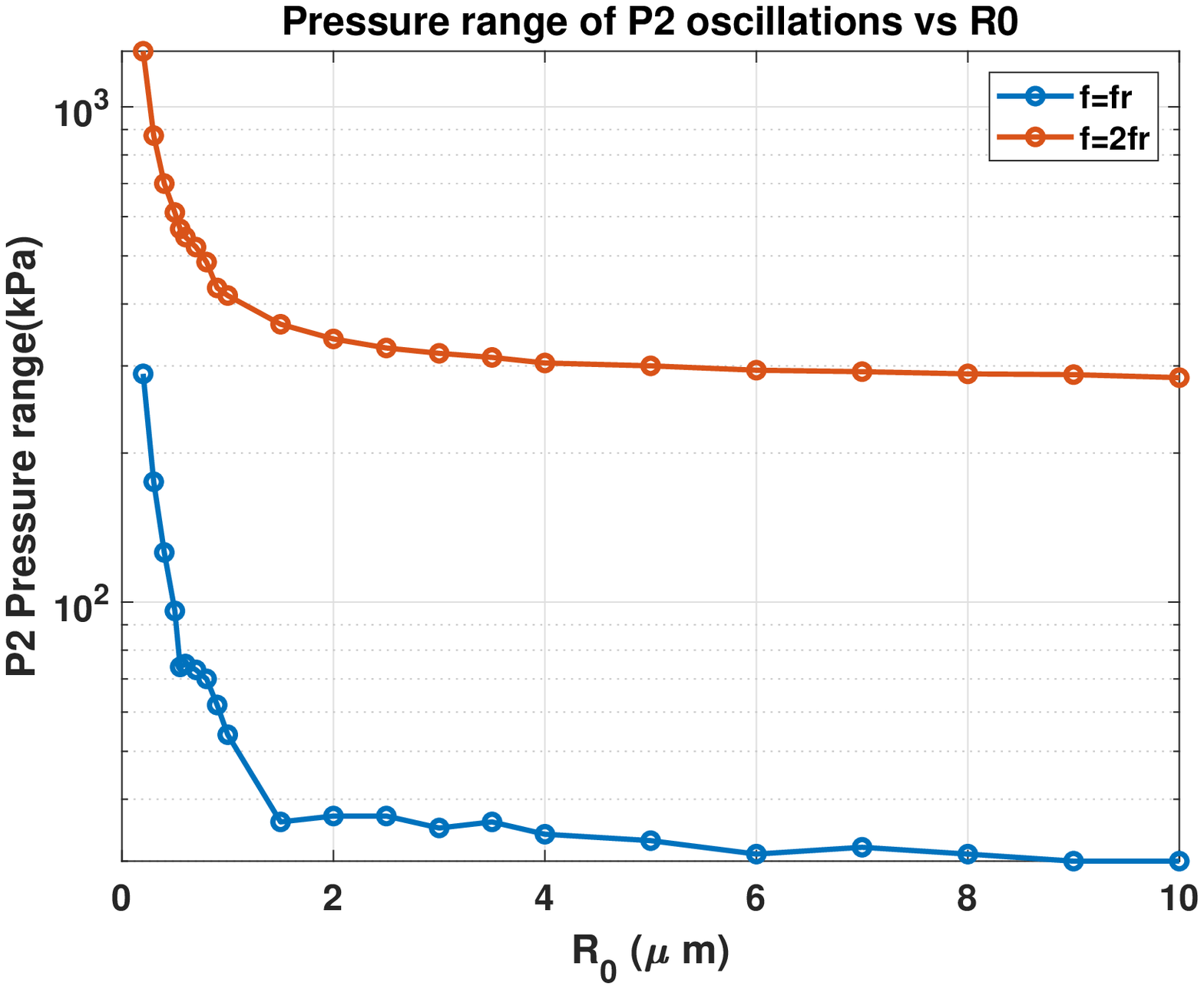}} \scalebox{0.33}{\includegraphics{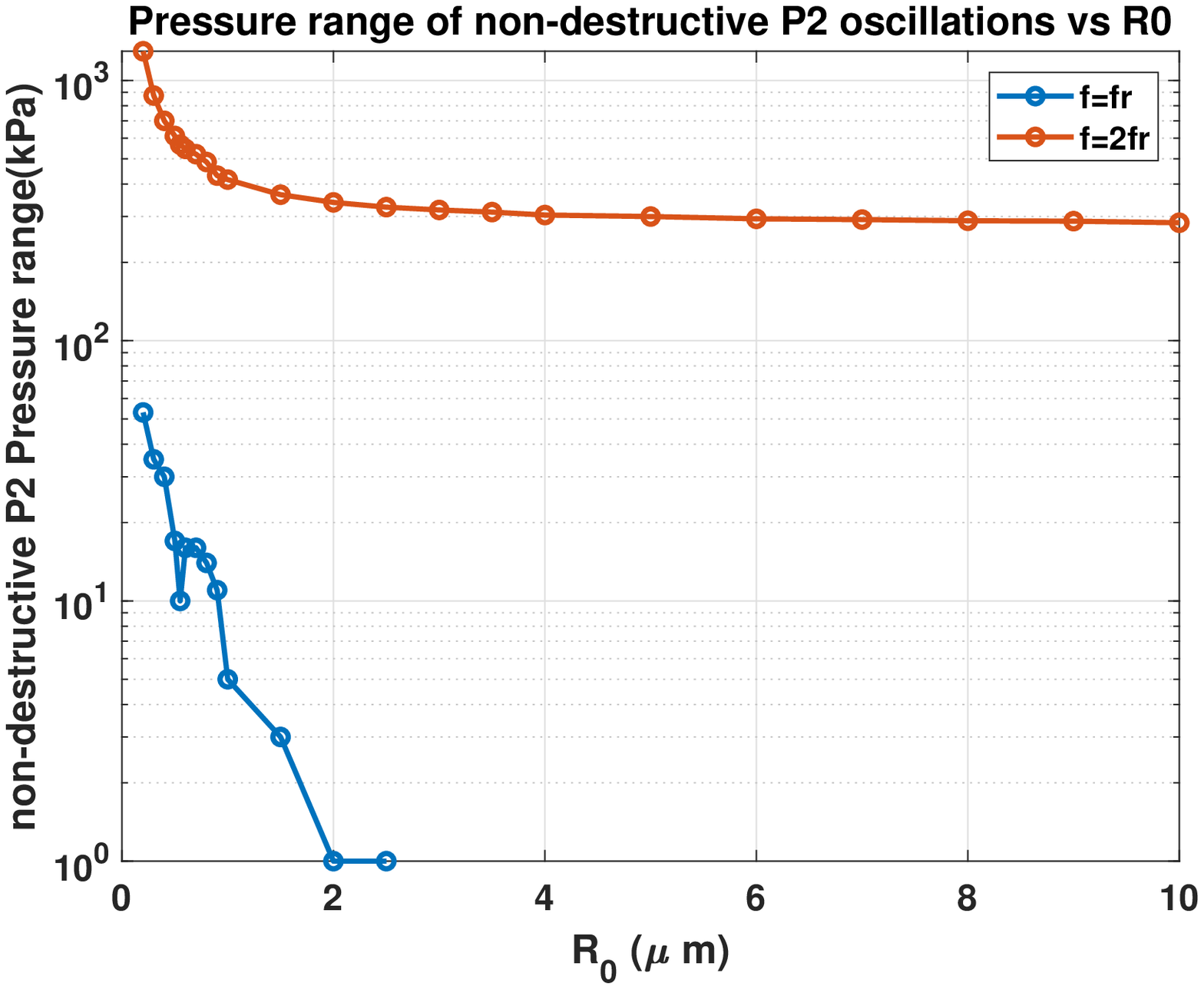}}\\
	\hspace{0.5cm} (a) \hspace{6cm} (b)\\
	\caption{a) Acoustic pressure range of P2 oscillations as a function of $R_0$, b) Non-destructive ($\frac{R}{R_0}\leq2$) acoustic pressure range of P2 oscillations as a function of $R_0$.}
\end{center}
\end{figure*}
\subsection{Period doubling and SH initiation, growth and saturation}

\begin{figure*}
\begin{center}
	\scalebox{0.43}{\includegraphics{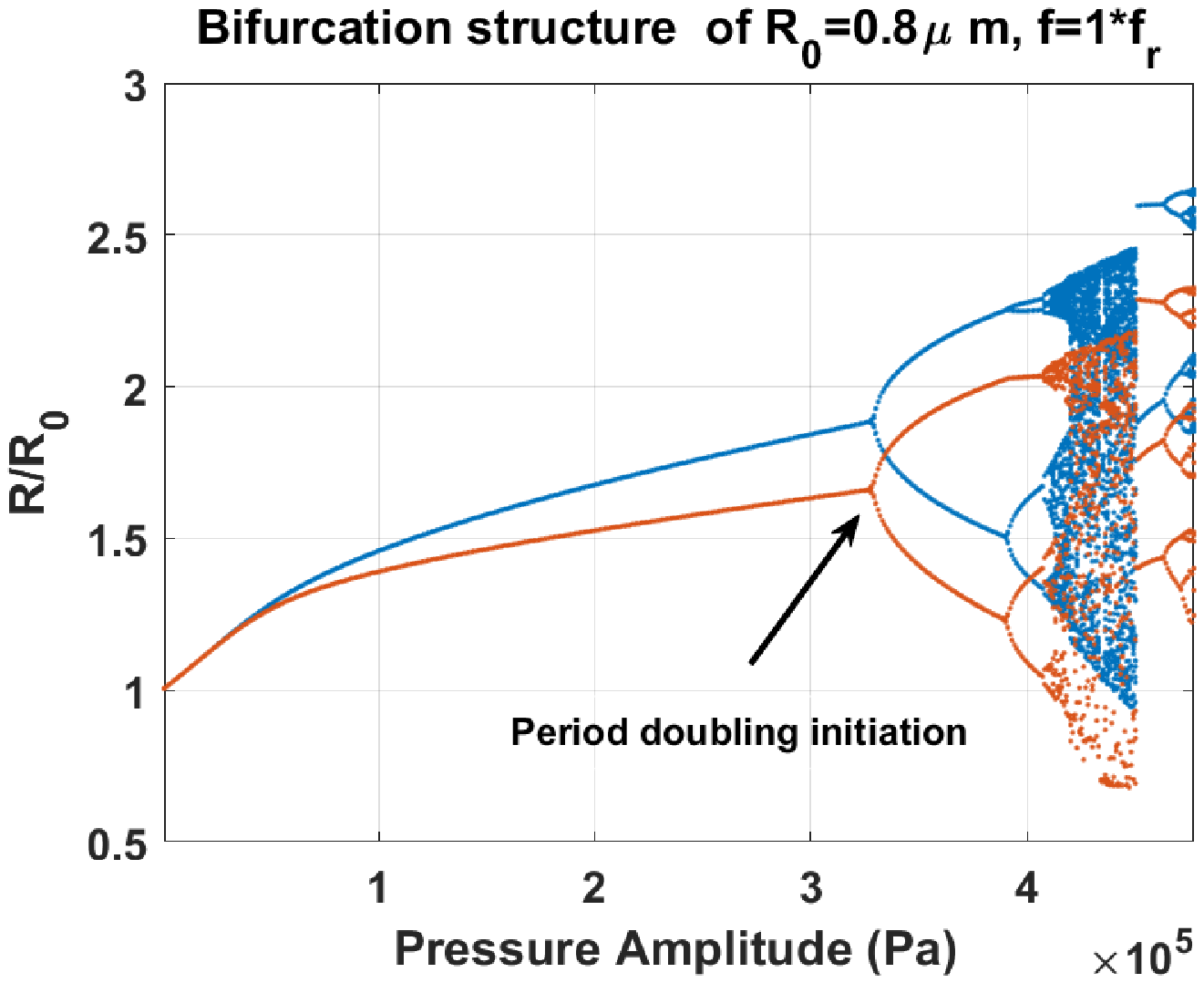}} \scalebox{0.43}{\includegraphics{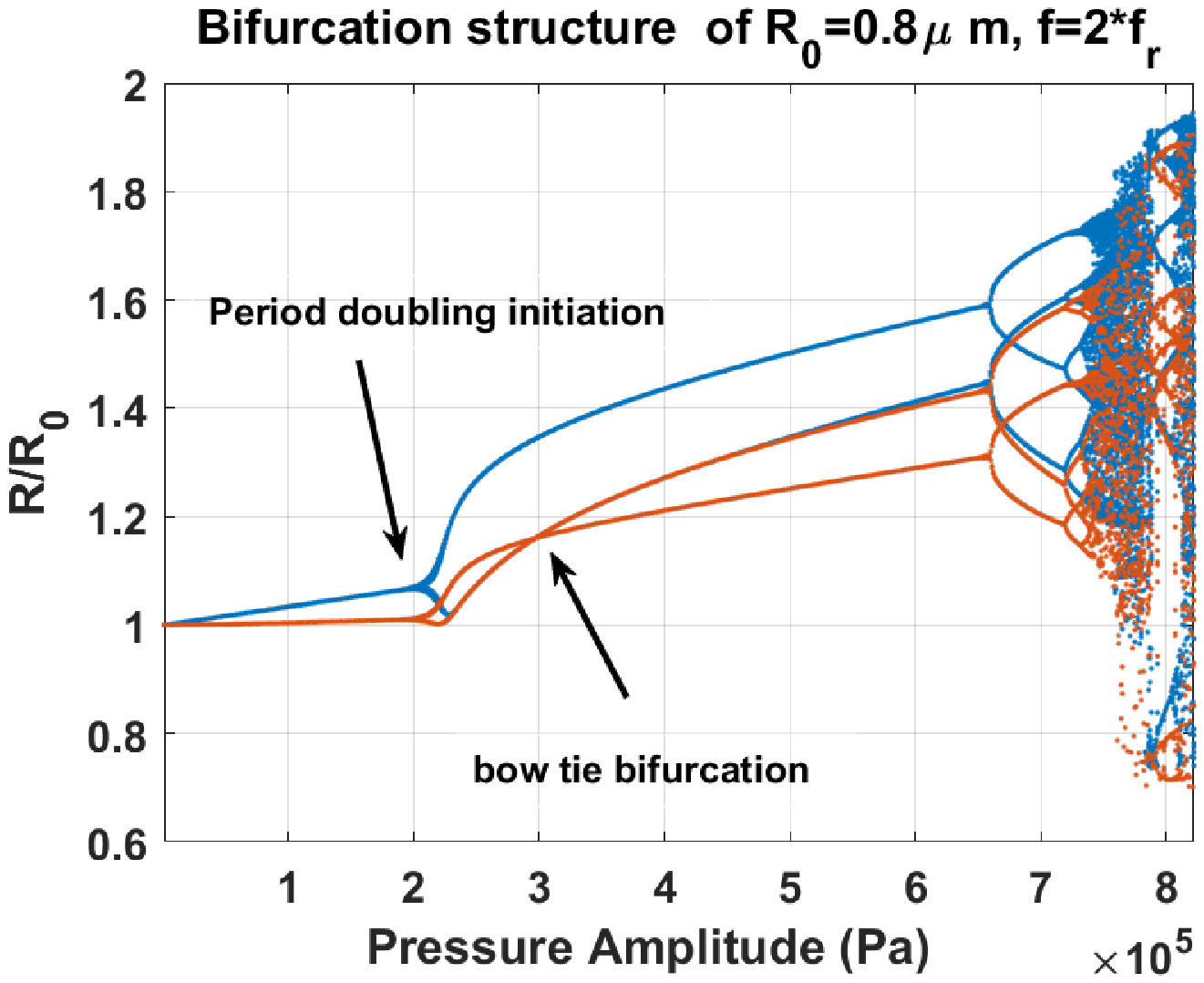}}\\
	\hspace{0.5cm} (a) \hspace{6cm} (b)\\
	\scalebox{0.43}{\includegraphics{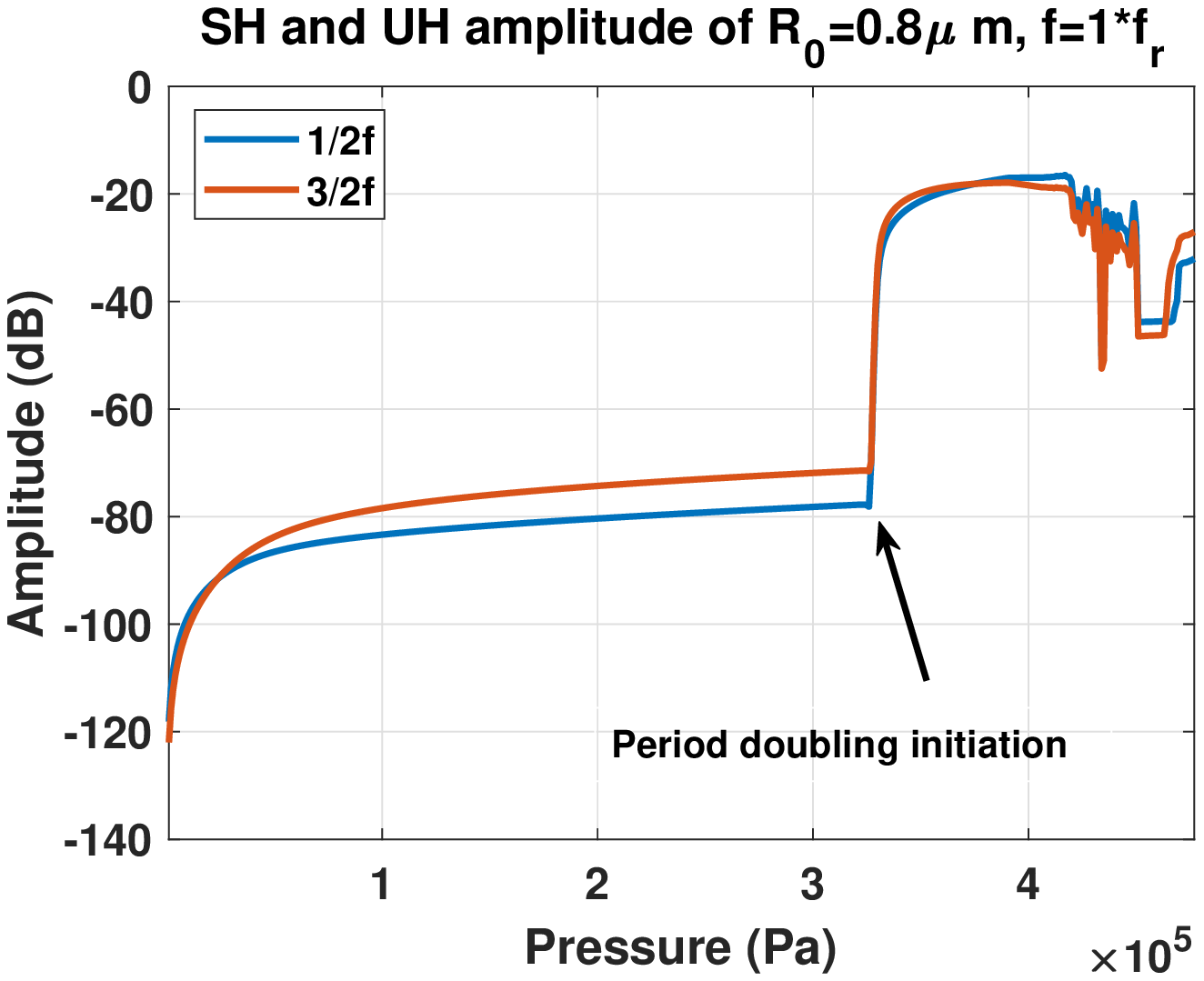}} \scalebox{0.43}{\includegraphics{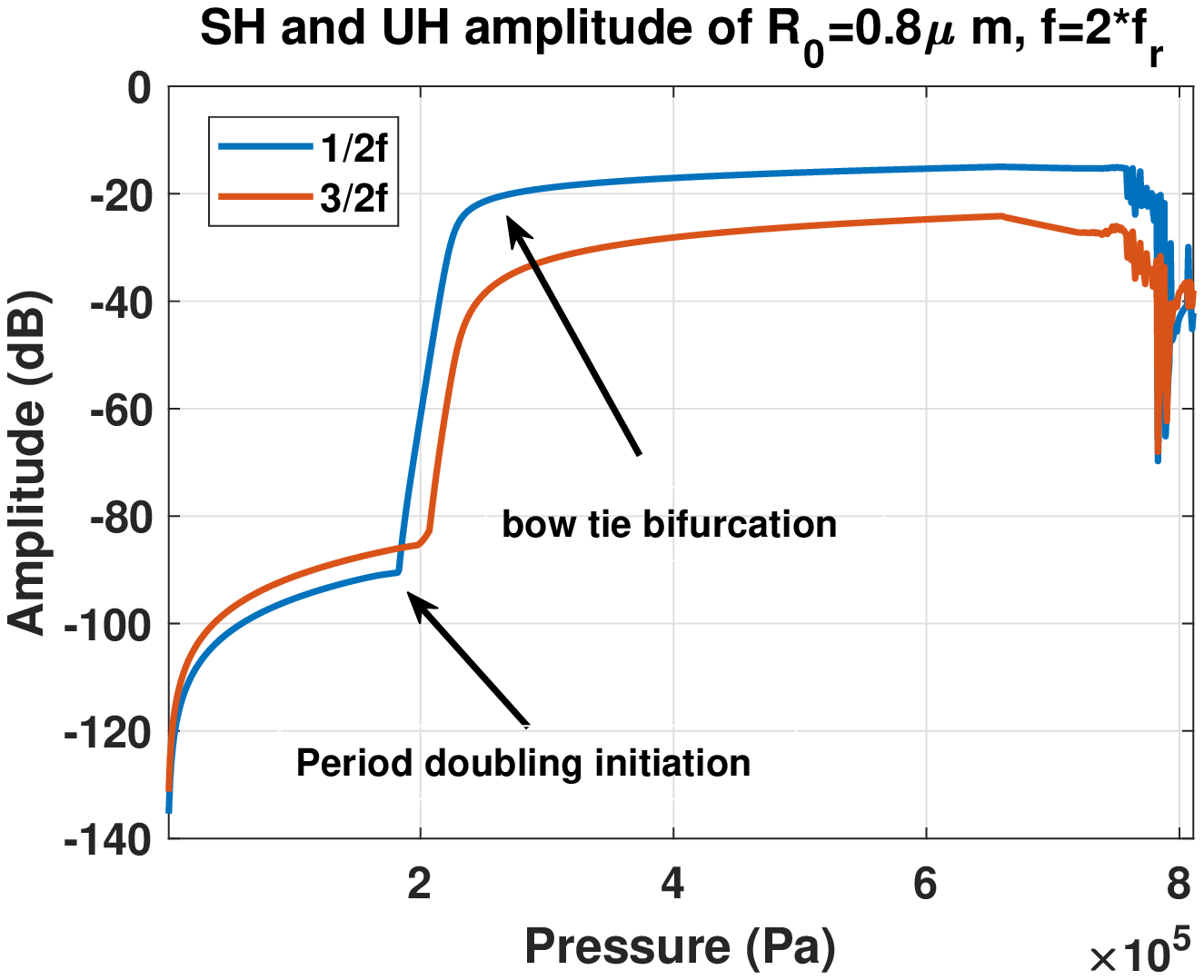}}\\
	\hspace{0.5cm} (c) \hspace{6cm} (d)\\
	\caption{Bifurcation structure of bubble with $R_0=800 nm$ driven by a) its resonance frequency ($f_r$), b)$2f_r$. SH and UH component of the backscattered signal when c)$f=f_r$, d)$f=2f_r$.}
\end{center}
\end{figure*}
In order to have a better understanding on the effect of period doubling  and chaos on the SH and UH emissions of the bubble oscillations, Figure 5 plots the bifurcation structure of $\frac{R}{R_0}$ as a function of pressure and the corresponding SH and UH amplitude of the backscattered pressure side by side. The bubble has an initial radius of 800 nm (this size has been chosen as a sample and because different stages of the dynamical process can be seen more clearly). The left column shows the dynamics of the bubble when $f=f_r$ and the right column represents the case of $f=2f_r$. Period doubling in both cases results in the initiation and fast growth of SH amplitudes; as pressure increases, the SH and UH components grow in magnitude and reach a saturation value. This behavior (initiation, growth and saturation) has been also observed experimentally [67].  The P2 oscillations undergo further period doubling cascades to chaos; the occurrence of chaotic oscillations is concomitant with a decrease in SH and UH amplitude which continue rapid fluctuation in amplitude for chaotic oscillations. In case of sonication with $f=f_r$ and when PD occurs, UH oscillations grow faster than the SH oscillations; however, for $f=2f_r$, the SH component of the backscattered pressure grows faster than the UH component and becomes stronger in magnitude.
\subsection{Different stages of the two main routes of period doubling and their dynamical properties}
\begin{figure*}
\begin{center}
	\scalebox{0.43}{\includegraphics{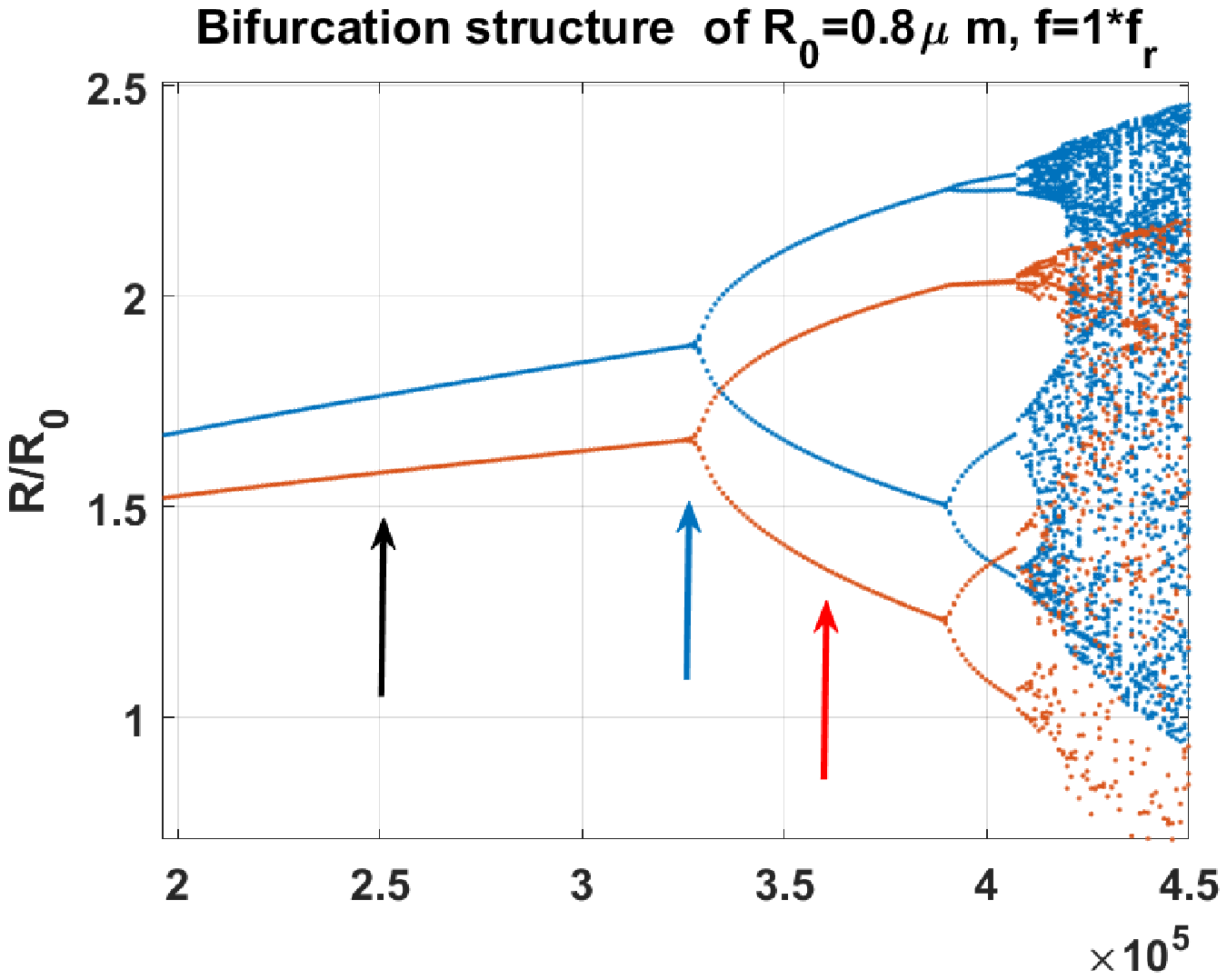}} \scalebox{0.43}{\includegraphics{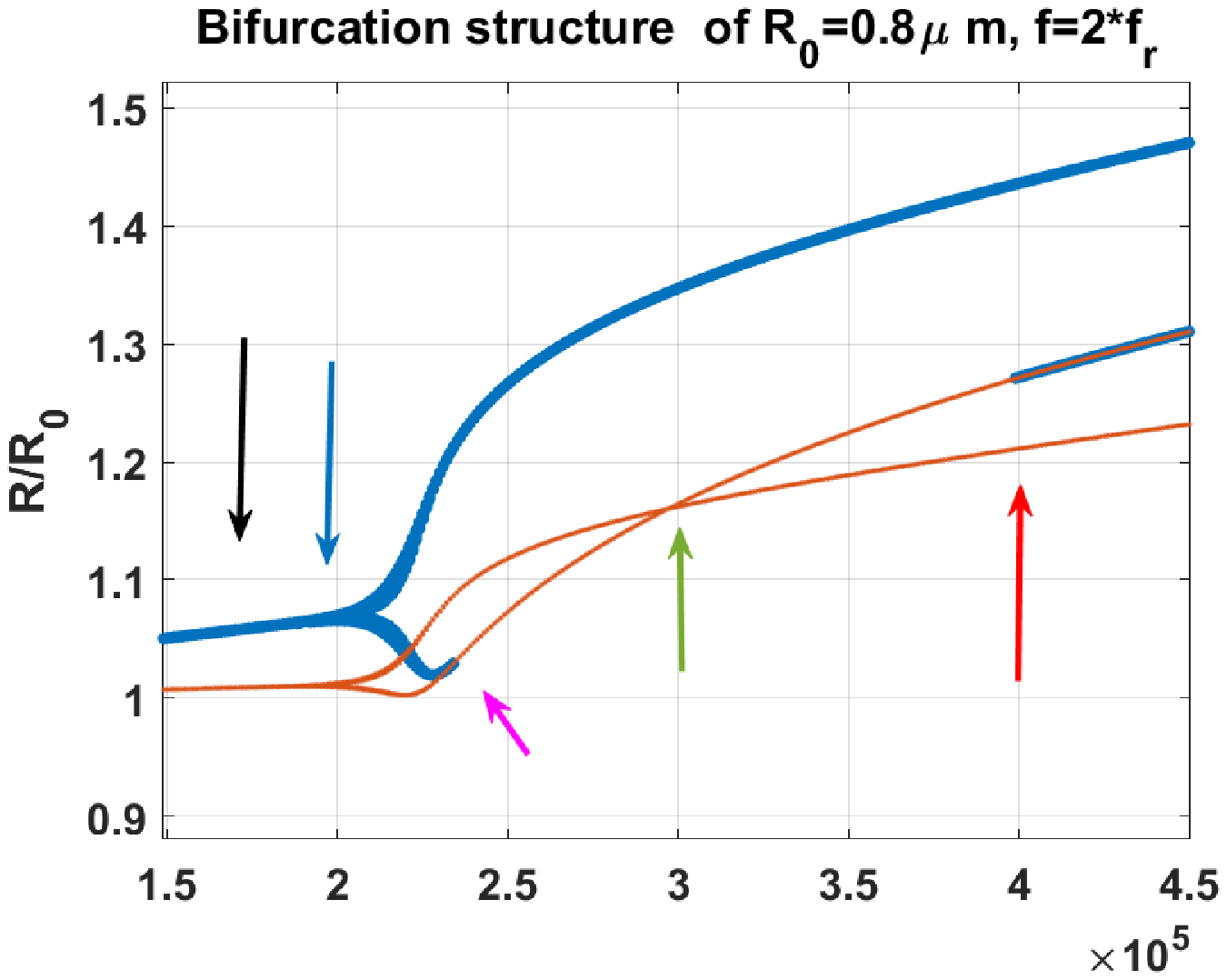}}\\
	\hspace{0.5cm} (a) \hspace{6cm} (b)\\
	\scalebox{0.43}{\includegraphics{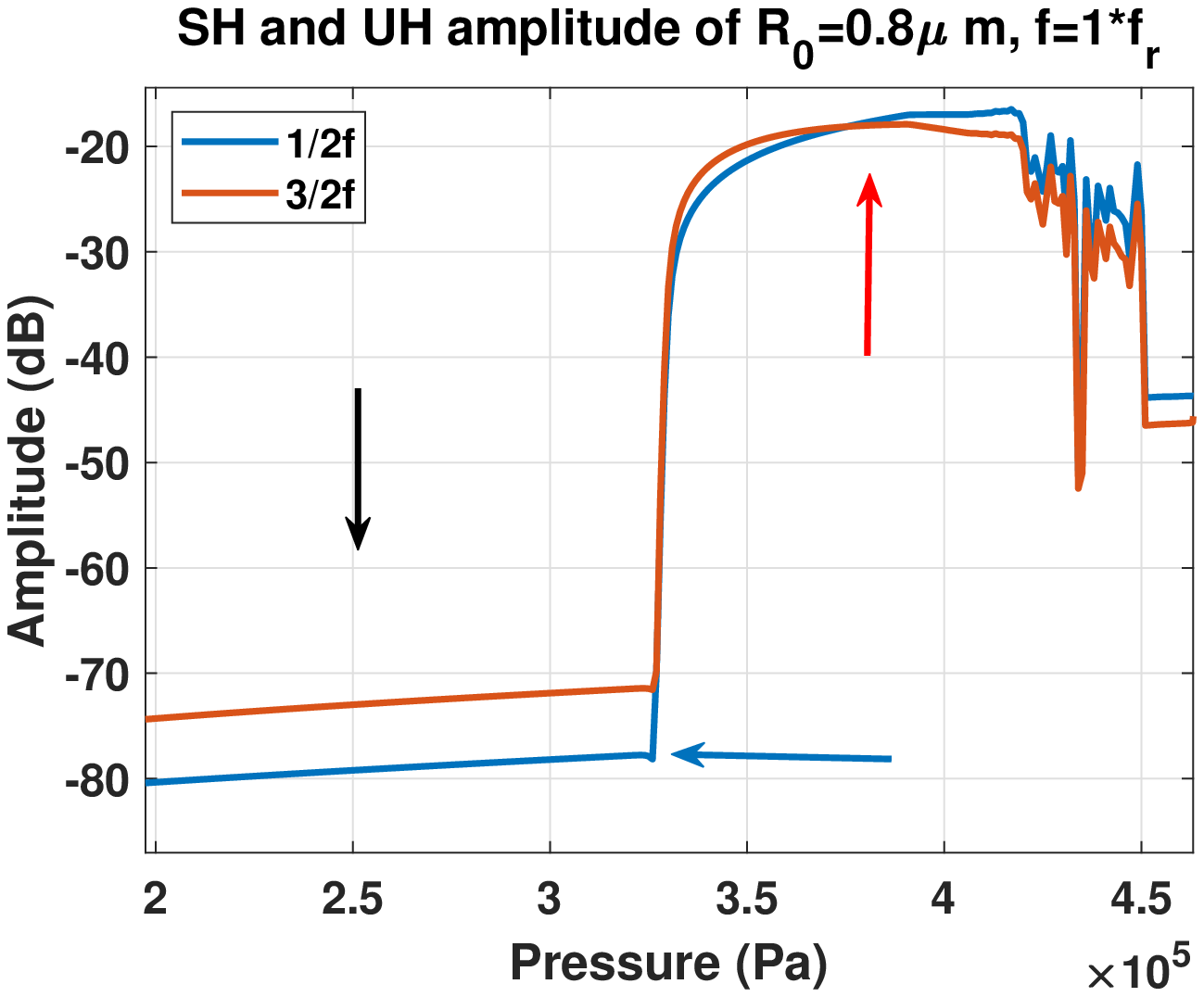}} \scalebox{0.43}{\includegraphics{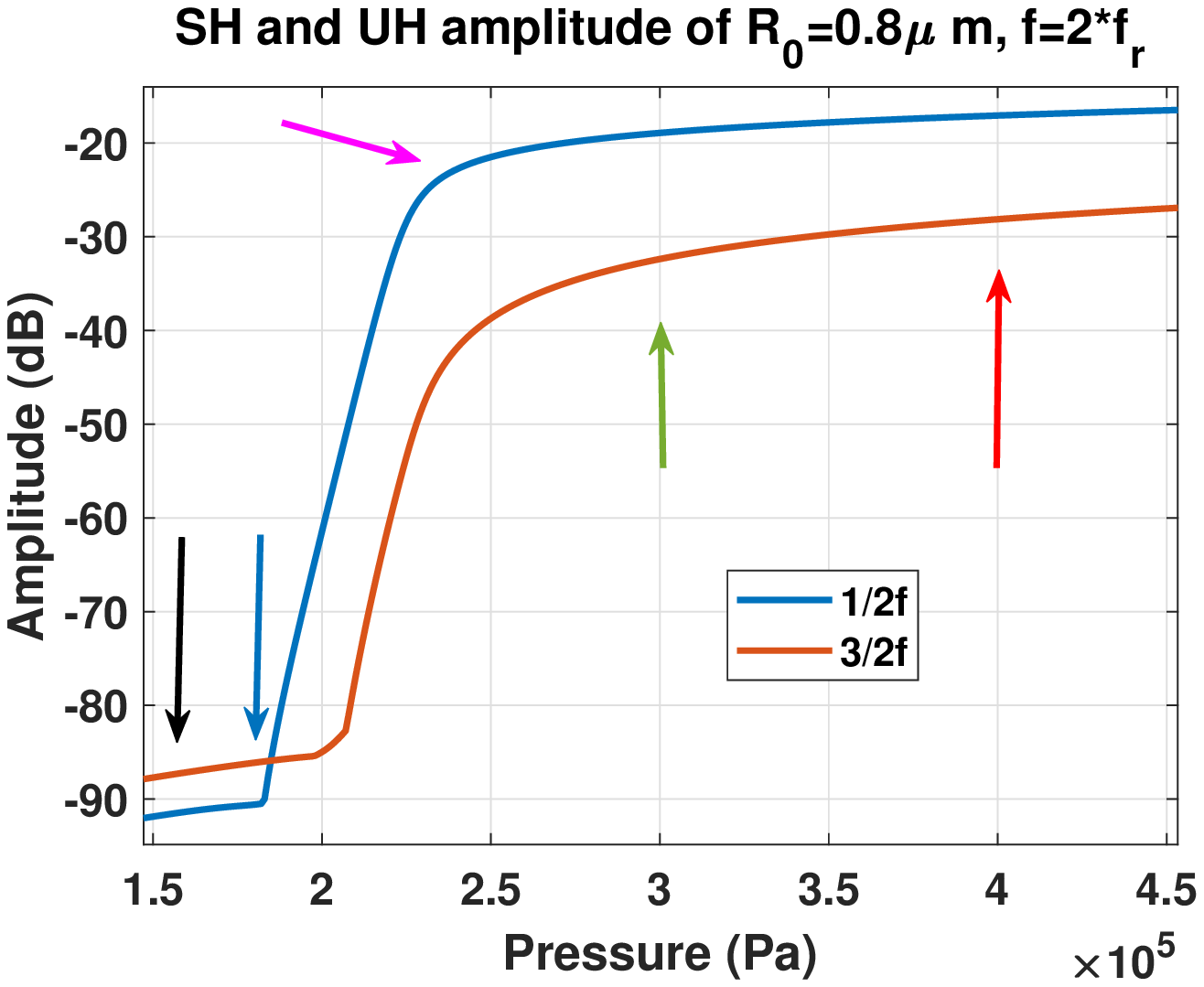}}\\
	\hspace{0.5cm} (c) \hspace{6cm} (d)\\
	\scalebox{0.43}{\includegraphics{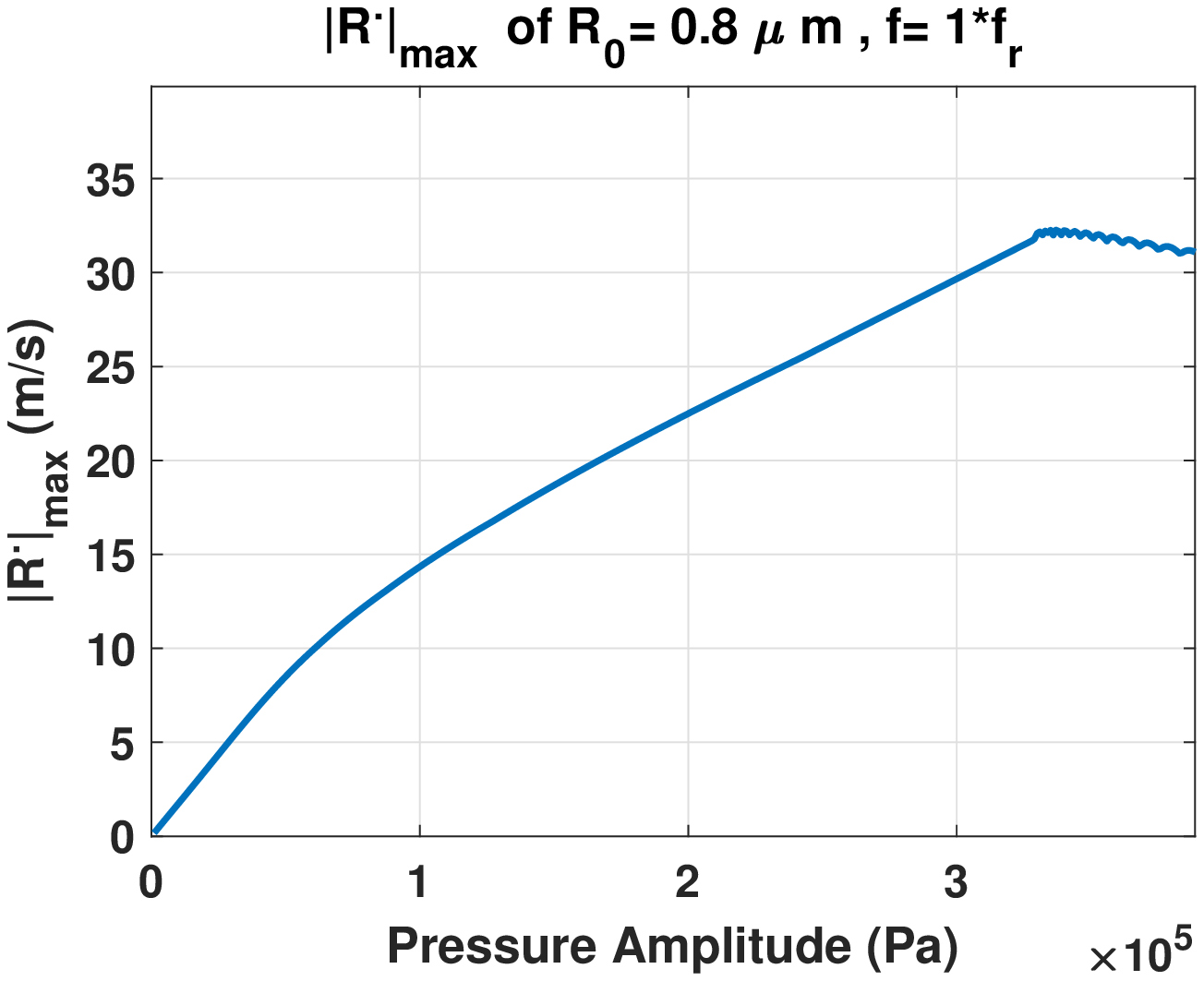}} \scalebox{0.43}{\includegraphics{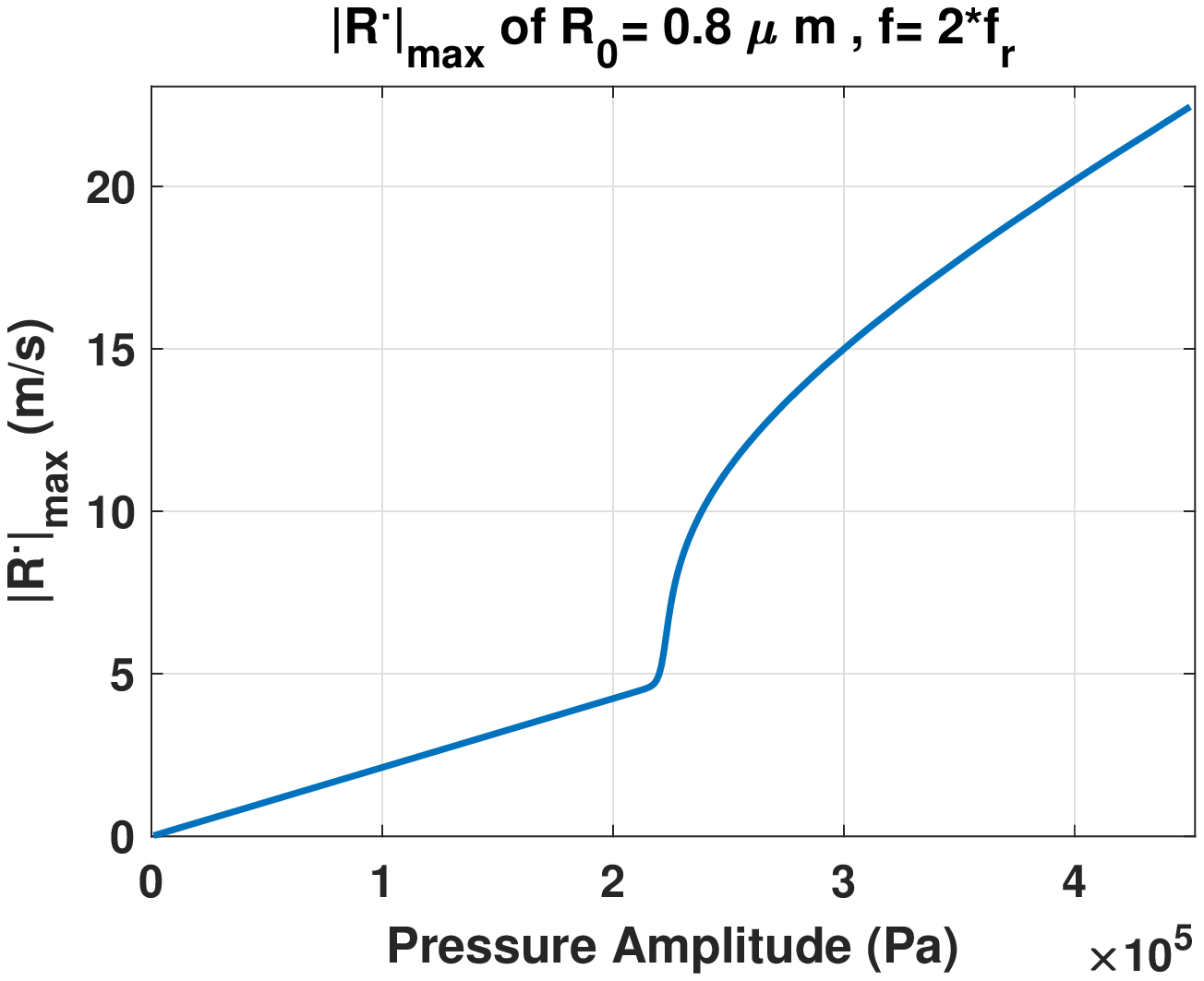}}\\
	\hspace{0.5cm} (e) \hspace{6cm} (f)\\
	\caption{ A closer look at the evolution of the P2 attractor when a) $f=f_r$ and b) $f=2f_r$. The SH and UH component of the signal when c) $f=f_r$, and d) $f=2f_r$. Wall velocity as a function of pressure when e) $f=f_r$, and f) $f=2f_r$ (Maximum wall velocity of oscillations with periods higher than 2 are not displayed)}
\end{center}
\end{figure*}
Focusing on a narrow pressure range allows us to better understand the mechanism of the two different PDs and the corresponding dynamics of SH and UH components of the backscattered signal (Figure 6). In this figure, the bubble has $R_0$=800 nm; the left column represents the case in which $f=f_r$ and the right column represents the case in which $f=2f_r$.\\ 
When bubble is sonicated with $f=f_r$ (Fig. 6a) the bifurcation structure of the bubble has 3 different oscillation regions; the linear oscillation stage (black arrow), the initiation of PD stage (blue arrow) and the P2 oscillations stage (red arrow). The bifurcation structures that are generated using the conventional and maxima methods have concomitant PD and P2 oscillations, indicating the bubble oscillation is a P2 oscillation with two maxima. The corresponding SH and UH components of the backscattered pressure are plotted in fig. 6b and can be categorized in 3 regions which can be described by the absence of SHs and UHs (black arrow), the initiation and fast growth of SH and UH backscatter power concomitant with period doubling in fig. 6a (blue arrow) and the region of UH and SH amplitude saturation (red arrow). Fig 6c shows the maximum amplitude of the wall velocity. The velocity increases monotonically with pressure elevation and undergoes a decrease concomitant with PD. The decrease of wall velocity concomitant with PD when the bubble is sonicated with a frequency close to $f_r$ is studied in detail in [15].\\
Fig 6d shows the bifurcation structure of the $\frac{R}{R_0}$ of the bubble when it is sonicated with $f=2f_r$; the dynamics of the PD bifurcation is different from the case of sonication with $f=f_r$ (Fig 6a). The bubble initially starts with period 1 (black arrow). The corresponding SH and UH amplitude in Fig 6d are weaker. At~ 165 kPa the bubble undergoes a PD which is concomitant in both bifurcation diagrams. As soon as PD occurs, the SH amplitude and wall velocity undergo a rapid increase (Fig 6d and Fig 6f).  As the pressure increases, one of the maxima in the bifurcation structure (red curve) disappears while the conventional bifurcation still keeps P2 oscillations (purple arrow). Just before the disappearance, the value of the maxima overlaps one of the solutions in the blue curve; this implies wall velocity of bubble oscillations are in phase with acoustic force once every two acoustic cycles; thus, SH resonance is generated. The disappearance of the second maxima (at ~245 kPa) is concomitant by a fast increase in the UH amplitude (purple arrow). Above this pressure, the growth rate of SH amplitude changes and starts to plateau. As the pressure increases, the two solutions in the conventional bifurcation diagram (blue curve) converge and at $\approxeq$ 297 kPa. At this pressure both solutions have one value; this results in only one point in the blue curve (green arrow). This is concomitant with further decrease of the growth rate of the SH and UH components of the backscattered pressure.\\
\begin{figure*}
\begin{center}
	\scalebox{0.3}{\includegraphics{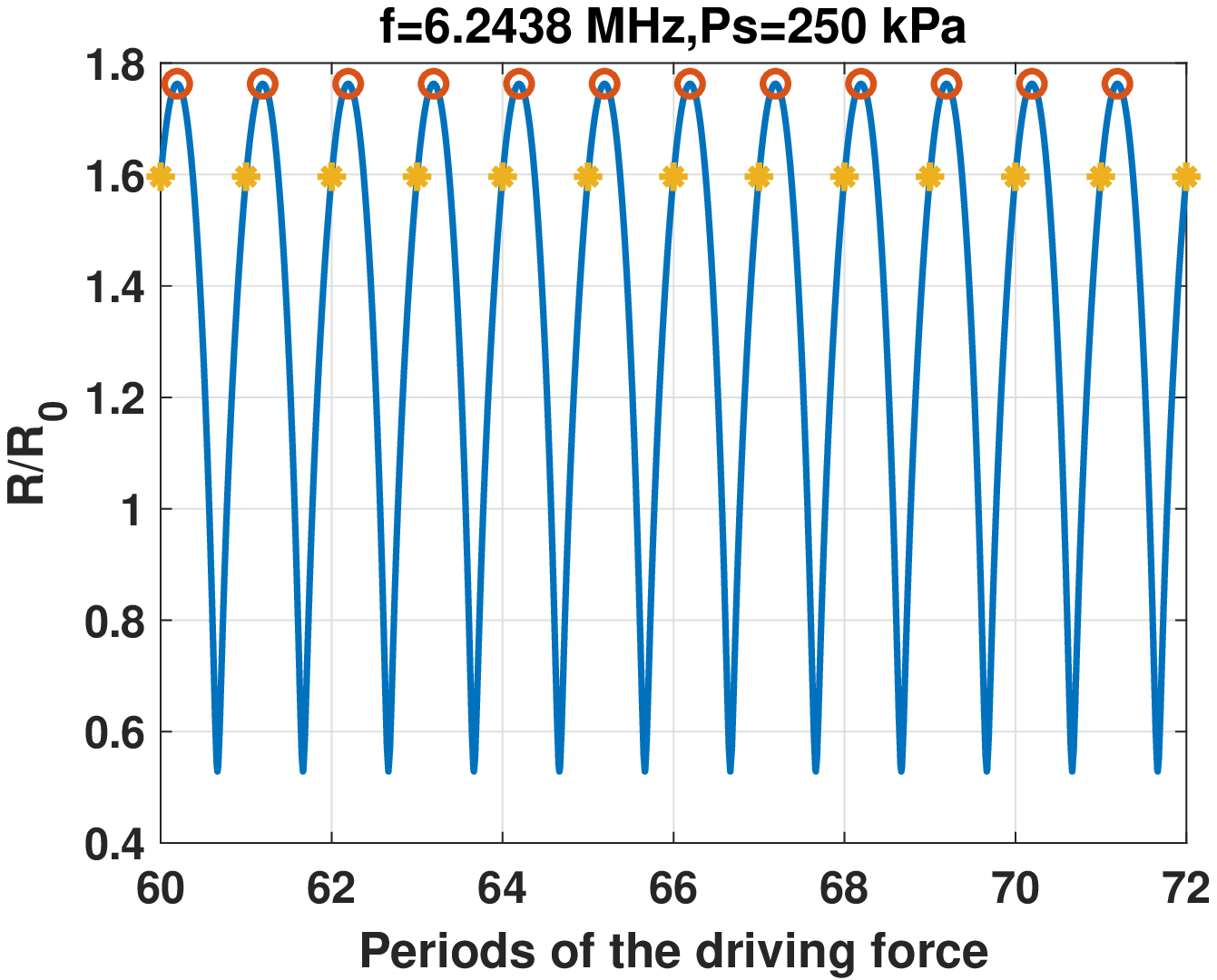}} \scalebox{0.3}{\includegraphics{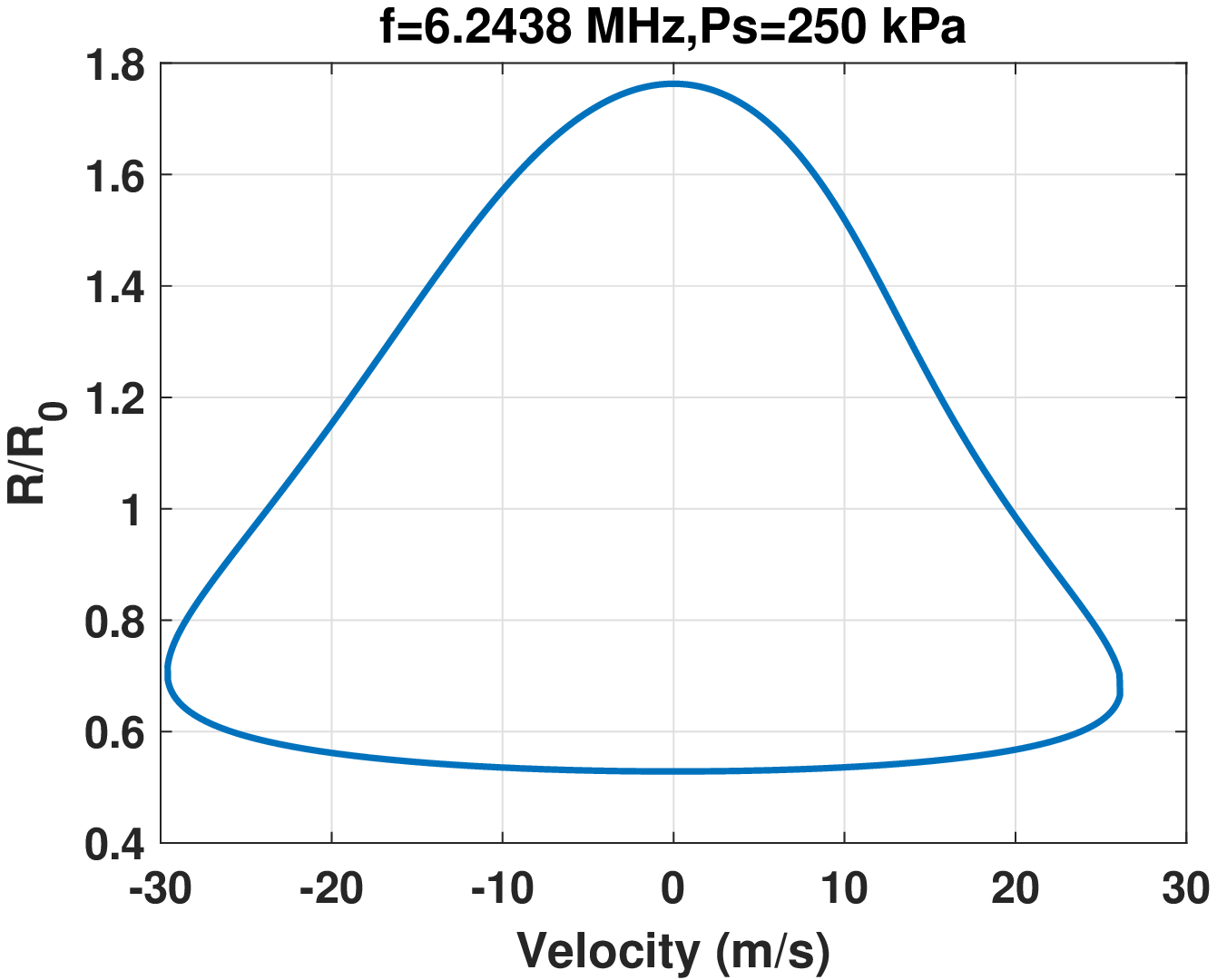}}\scalebox{0.3}{\includegraphics{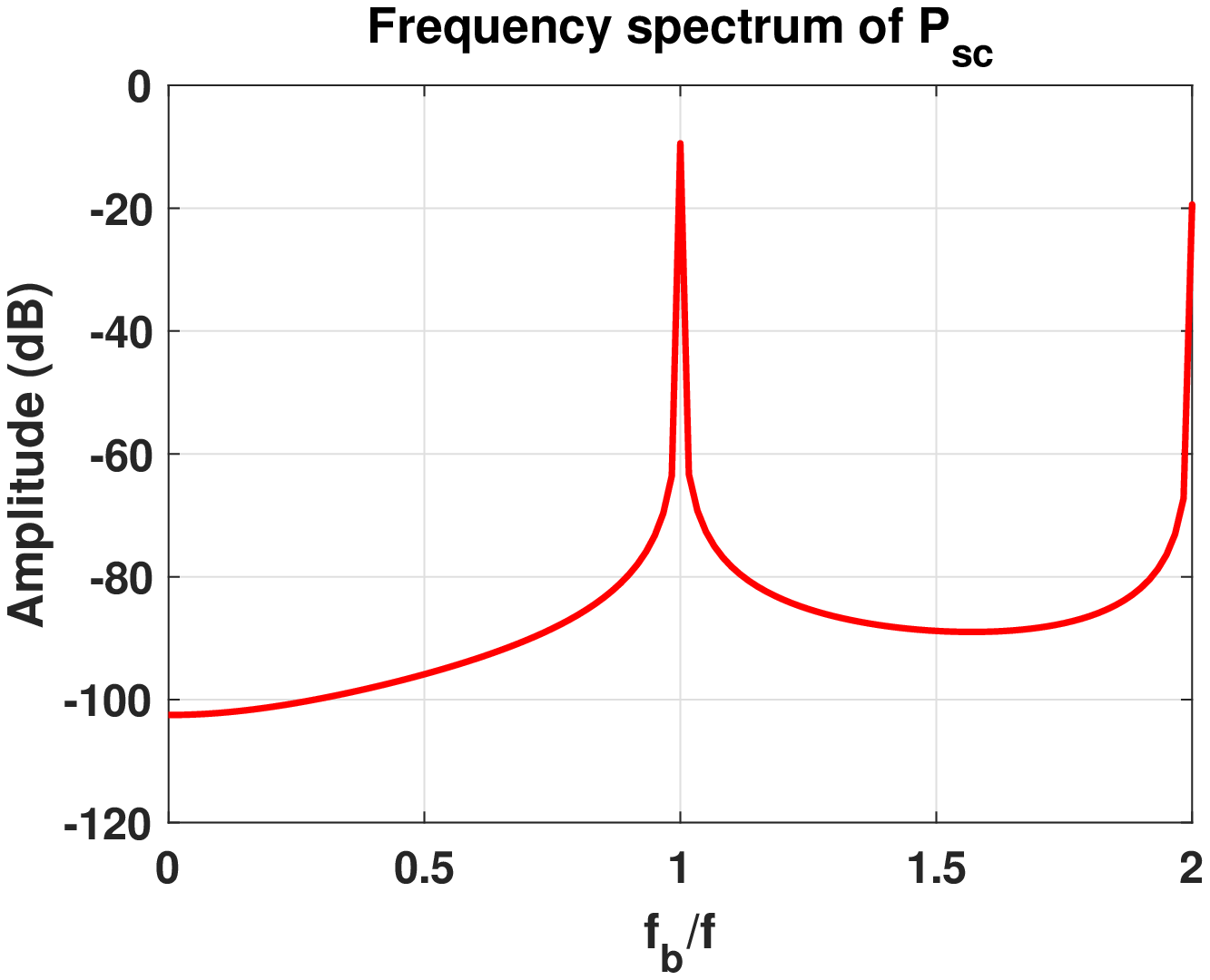}} \\
	(a) \hspace{4cm} (b) \hspace{4cm} (c)\\
	\scalebox{0.3}{\includegraphics{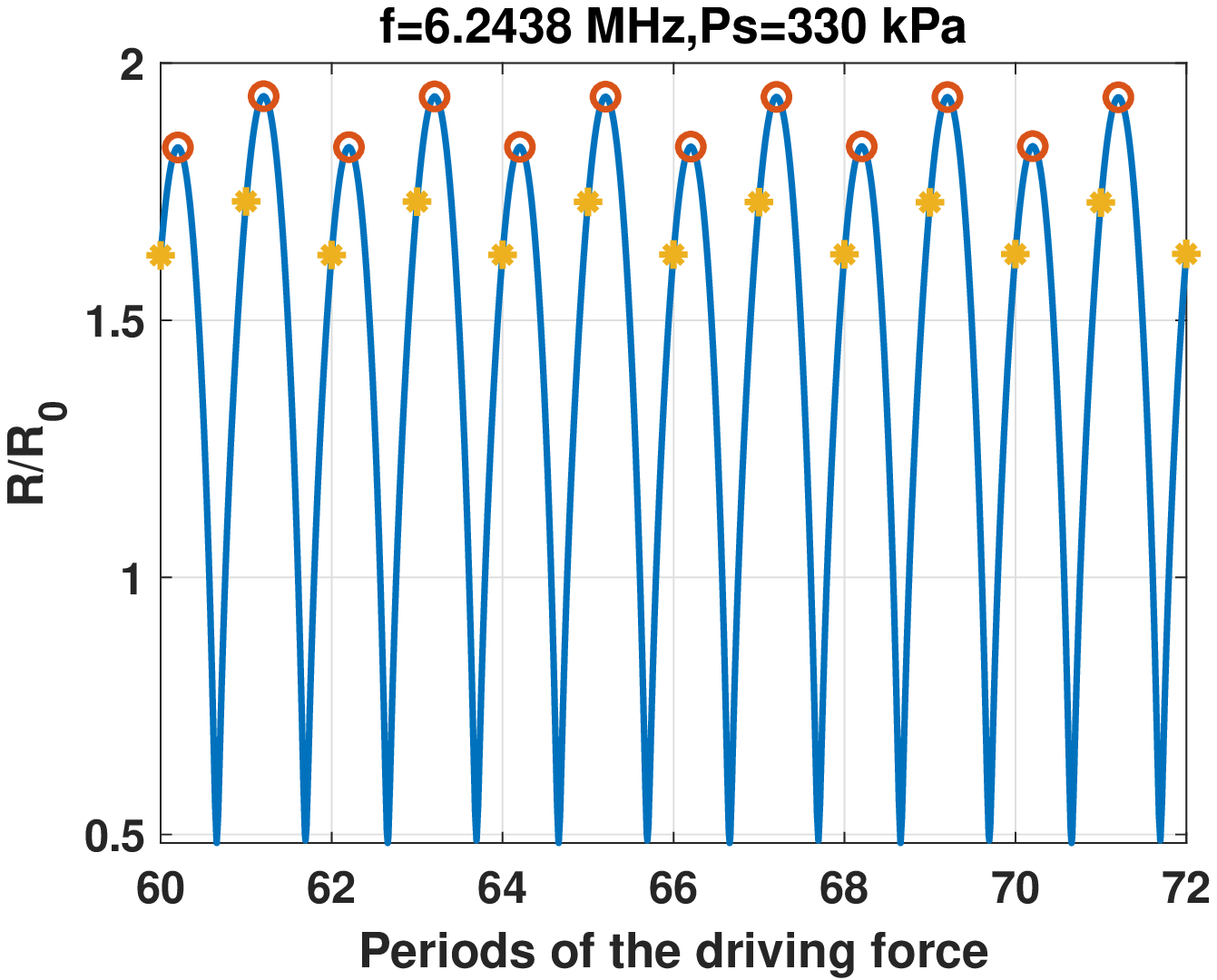}} \scalebox{0.3}{\includegraphics{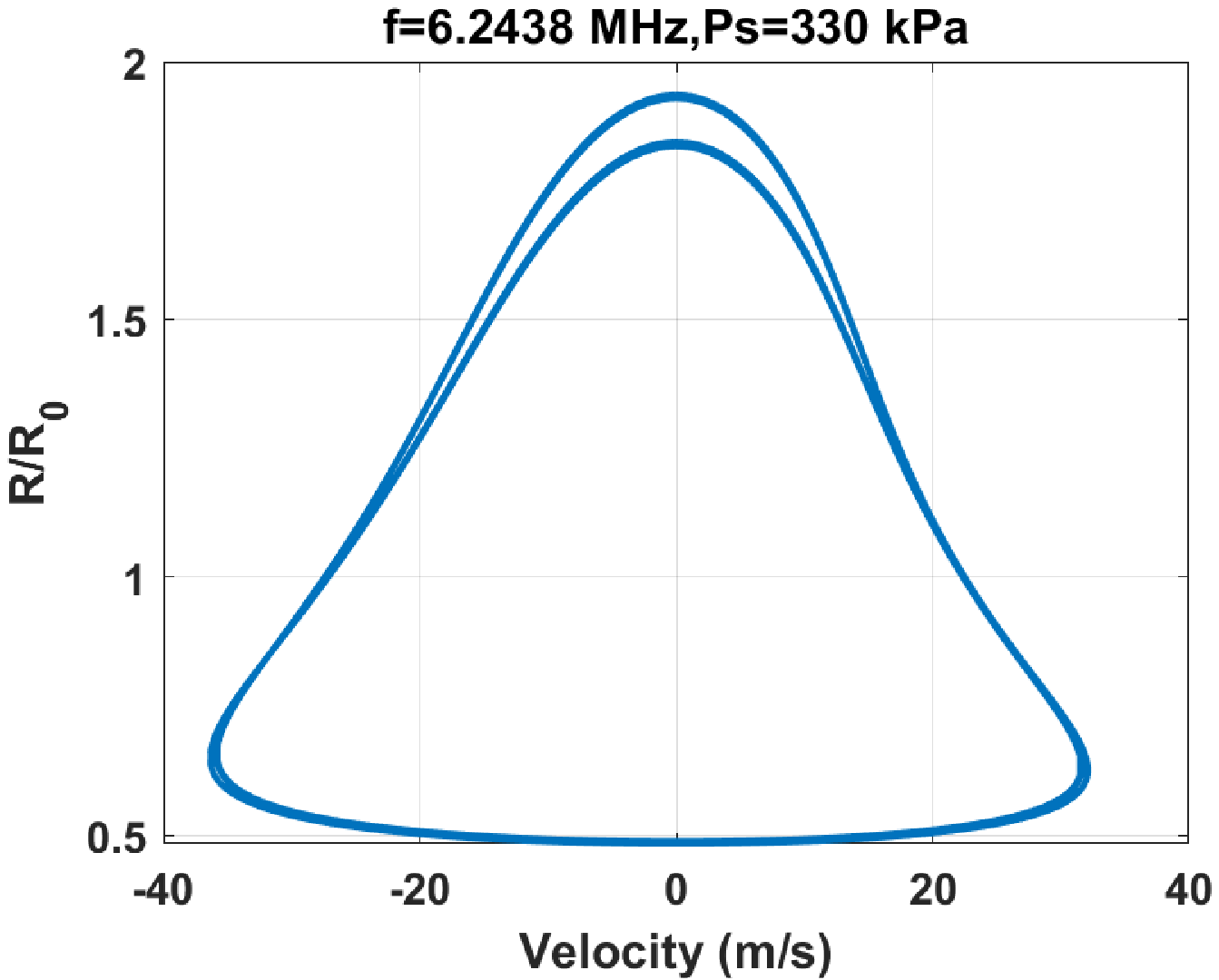}}\scalebox{0.3}{\includegraphics{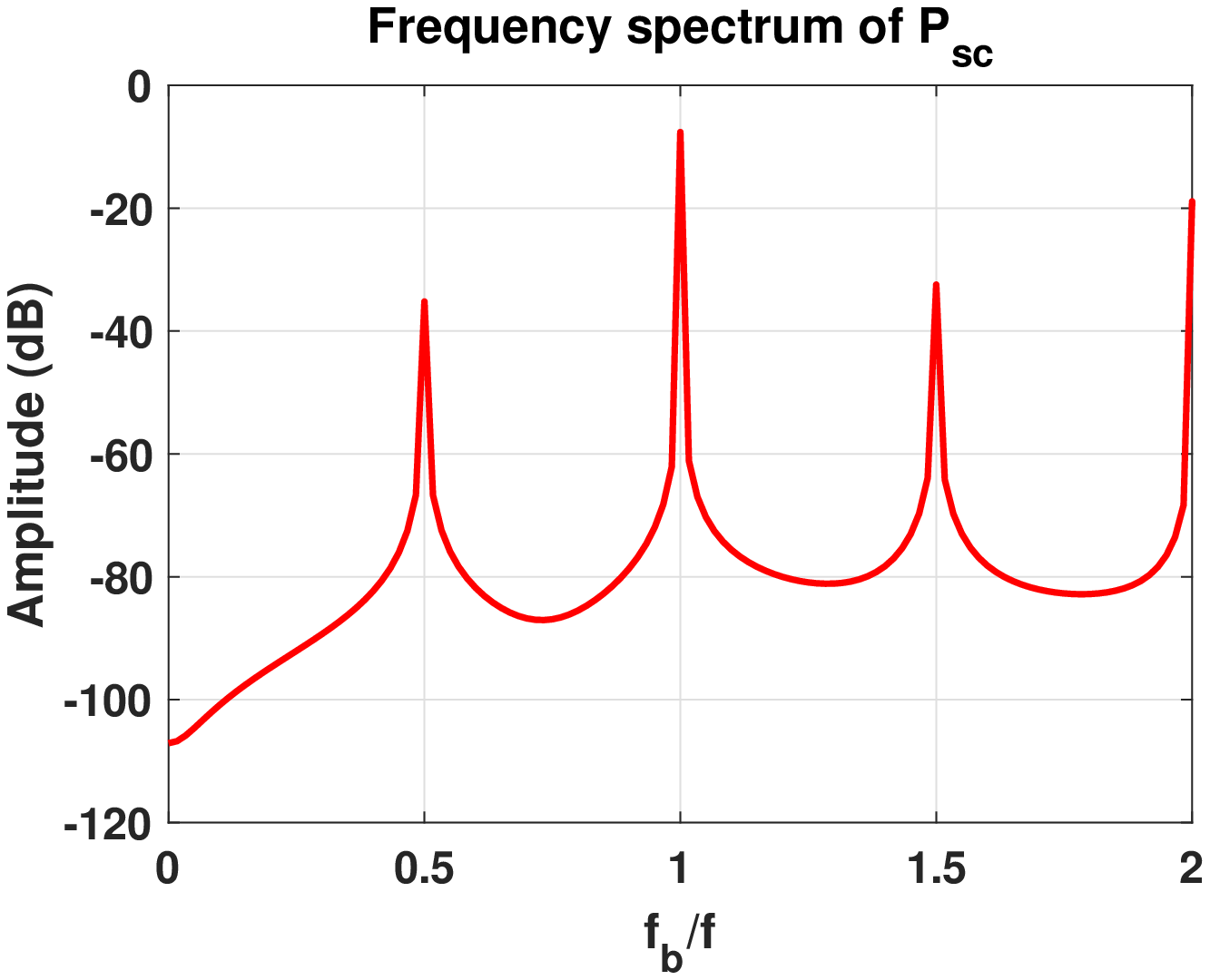}} \\
	(d) \hspace{4cm} (e) \hspace{4cm} (f)\\
	\scalebox{0.3}{\includegraphics{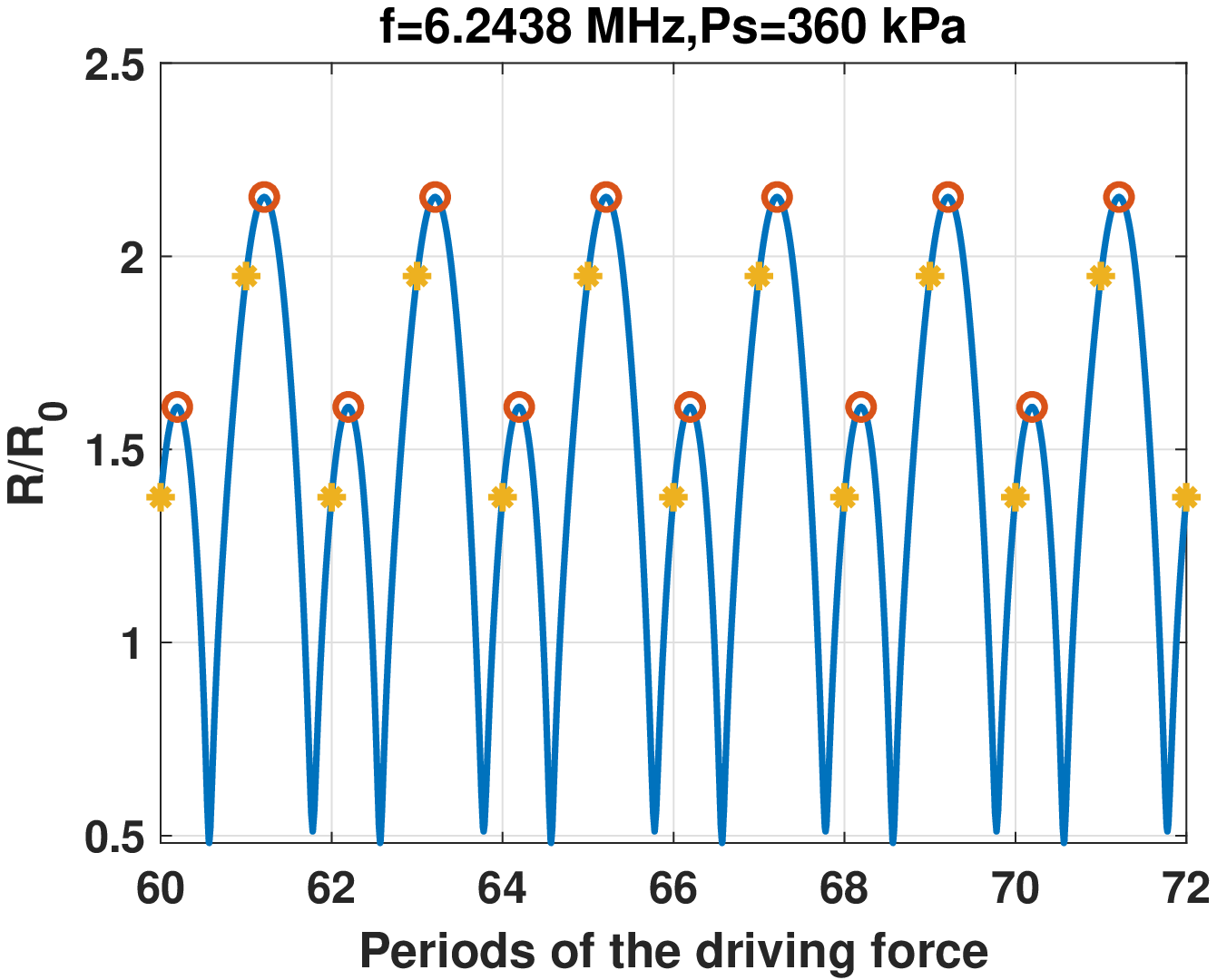}}\scalebox{0.3}{\includegraphics{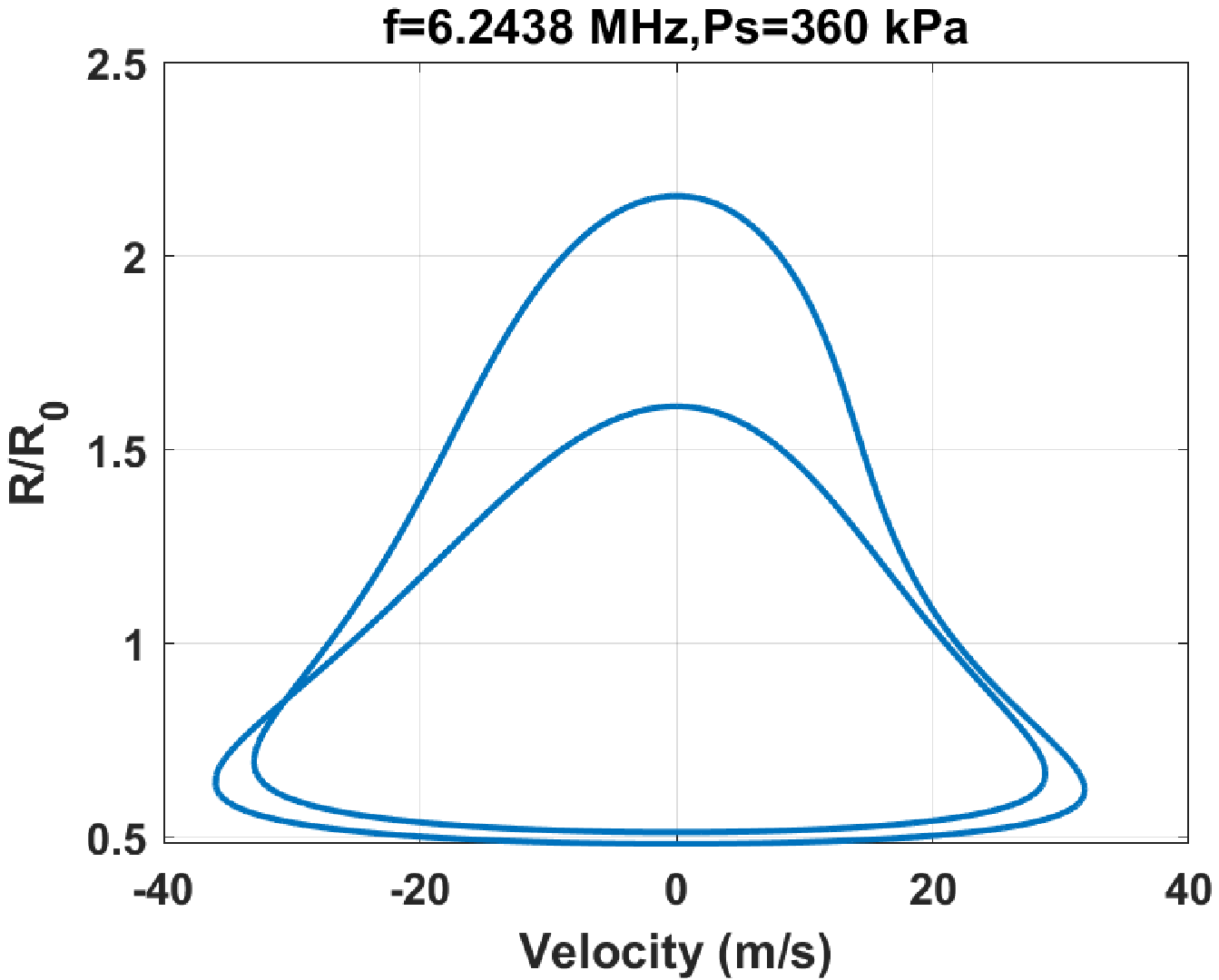}}\scalebox{0.3}{\includegraphics{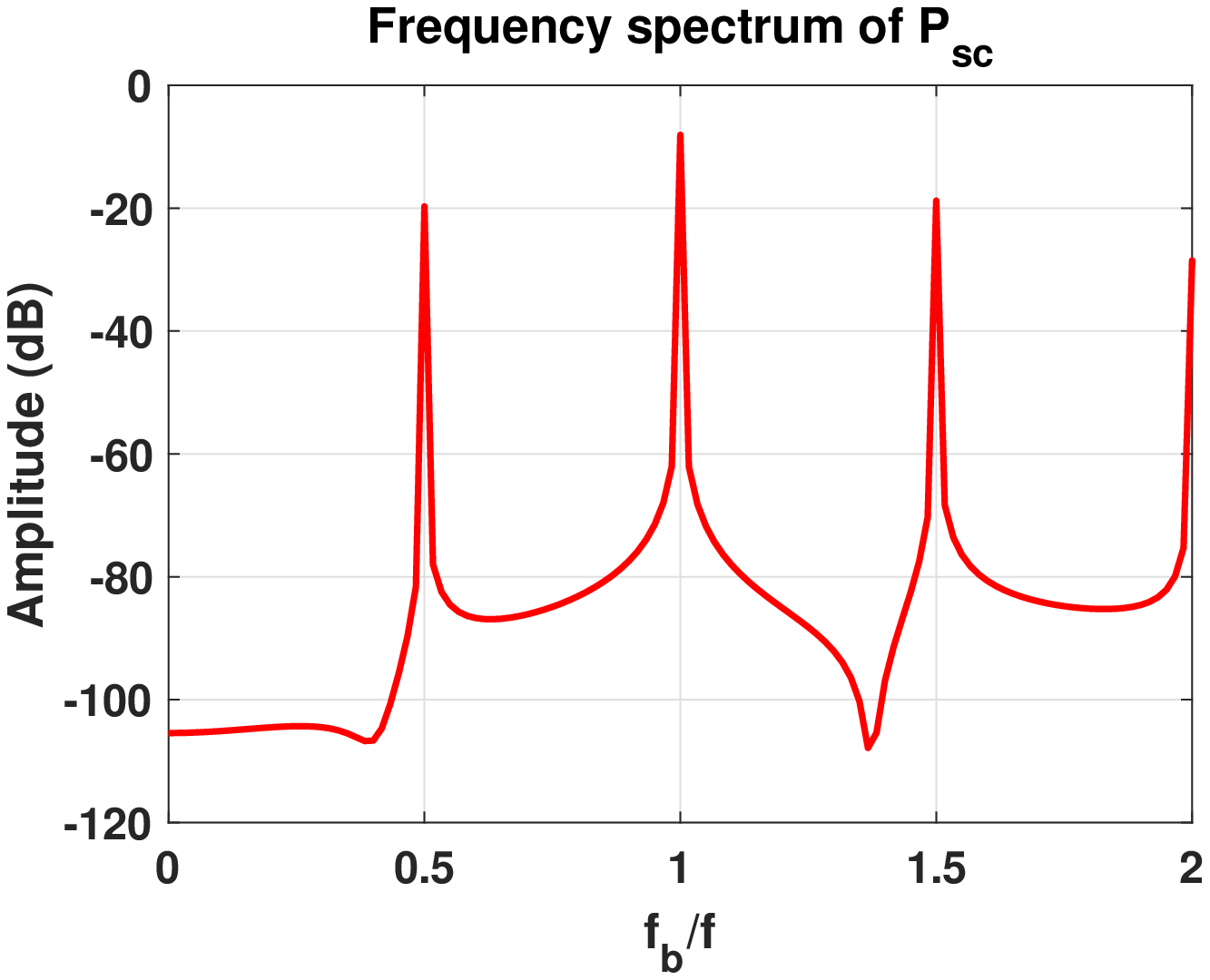}} \\
	(g) \hspace{4cm} (h) \hspace{4cm} (i)\\
	\caption{ Evolution of the dynamics of the bubble with $R_0=800 nm$ when $f=f_r$. Diagrams are plotted for three difference pressures of interest (250, 330 and 360 kPa, see Figure 6a). a) Radial oscillations, b) phase portrait diagram and c) frequency spectrum of the $P_{sc}$ when $P_s$=250 kPa. The d) radial oscillations, e) the phase portrait diagram and f) the frequency spectrum of the $P_{sc}$ are plotted when $P_s$=330kPa.The g) radial oscillations, h) the phase portrait diagram and i) the frequency spectrum of the $P_{sc}$ are plotted when $P_s$=360kPa.}
\end{center}
\end{figure*}
To more thoroughly examine the dynamics of PD when $f=f_r$, figure 7 shows the time-series, phase portraits and the frequency spectra of the backscattered pressure at three different pressures (black, blue, and red arrows in fig 6a).  When $P_A=250$ kPa, the oscillations are P1; fig 7a shows that the signal has one maximum (red circle) and $\frac{R}{R_0}$ has one single value at the end of each period of the acoustic driving force. The phase portrait (constructed over the last 40 cycles of a 200 cycle pulse) shown in fig. 7b is a bell shape orbit consisting of only one loop. There is no distinct SH component in the frequency spectrum of the Psc shown in fig. 7c.\\ 
At $P_A=330$ kPa (fig 7d), oscillations are of P2. There are two maxima and two distinct values for $\frac{R}{R_0}$ at the end of each driving period (shown in red and yellow circles respectively). The phase portrait in fig 7e, consists of two bell shape orbits with one enclosing the other. The frequency spectrum of the $P_{sc}$ in fig 7f now has distinct SH and UH components with UH component stronger than the SH component. As pressure increases the separation between the two distinct peaks of the $\frac{R}{R_0}$ curve increases (fig 7g), as do the two orbits of the phase portrait (fig 7h). In this case, the SH and UH components of the $P_{sc}$ grow stronger with increasing pressure.\\ 
When sonicated with $f=2f_r$, the period doubling can be analyzed in 5 pressure ranges (indicated by the black, blue, purple, green and red arrows in fig 6b).\\ Fig 8 shows the time series, phase portraits and the frequency spectrum of the $P_{sc}$ of a bubble with $R_0=800 nm$ when $f=2f_r$. Fig 8a-c shows the time series, phase portrait and the frequency spectrum of the $P_{sc}$ when $P_A$=205 kPa; the oscillations are of P1, the phase portrait is an ellipsoidal orbit of only one loop and the frequency spectrum lacks any distinct SH or UH component.\\
Right after the generation of PD at $P_A$=205 kPa, the radial oscillations have two maxima (red circles) and $\frac{R}{R_0}$ has two distinct values at the end of each period (yellow circles). When PD occurs radial oscillations are much smaller ($<$8 $\%$ expansion ratio) when $f=2f_r$ (fig 8d) compared to when $f=f_r$ with 195 $\%$ expansion ratio (fig 7d). The phase portrait in fig 8e consists of two ellipsoidal orbits with the bigger one giving birth to the smaller orbit. The frequency spectrum of the $P_{sc}$ in fig 8f shows a distinct SH component indicating the generation of SH oscillations.\\
When pressure is increased one of the maxima of the $\frac{R}{R_0}$ oscillations disappear while the oscillations remain P2 oscillations. Fig 8g shows a representative $\frac{R}{R_0}$ time series of this stage of oscillations (purple arrow in fig 6b) when $P_A$=240 kPa.  The radial oscillations have only one maximum however there are two distinct values (yellow circles) for $\frac{R}{R_0}$ at the end of every period. The phase portrait has an interesting heart –like shape, which is rotated by -90 degrees around the y-axis shown in fig 8h. The frequency spectrum of $P_{sc}$ illustrates distinct SH and UH peaks in fig 8i.\\
\begin{figure*}
\begin{center}
	\scalebox{0.3}{\includegraphics{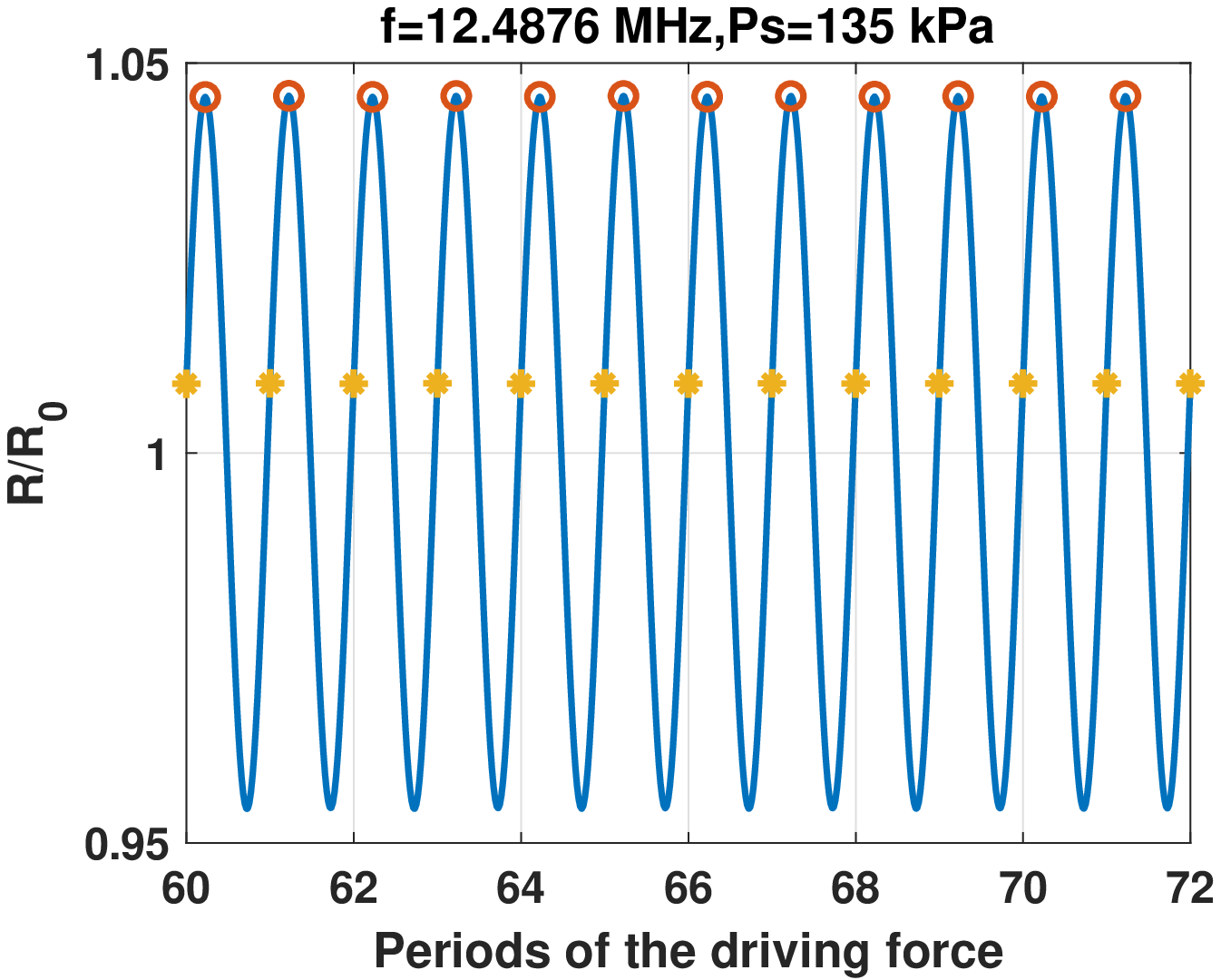}} \scalebox{0.3}{\includegraphics{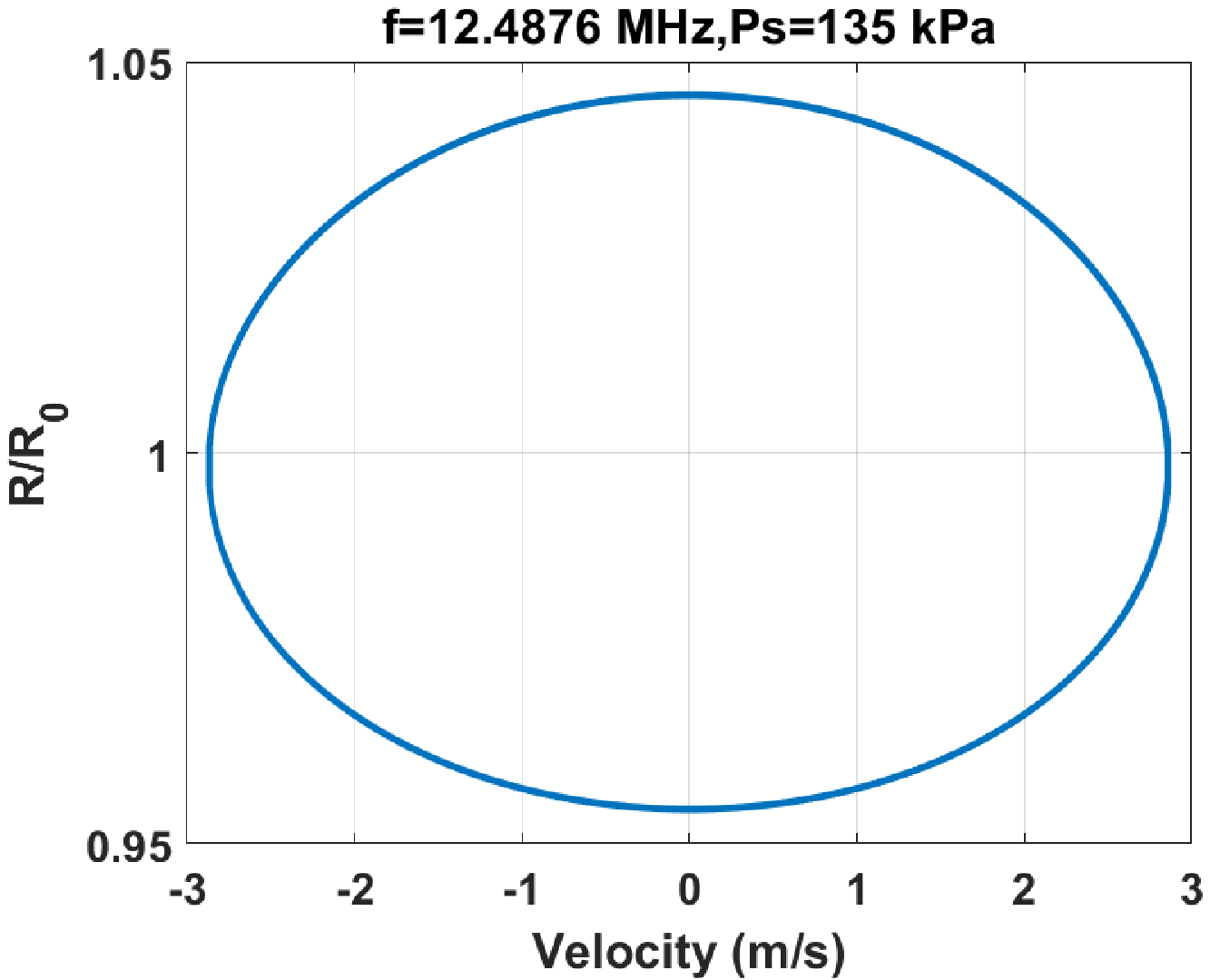}}\scalebox{0.3}{\includegraphics{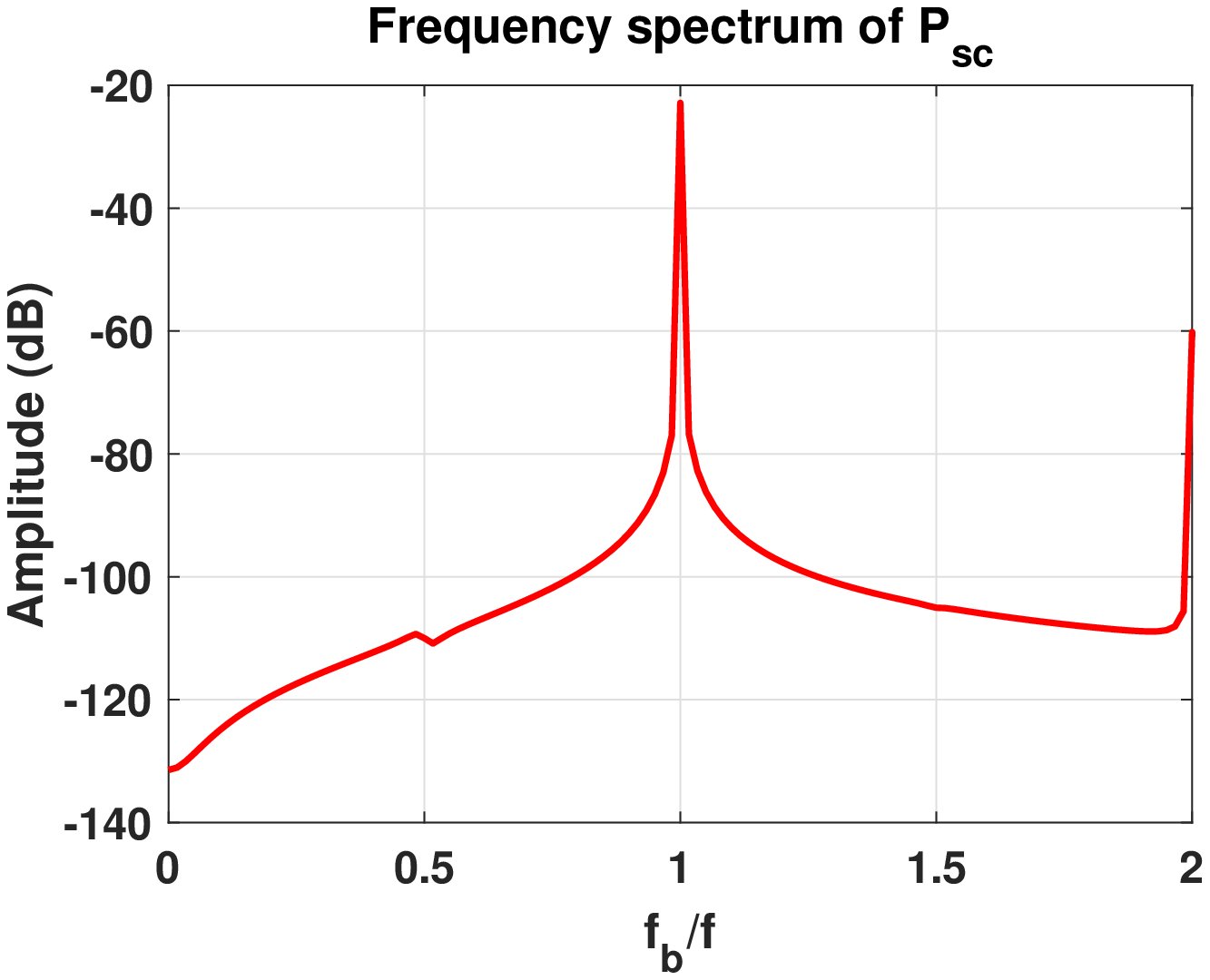}} \\
	(a) \hspace{4cm} (b) \hspace{4cm} (c)\\
	\scalebox{0.3}{\includegraphics{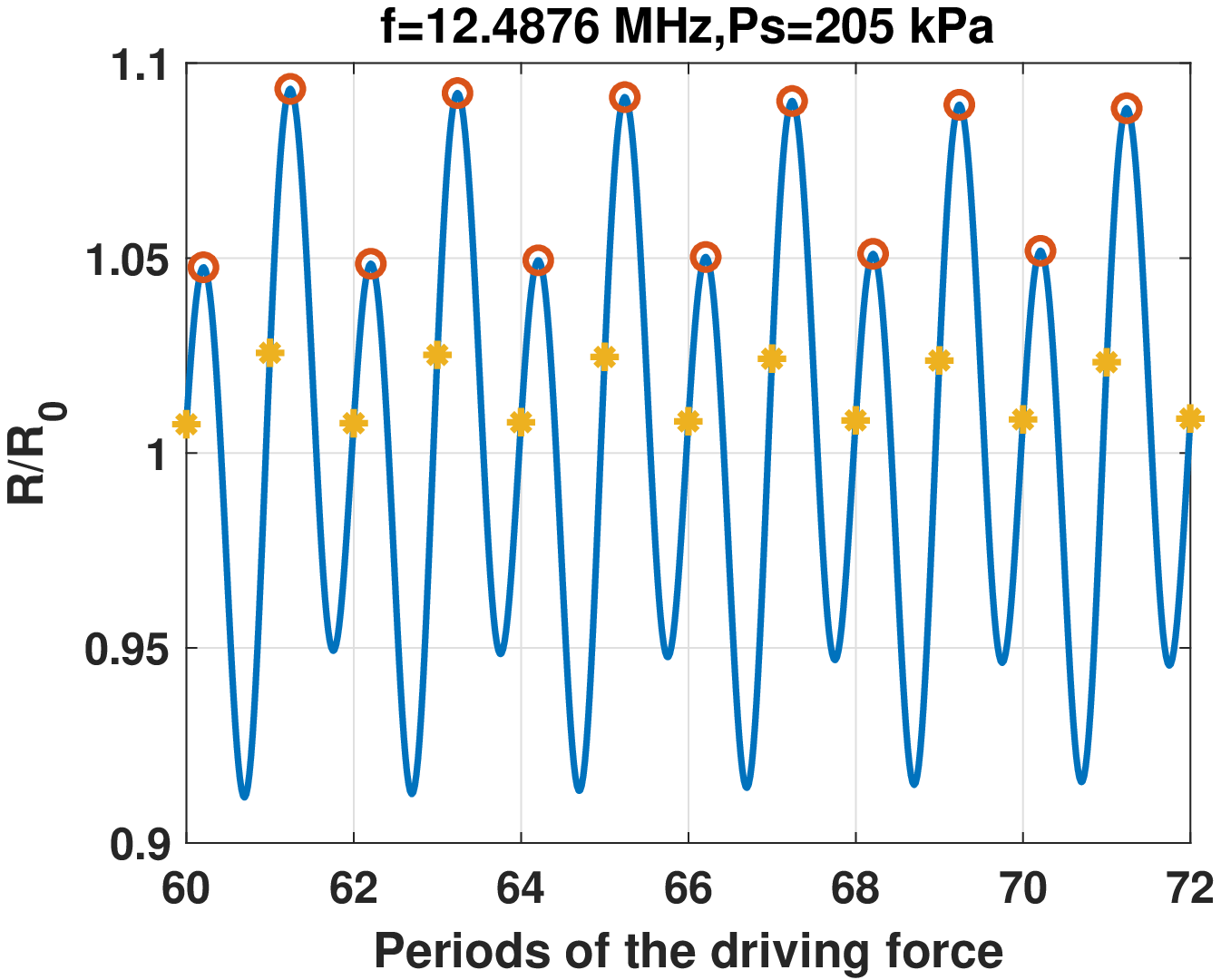}} \scalebox{0.3}{\includegraphics{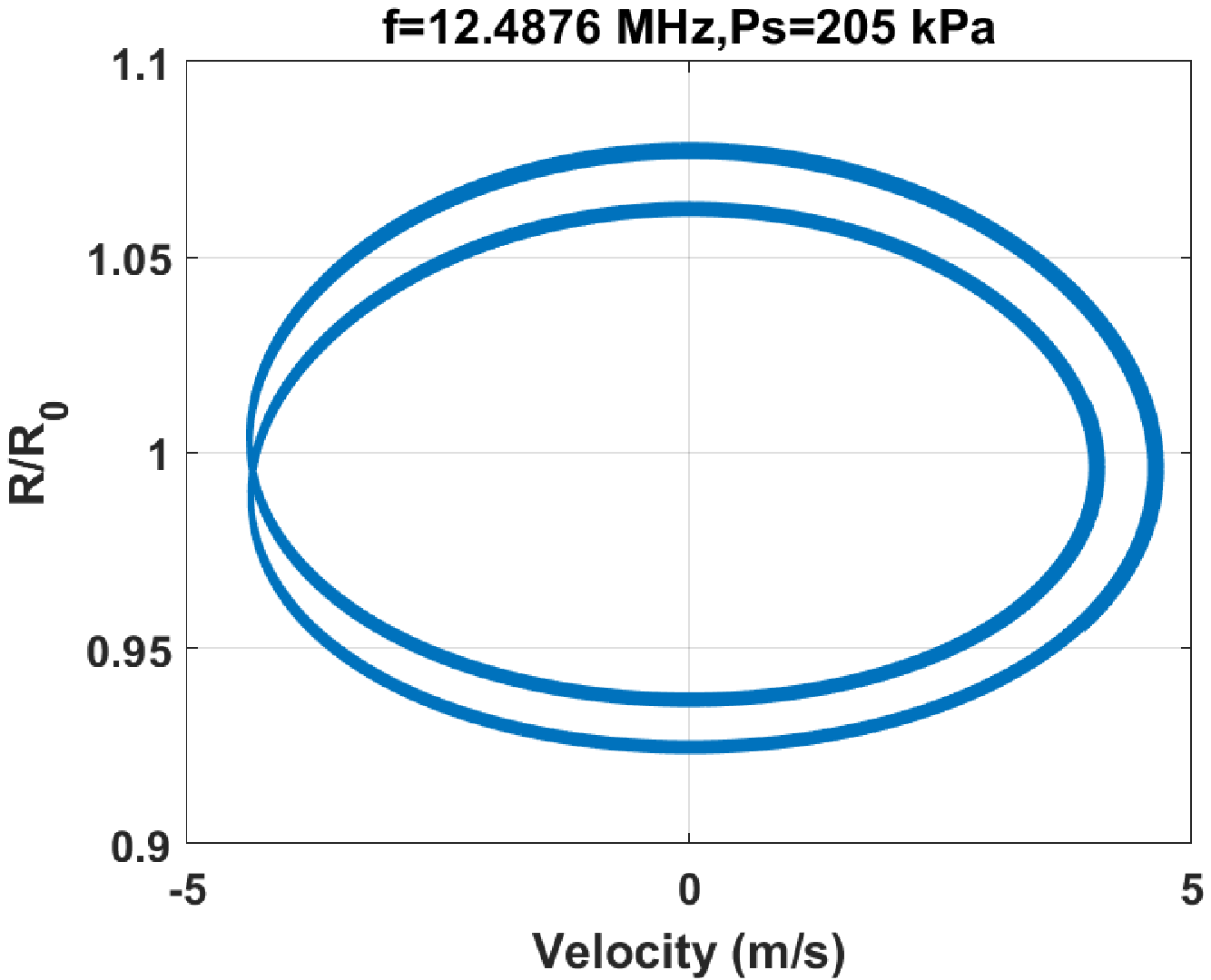}}\scalebox{0.3}{\includegraphics{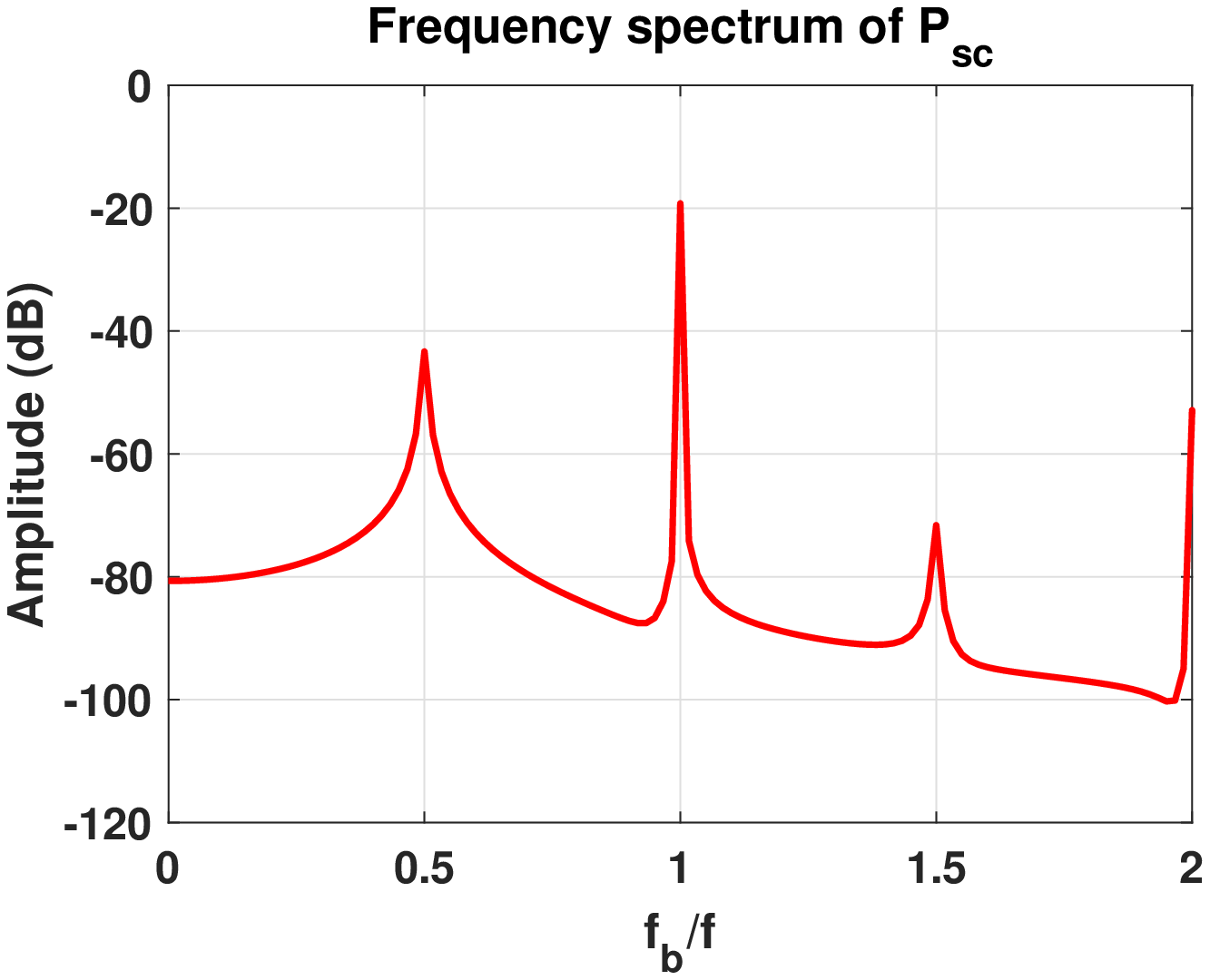}} \\
	(d) \hspace{4cm} (e) \hspace{4cm} (f)\\
	\scalebox{0.3}{\includegraphics{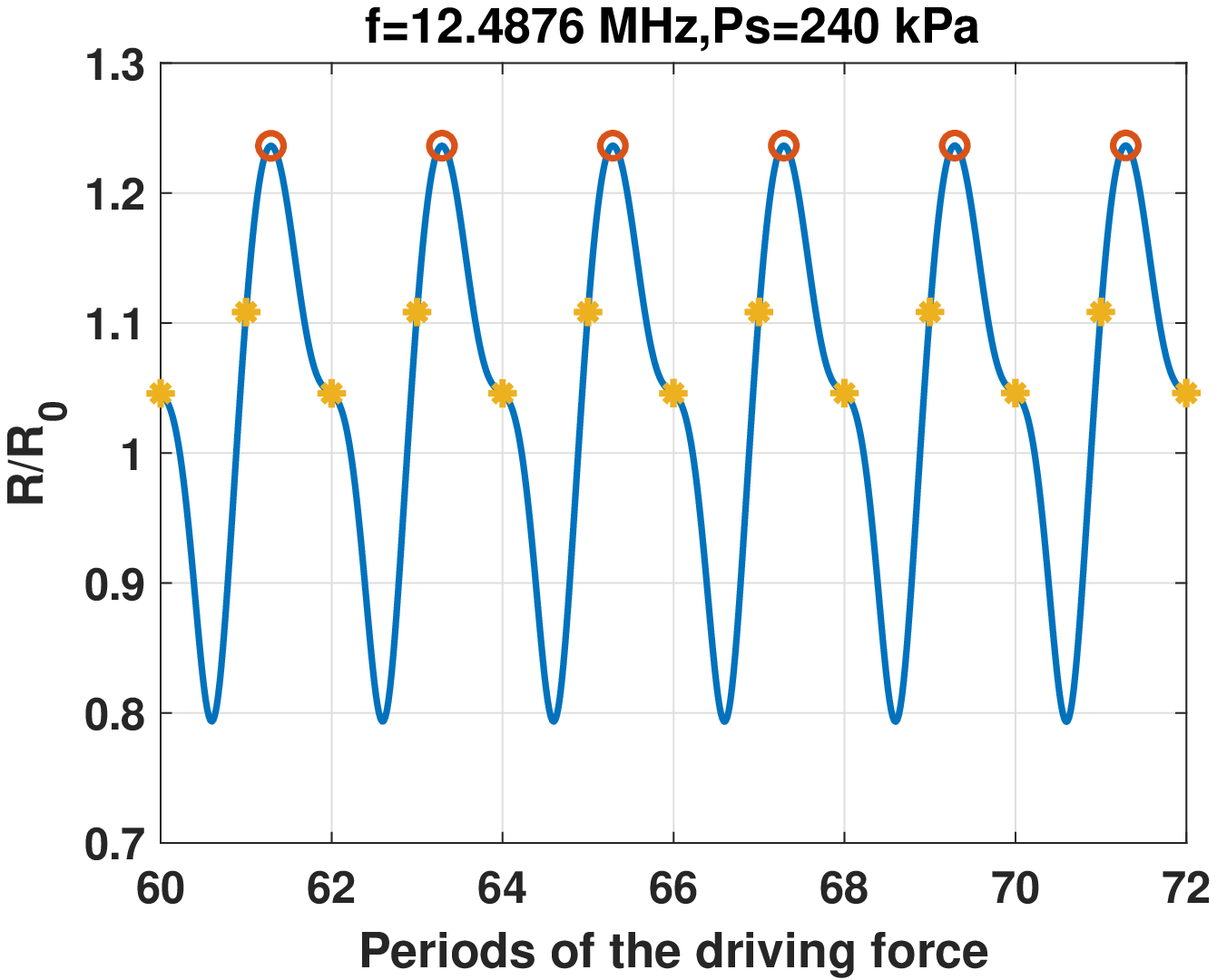}}\scalebox{0.3}{\includegraphics{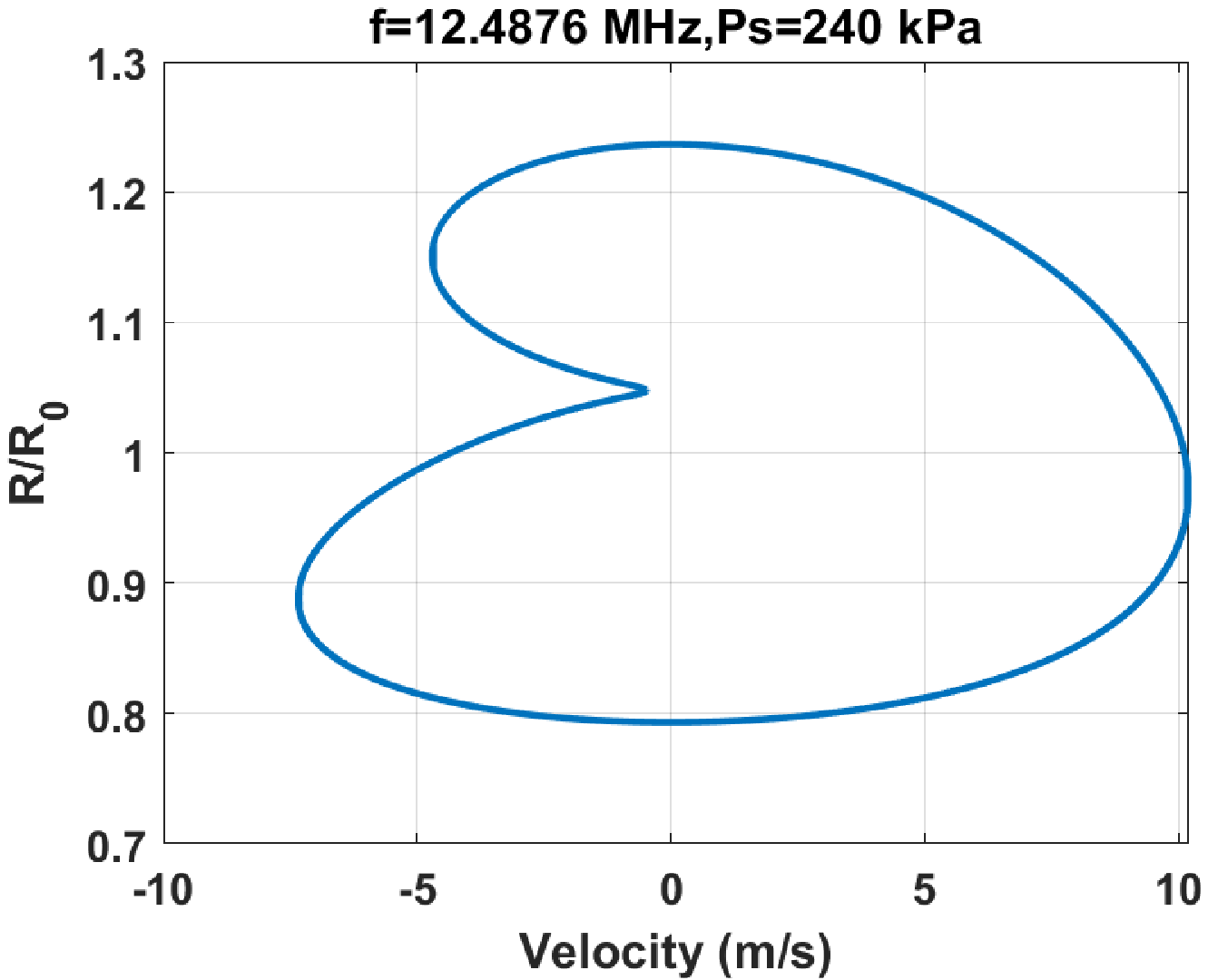}}\scalebox{0.3}{\includegraphics{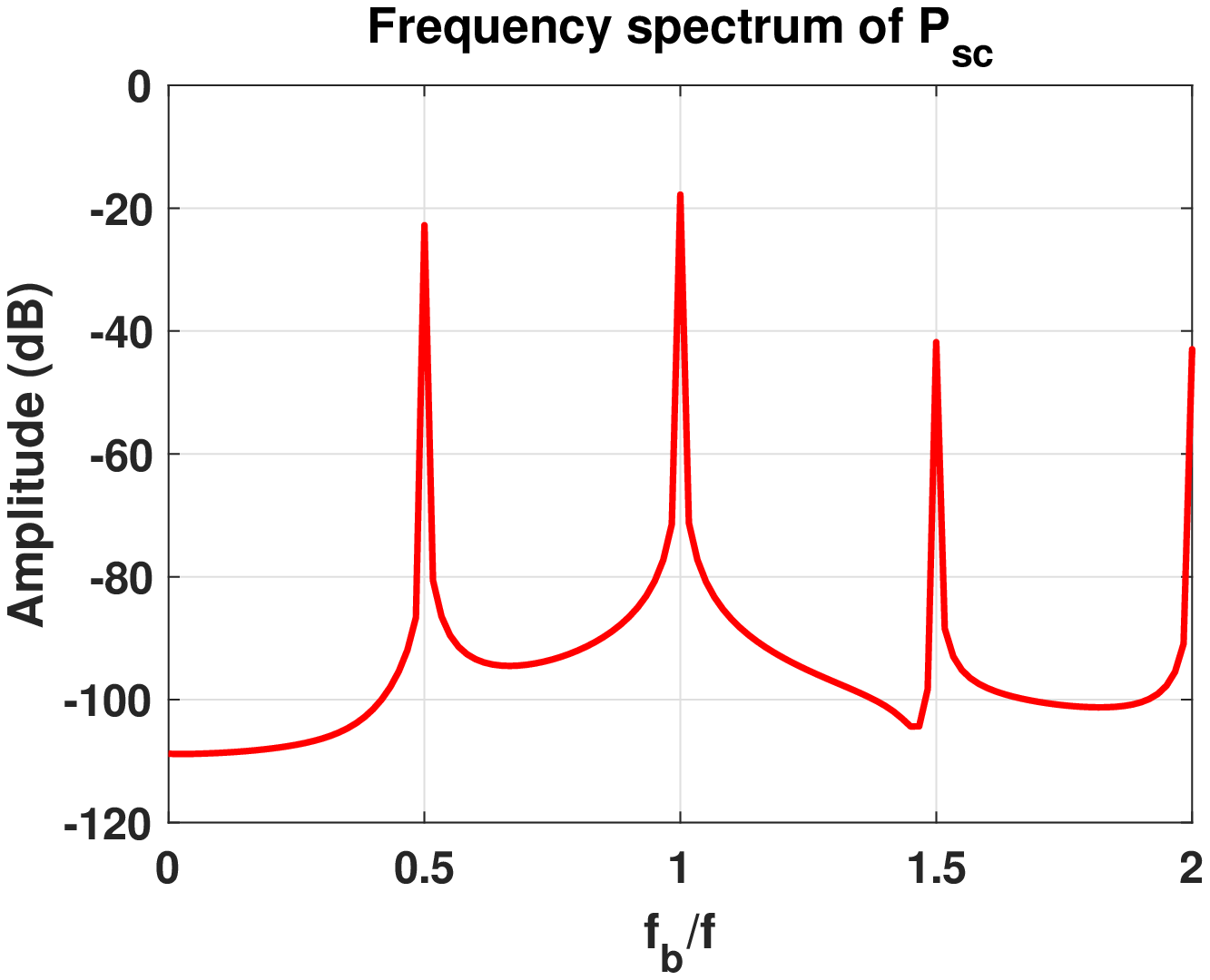}} \\
	(g) \hspace{4cm} (h) \hspace{4cm} (i)\\
	\scalebox{0.3}{\includegraphics{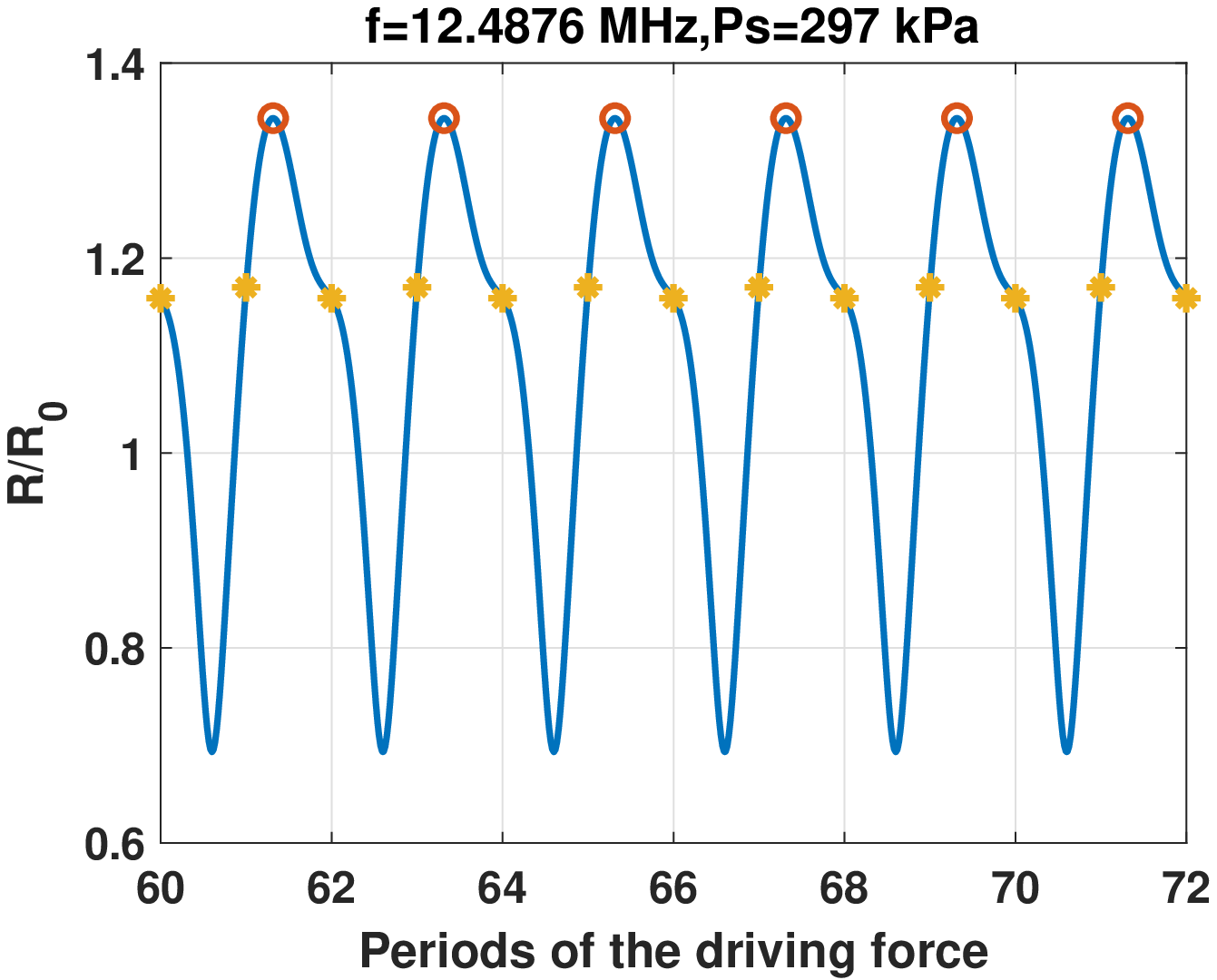}}\scalebox{0.3}{\includegraphics{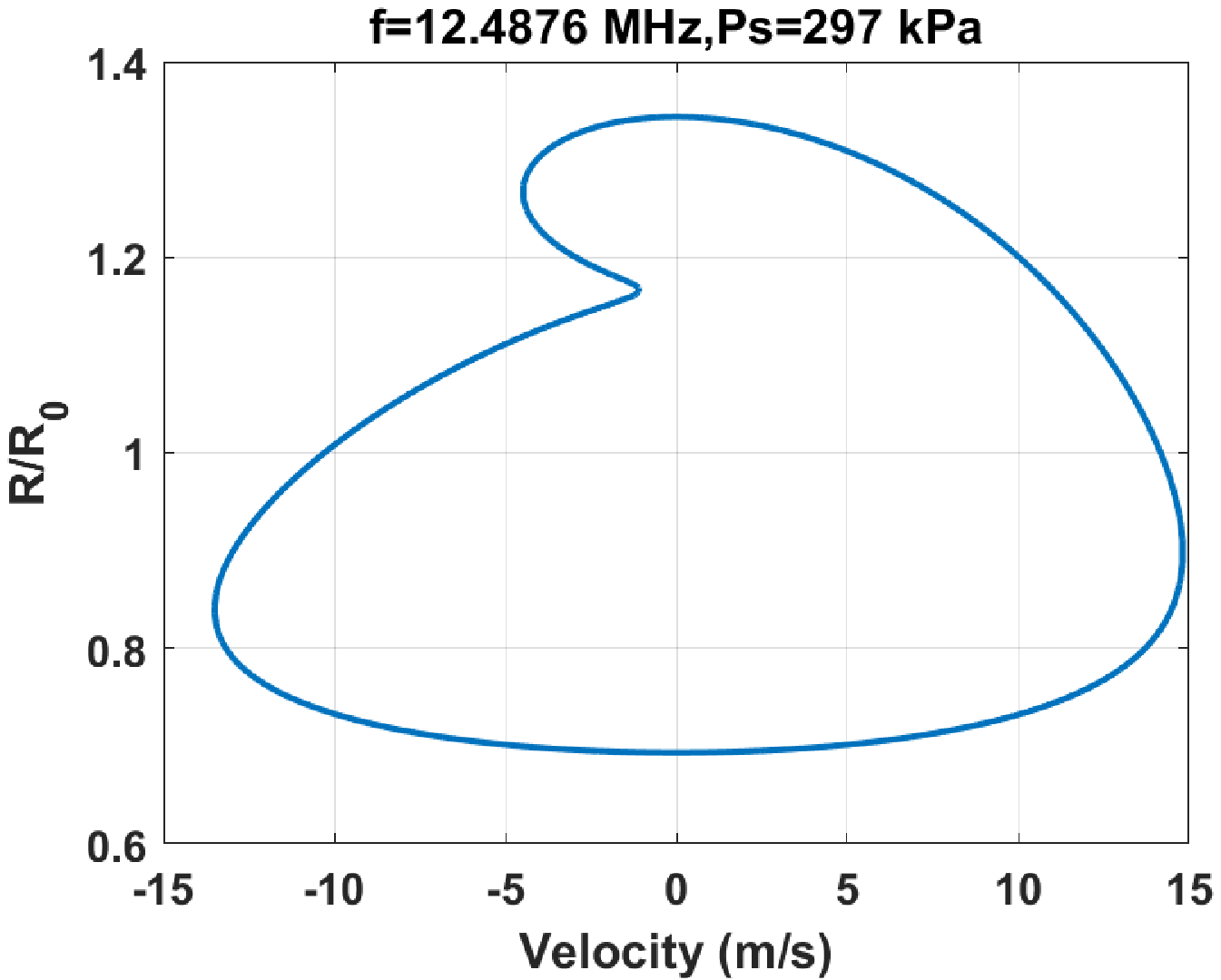}}\scalebox{0.3}{\includegraphics{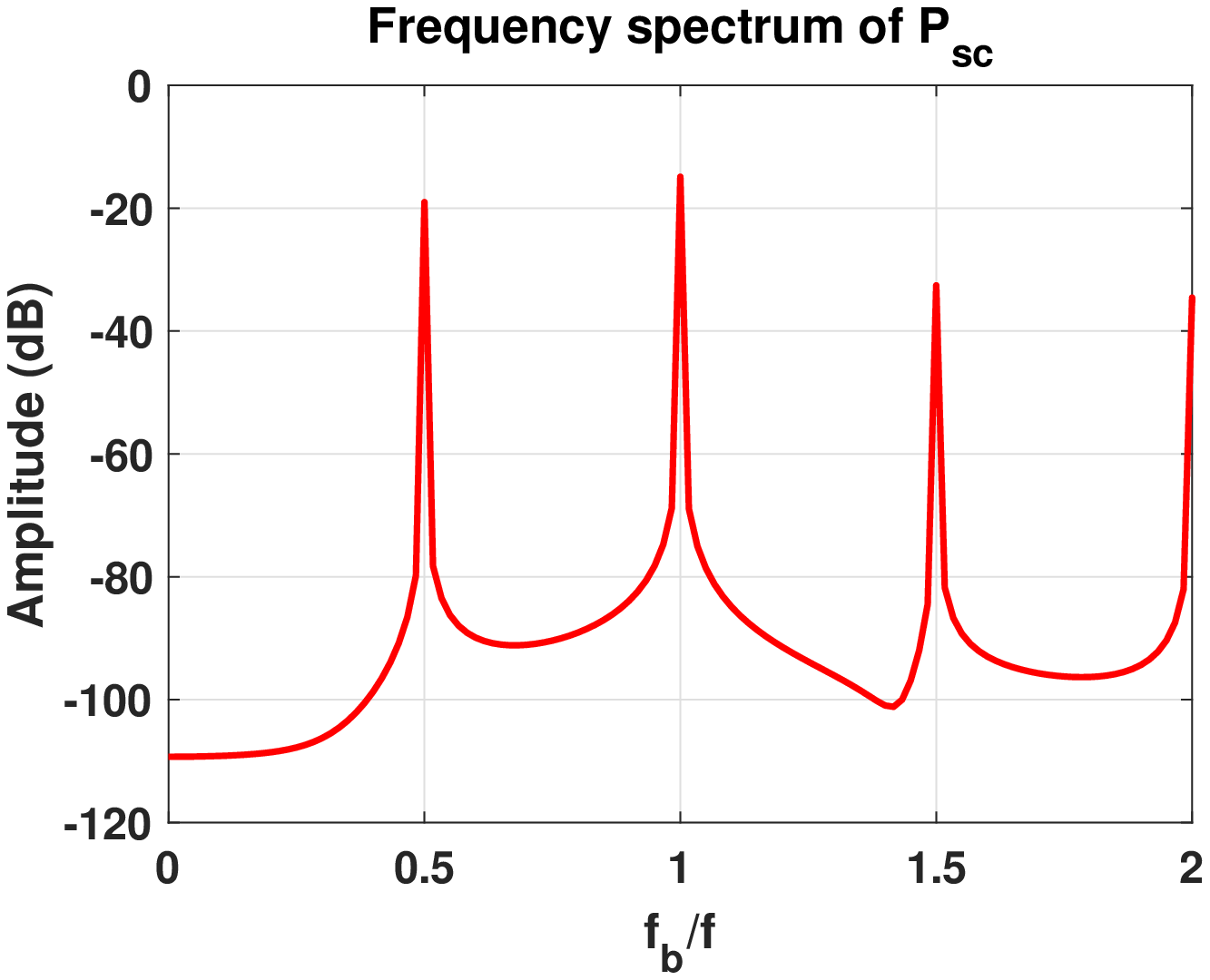}} \\
	(j) \hspace{4cm} (k) \hspace{4cm} (l)\\
	\scalebox{0.3}{\includegraphics{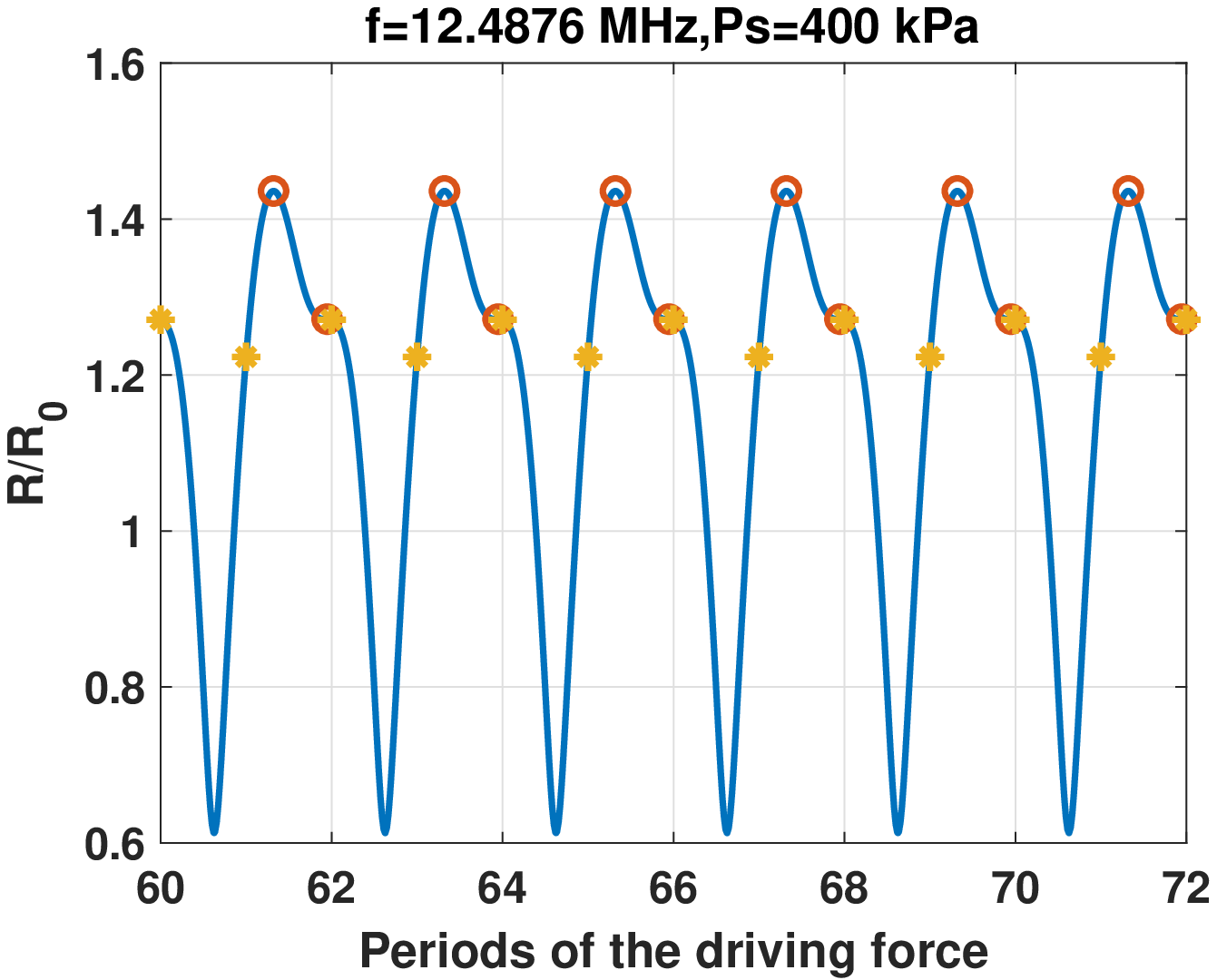}}\scalebox{0.3}{\includegraphics{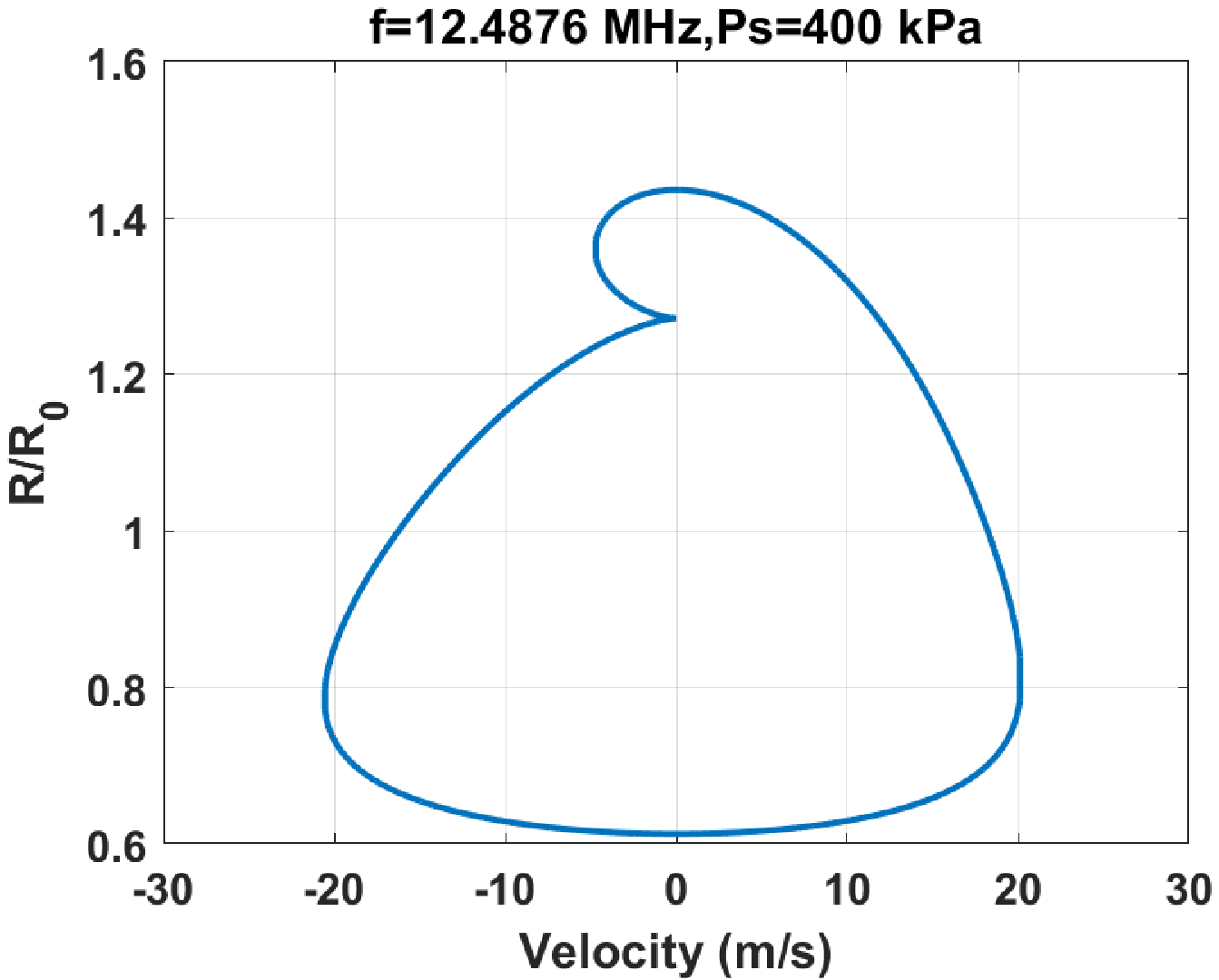}}\scalebox{0.3}{\includegraphics{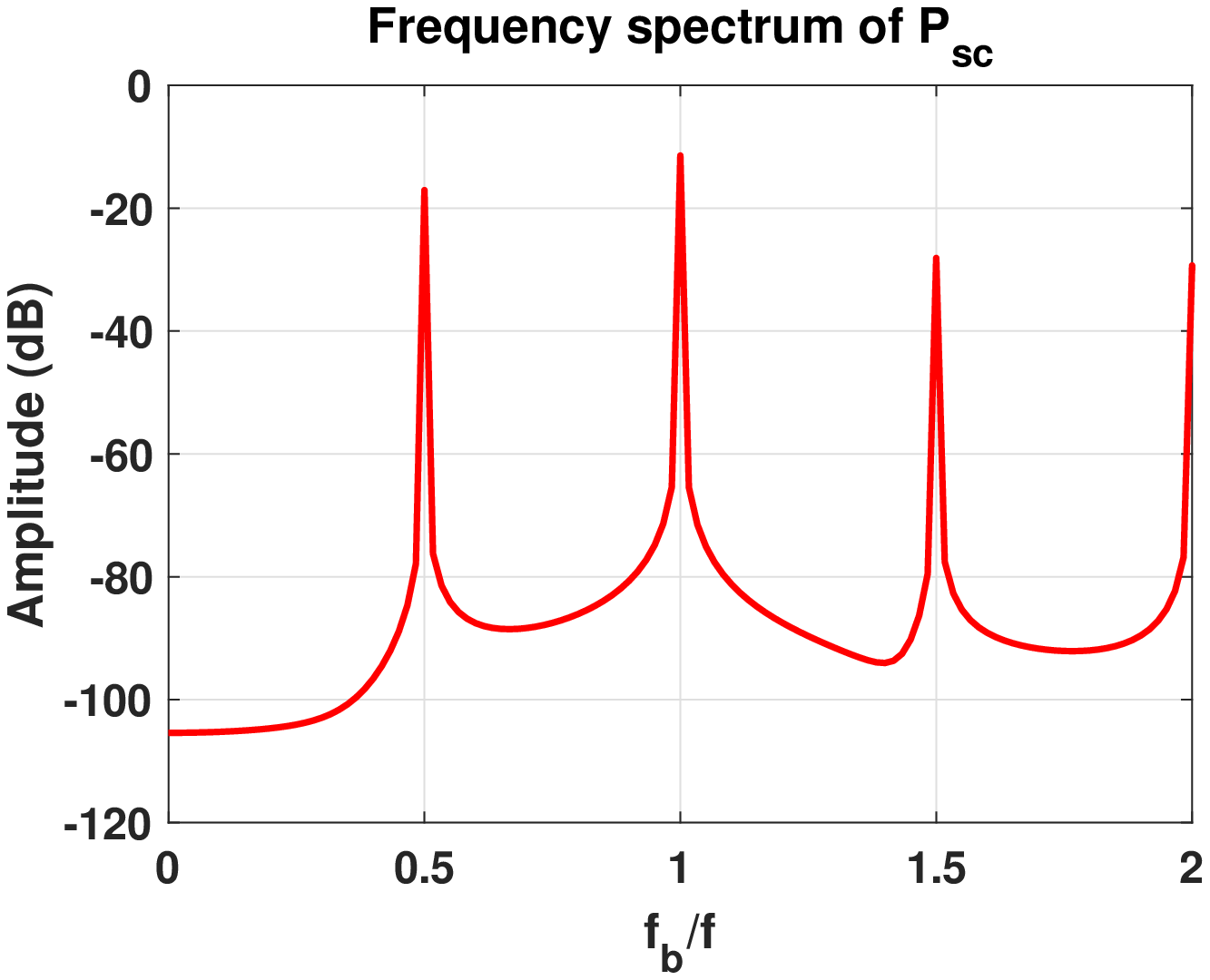}} \\
	(m) \hspace{4cm} (n) \hspace{4cm} (o)\\
	\caption{ Evolution of the dynamics of the bubble with $R_0=800 nm$ when $f=2f_r$. Diagrams are plotted for four difference pressures of interest (see Figure 6b). Radial oscillations are plotted in the left column, phase portrait diagrams in the middle column and the frequency spectrum of the $P_{sc}$ in the right column. The top row is for $P_s$=135 kPa, and the rows after for 205, 240, 297 and 400 kPa.}
\end{center}
\end{figure*}
As the pressure further increases, the yellow circle with lower amplitude in the radial oscillation curve (fig. 8a) grows quicker than the initially higher amplitude yellow circle. Thus, at a pressure that is shown by green arrow in fig 6b, the conventional bifurcation diagram only shows one point as the two solutions have the same amplitude. To shed a better light on the dynamics of the bubble shown by the green arrow in fig 6b, the time series of $\frac{R}{R_0}$ as a function of periods is shown in fig 8j. The signal has one maxima and repeats its shape once every two acoustic cycles; the amplitude of $\frac{R}{R_0}$ (yellow circle) at each period is the same however; one  yellow circle is located at the growth stage (positive wall velocity) while the second one is located at the collapse stage of the oscillations (negative wall velocity). The phase portrait still is in a heart shape form with one loop and SH and UH components are grown stronger as a function of pressure (fig 8k and fig 8l respectively).\\ 
\begin{figure*}
\begin{center}
	\scalebox{0.43}{\includegraphics{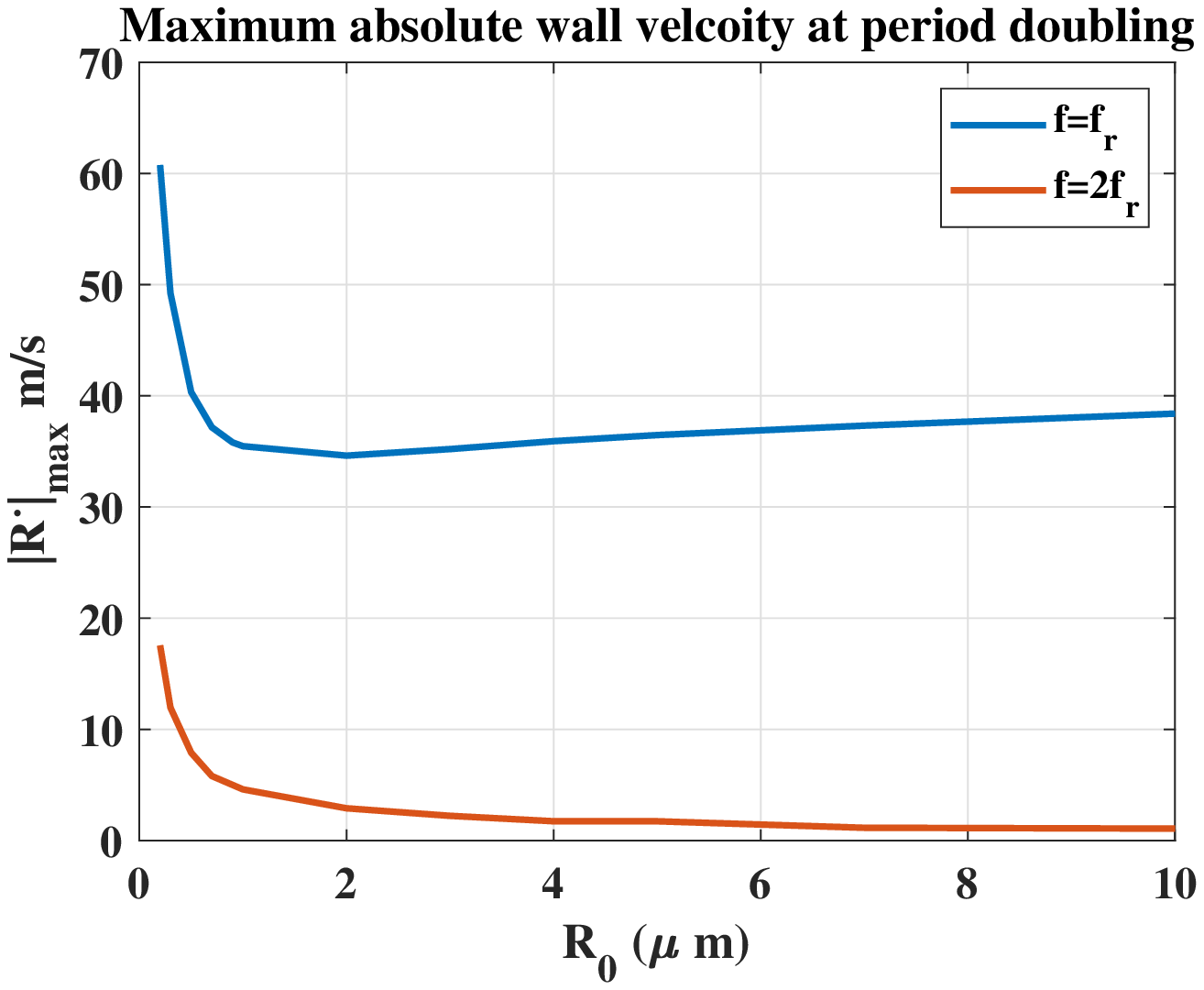}}\\
	(a) 
	\caption{Maximum wall velocity amplitude ($|\dot{R(t)}|_{max}$) when period doubling occurs as a function of $R_0$. Blue is for $f=f_r$, and red denotes the case of $f=2f_r$.}
\end{center}
\end{figure*}

\begin{figure*}
\begin{center}
	\scalebox{0.43}{\includegraphics{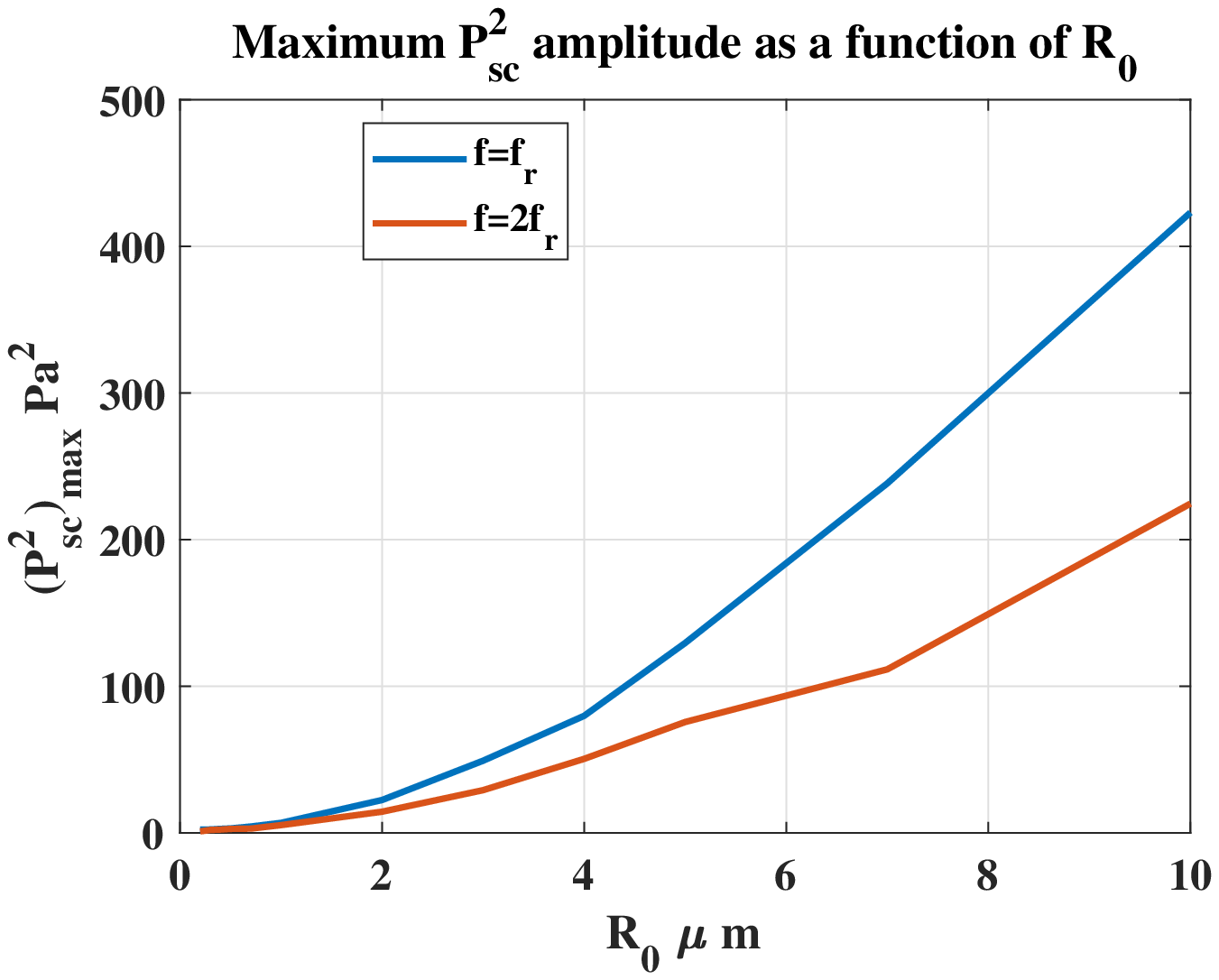}} \scalebox{0.43}{\includegraphics{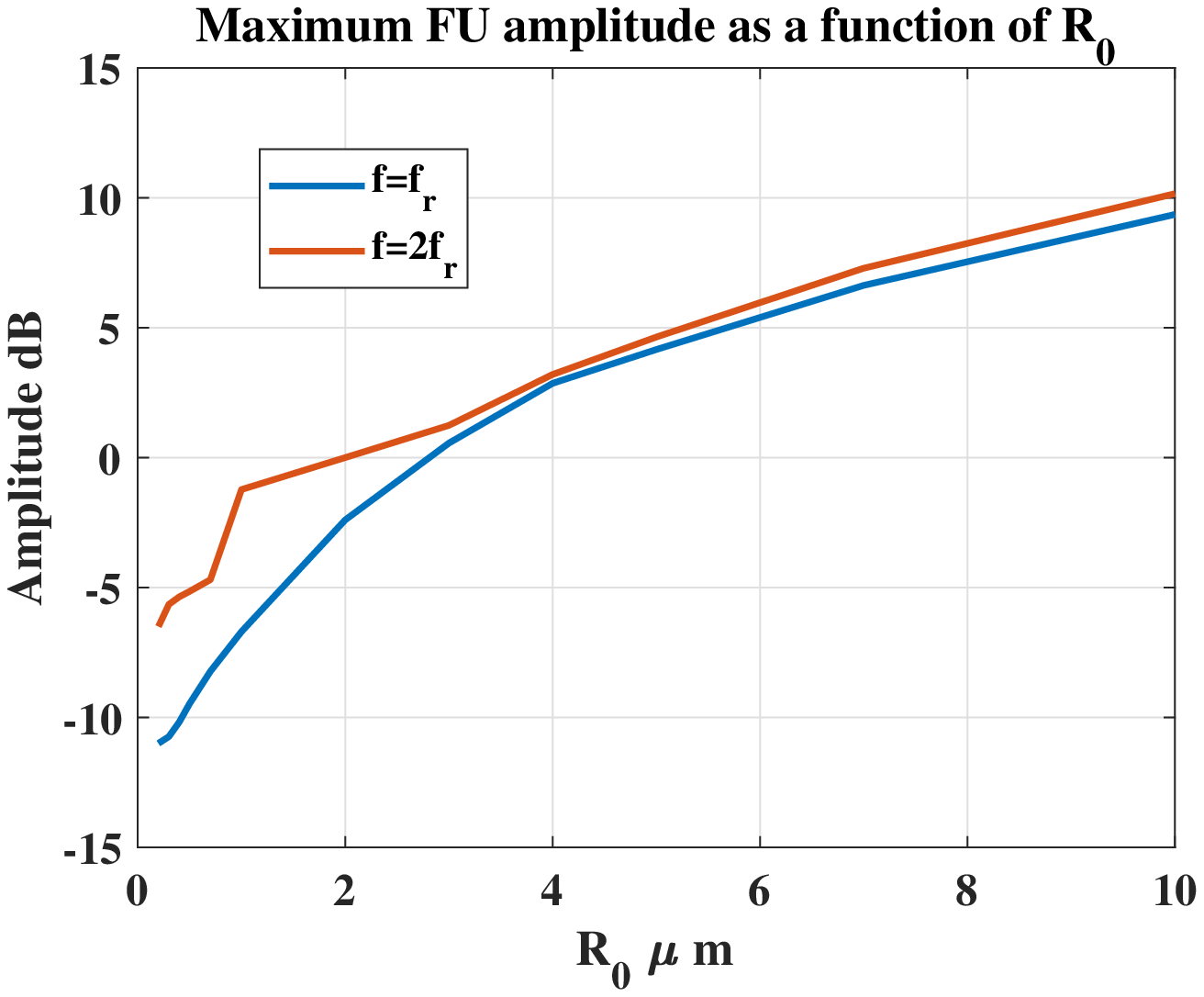}}\\
	\hspace{0.5cm} (a) \hspace{6cm} (b)\\
	\scalebox{0.43}{\includegraphics{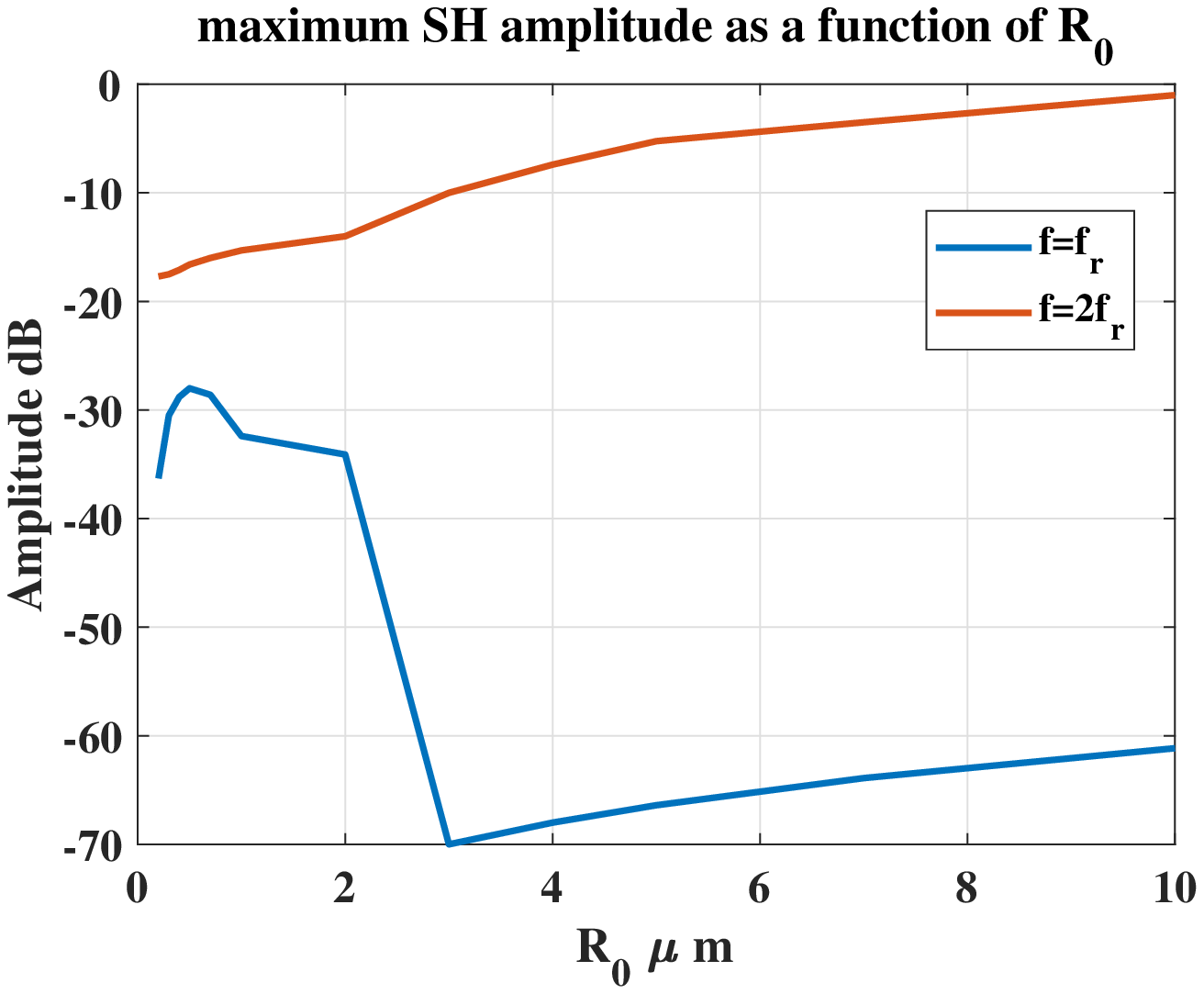}} \scalebox{0.43}{\includegraphics{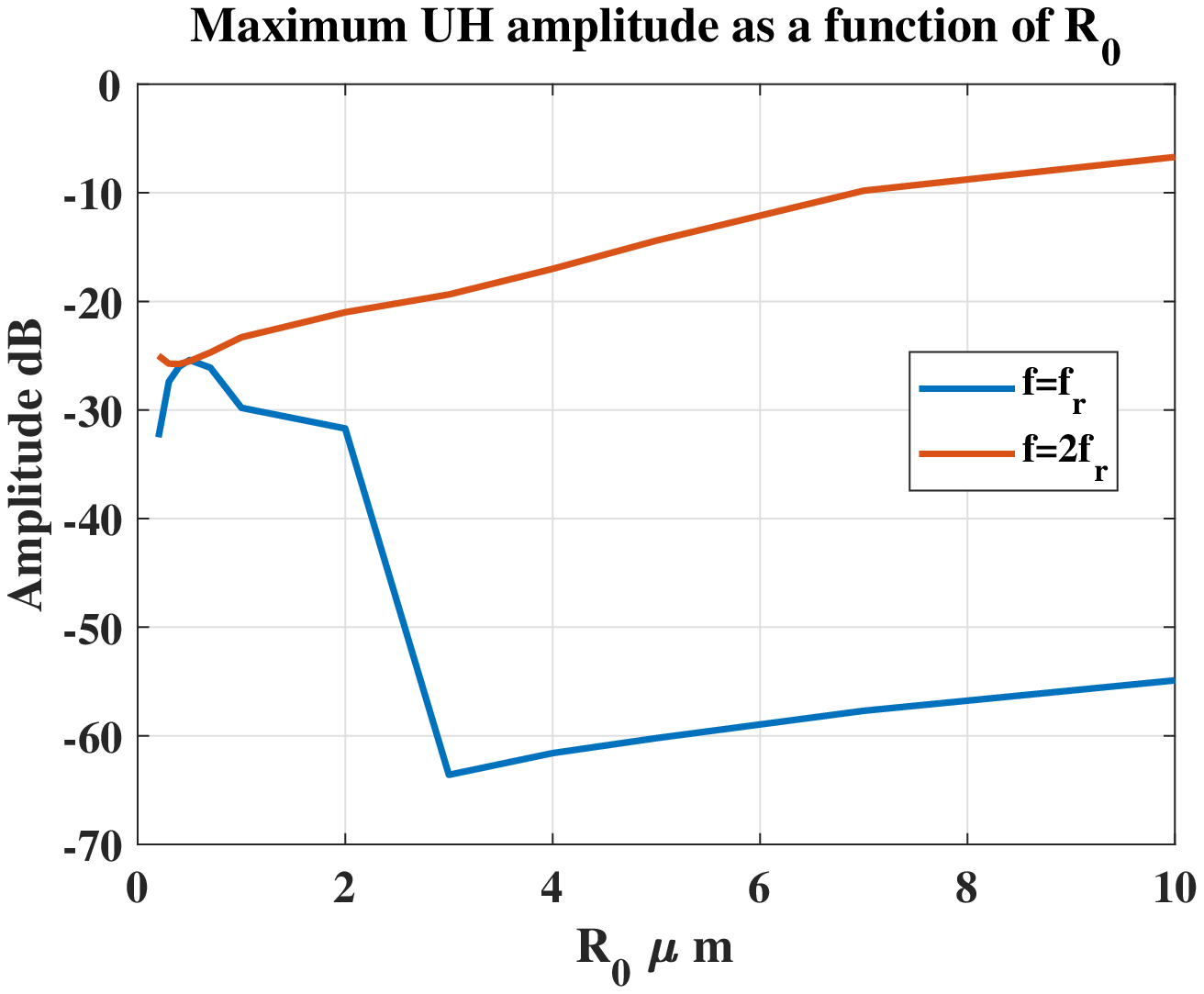}}\\
	\hspace{0.5cm} (c) \hspace{6cm} (d)\\
	\caption{Maximum possible non-destructive: a)$P_{sc}^2$, b) Fu amplitude, c) SH amplitude and d) UH amplitude as a function of $R_0$.}
\end{center}
\end{figure*}
\begin{figure*}
\begin{center}
	\scalebox{0.43}{\includegraphics{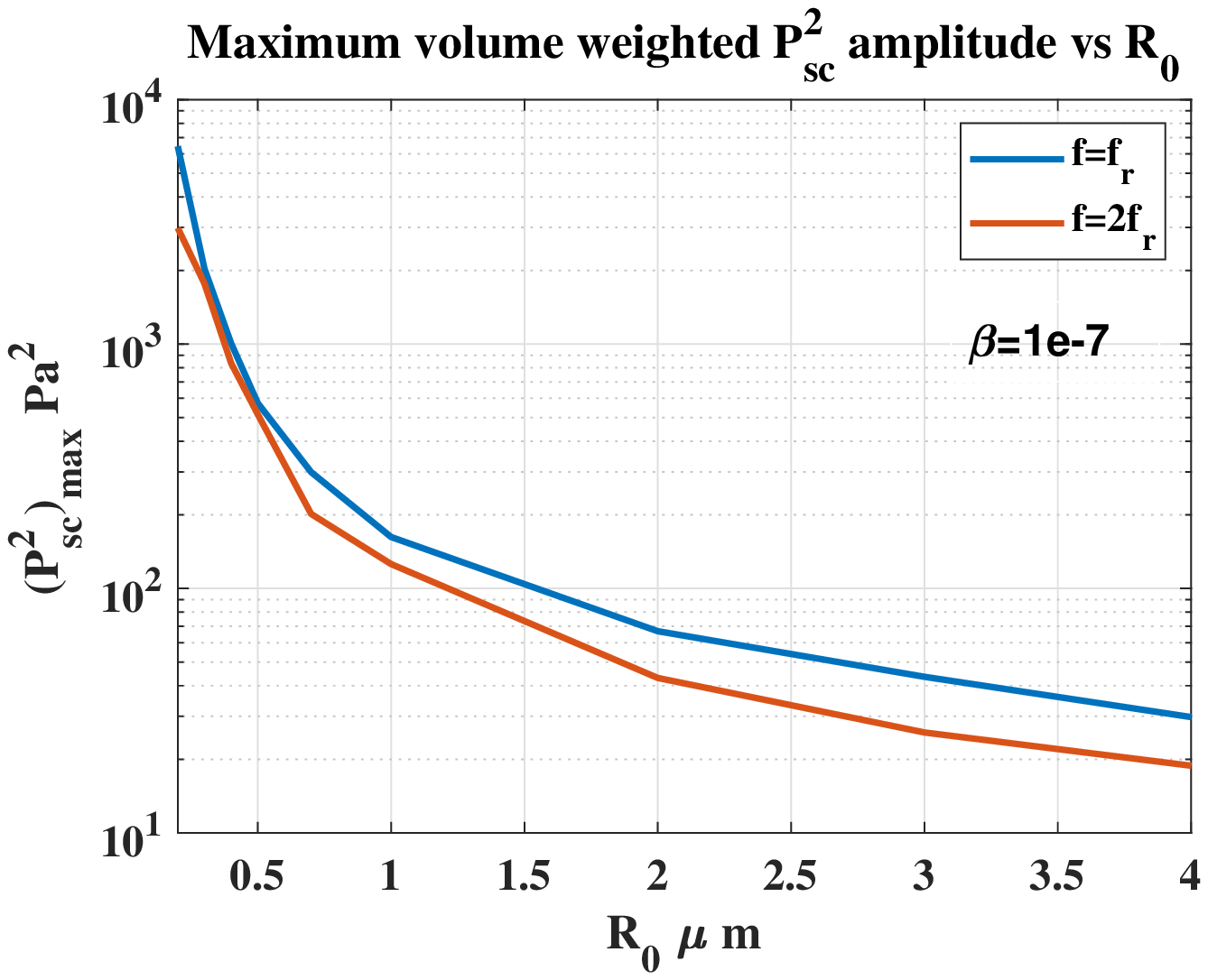}} \scalebox{0.43}{\includegraphics{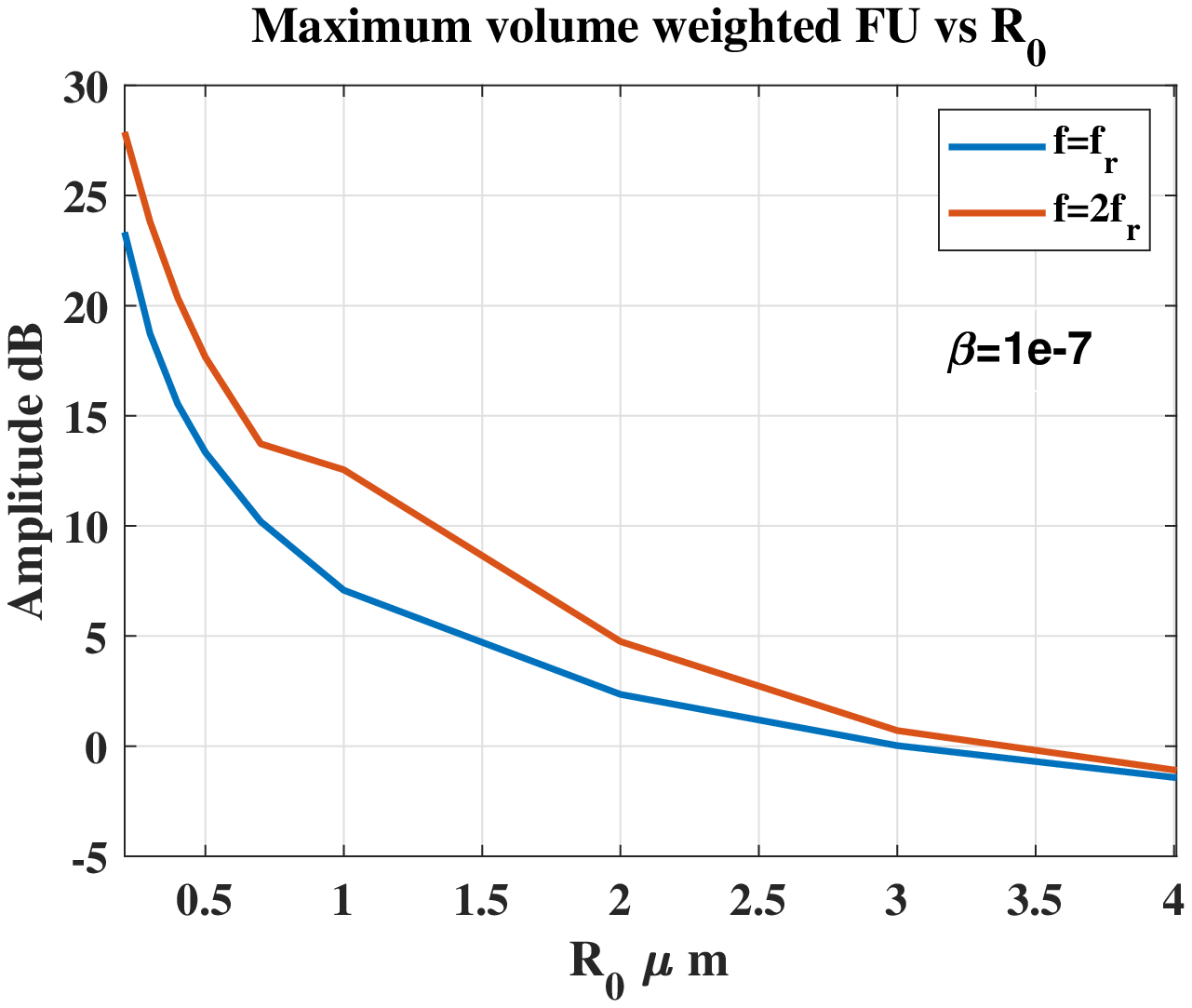}}\\
	\hspace{0.5cm} (a) \hspace{6cm} (b)\\
	\scalebox{0.43}{\includegraphics{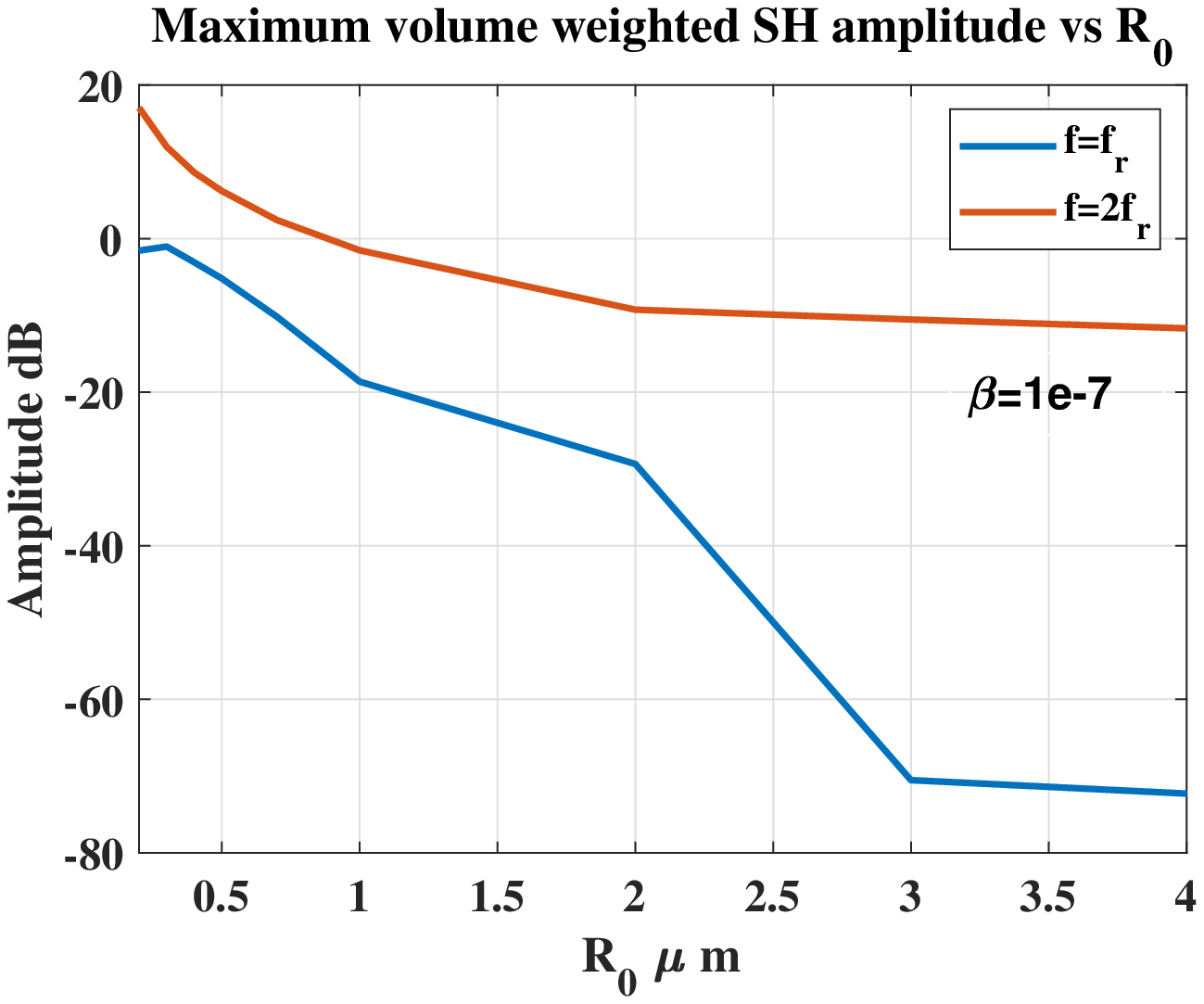}} \scalebox{0.43}{\includegraphics{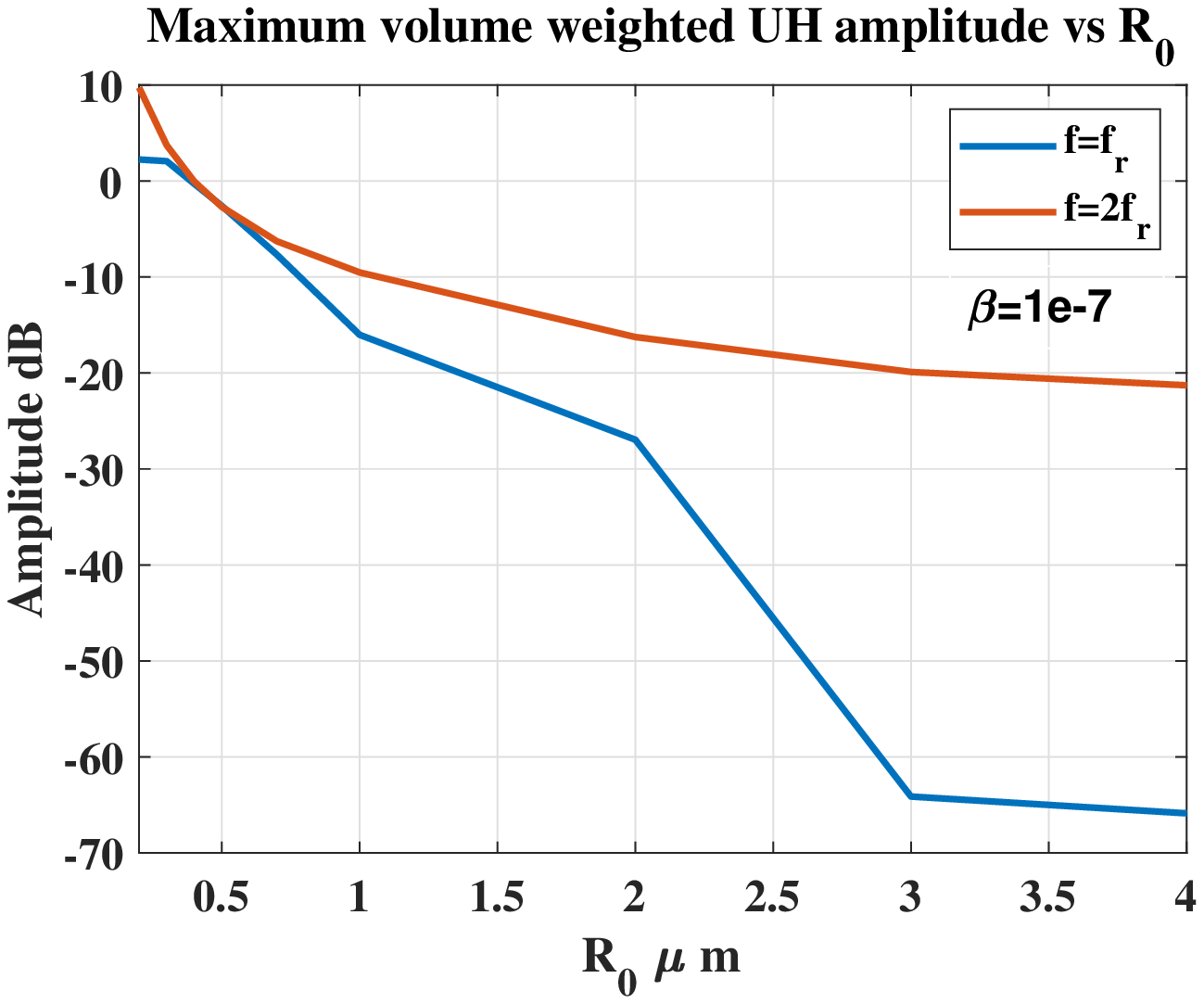}}\\
	\hspace{0.5cm} (c) \hspace{6cm} (d)\\
	\caption{ Gas volume weighted ($\beta$=1e-7) maximum possible non-destructive: a) $P_{sc}^2$, b) Fu amplitude, c) SH amplitude and d)UH amplitude as a function of $R_0$.}
\end{center}
\end{figure*}
As the pressure increases, the second maxima re-appears. Fig 8m shows that the $\frac{R}{R_0}$ oscillations of the bubble ($P_A$=400 kPa) have two maxima (red circles) and $\frac{R}{R_0}$ represent two distinct values after every period. One of the maxima (red circles) is located exactly on one of the (yellow circles) indicating velocity becomes zero once every two acoustic cycles and right at the end of the acoustic driving period. This can be interpreted as full SH resonance with a fully developed SH component in fig 8o.  
\subsection{Absolute wall velocity of the two main period 2 oscillations }
Wall velocity is an important measure of the bubble collapse and strongly contributes to the bubble backscatter in (Eq. 2). To compare the magnitudes of the bubble wall velocities, the absolute wall velocities for bubbles of initial sizes between 400nm-20 $\mu m$ were plotted alongside their bifurcation diagram as a function of pressure (Appendix A: Fig A.1 and Fig A.2). The absolute value of the bubble wall velocity when PD occurs and the maximum achievable P2 absolute wall velocities were extracted from Fig A1 and A2 and are plotted as a function of size in Figure 9.
Figure 9 depicts the bubble absolute wall velocity when PD occurs as a function of size. The blue line shows the case of f=fr and the red line represents $f=2f_r$. For the micron size bubbles when $f=f_r$, the wall velocity is $\approx$ 34-40 m/s when PD occurs. For nanobubbles the PD wall velocity rapidly increases as the size decreases, reaching velocities greater than 60 m/s for the 400 nm sized bubble. The same trend is seen when $f=2f_r$; the PD wall velocity is less than $\approx$ 6 m/s for micron size bubbles and rapidly increases for nanobubbles approaching 20 m/s for the 400 nm bubble size. In comparison to bubbles sonicated by $f=f_r$, bubbles sonicated with $f=2f_r$ exhibit much smaller wall velocities when PD occurs (e.g. for micron size bubbles 6-36  times smaller and for nanobubbles about 3-5 times smaller). This suggests that bubbles sonicated with $f=2f_r$ are more likely to sustain P2 non-destructive oscillations. 
\subsection{Analysis of the backscatter signal for the two types of Period two oscillations}
To identify the maximum achievable non-destructive backscattered signal strength and it's fundamental (FU), $\frac{1}{2}$ order subharmonic (SH) and $\frac{3}{2}$ order ultraharmonic (UH) components, the maximum nondestructive ($\frac{R}{R_0}$$\leq$2) values for each bubble size when $f=f_r$ and $f=2f_r$ were plotted (fig 10). These values are extracted from Figs A.3-A.6 in Appendix A.\\ 
The maximum value of $P_{sc}^2$ ($P_{sc}^2$ is used instead of $P_{sc}$ to better relate to the signal intensity) and the fundamental component of the $P_{sc}$ are stronger when the bubble is sonicated with $f=f_r$ (fig 10a-b), with bigger bubbles scattering stronger than smaller bubbles. However, the maximum non-destructive SH and UH amplitude of $P_{sc}$  are stronger when $f=2f_r$ (fig 10c-d).\\ 
When $f=2f_r$ nondestructive $SH_{max}$ and $UH_{max}$ are greater for bigger bubbles; however, when $f=f_r$ nondestructive $SH_{max}$ and $UH_{max}$, of the 1 micron bubble ($R_0$=0.5 $\mu m$) exhibits the strongest possible nondestructive $SH_{max}$(-28 dB) and $UH_{max}$ ($\approx$ -25.5 dB). Nondestructive $SH_{max}$ and $UH_{max}$ are highest for bubbles with (300 nm$<$$R_0$$<$2 $\mu m$) when $f=f_r$. As is discussed in fig 4, when $f=f_r$, only a small range of bubble sizes can undergo non-destructive PD and consequently SH and UH emissions when $\frac{R}{R_0}$$\leq$2.\\
Gas volume plays an important role in many applications. For a given volume a greater number of smaller bubbles are possible. The volume fraction is given by $\beta = \frac{4}{3} \pi NR^3$ where $\beta$ is the volume fraction that the gas occupies, N is the number of bubbles per unit volume and $R$ is the radius of the bubbles at time t. To consider the effect of gas volume on the maximum achievable signal from bubbles, the results of figure 10 are normalized for a case of $\beta$=$10^{-7}$ and, the signal intensities were calculated for a 1 mm cube in figure 11 (assuming negligible interaction between bubbles). Results of the bubbles bigger than 8 micron in diameter ($R_0=4 \mu m$) were omitted as the maximum vessel diameter in biomedical applications is approximately 8 $\mu m$ [35].\\ Fig 11 shows that smaller bubbles (e.g. nanobubbles) have the potential to provide stronger signals in specific situations (e.g. when $f=2f_r$ and $\beta$=$10^{-7}$). If one assumes monodispersity and that all signals arrive in phase, then the 400 nm bubbles produce $SH_{max}$ and $UH_{max}$ of 15 dB and 18 dB higher than the 3 $\mu m$ bubbles when controlled for gas volume. 
\section{Summary and conclusion} 
Period doubling (PD) and chaos are one of the well-known characteristics of nonlinear dynamical systems including bubble oscillators [1-27].  PD results in bubble subharmonic oscillations which are of great importance in applications including but not limited to contrast-enhanced ultrasound imaging [54-56], monitoring therapeutic applications of ultrasound [50-52], non-destructive testing [53,64-66], sonoluminescence [41] and other applications.  The nonlinear and chaotic dynamics of the bubble oscillator have been the subject of many studies [1-27].\\ However, the two main routes of PD in the bubble oscillator have not been studied in detail. Because of the importance of bubble SH oscillations, comprehensive knowledge of the mechanisms of PD and SH oscillations can help in optimizing current applications or explore new potential parameters to be used in applications. In this work the bifurcation structure of the $\frac{R}{R_0}$ of the bubble oscillator was studied as a function of pressure; two important cases of sonication with linear resonance frequency ($f_r$) and subharmonic resonance frequency ($2f_r$) were studied in detail for bubble sizes of 400nm up to 20 microns. The SH, ultraharmonic (UH) and fundamental (FU) components of the backscattered pressure, as well as the $\frac{R}{R_0}$ vs time and phase portraits of the signals, were analyzed. The findings of this study can be summarized as follows:\\
1-	When $f=f_r$, the $\frac{R}{R_0}$ oscillations of the bubble increases monotonically with pressure, and above a pressure threshold PD occurs. This is concomitant with the appearance of SH oscillations.\\ 
2-	When $f=f_r$, the occurrence of PD is most likely concomitant with bubble destruction as $\frac{R}{R_0}$ is very close to, or above, 2. Only bubbles with sizes between 400nm and 5 microns can sustain stable P2 oscillations ($\frac{R}{R_0}$$\leq$2) and only for a narrow pressure range. The 400 nm bubble has the widest pressure range for non-destructive SH oscillations ($\approx$ 60 kPa)\\
3-	For the majority of the bubble sizes studied PD occurs at lower pressures when $f=2f_r$ compared to $f=f_r$. The difference between the two pressure thresholds is ~180-190 kPa for bubbles bigger than 1 micron. As the bubble size decreases, the two pressure thresholds converge and become approximately equal when $R_0$=200 nm.\\
4-	When the bubbles are sonicated with $2f_r$, PD is more likely to result in non-destructive oscillations as $\frac{R}{R_0}$ $<$1.2. Even as pressure increases, the $\frac{R_{max}}{R_0}$ of the P2 oscillations does not exceed 2. \\
5-	 When $f=2f_r$ and for bubble sizes greater than 740 nm the evolution of the dynamics from P1 to P2 is through a period-doubling that looks like a bow-tie shape. To our best knowledge, this type of period doubling is first reported in this paper. \\
6-	When $f=2f_r$, the period doubling happens over a much wider pressure range when compared to $f=f_r$. This makes the period doubling shape to be elongated in the phase portrait diagrams. Due to the lower oscillation amplitude and gentle bubble collapse (lower bubble wall velocities), the bubble can sustain stable P2 oscillations for a longer duration and over a broader range of acoustic pressures.\\
7-	The occurrence of PD is concomitant with the initiation of the growth of SH and UH component of the signal. When $f=f_r$, the UH component of the signal undergoes the initiation first; however, when $f=2f_r$ the SH component of the signal grows first.\\ 	
8-	Different stages of the PD of the bow-tie bifurcation correspond to the initiation (start of the PD), growth (PD) and saturation (at bow-tie point when two red curves overlap) and then overlap of one of the maxima with one of the solutions of the conventional bifurcation method) of the SH and UH signals. The phenomena of initiation, growth and saturation of subharmonics have also been confirmed by experimental observations [68].\\
9-	When $f=f_r$, the occurrence of PD is concomitant with a decrease in bubble wall velocity; however, when $f=2f_r$, the bubble wall velocity undergoes a rapid increase as soon as PD occurs.\\
10-	 For bubble sizes $>$ 600 nm when PD occurs wall velocity is approximately 30-45 m/s; however, when $f=2f_r$, the PD wall velocity is less than 10 m/s. This is another reason as to why P2 oscillations are more likely non-destructive when $f=2f_r$.\\
11-	When sonicated with 2fr, the phase portrait of the P2 attractor differs from the P2 attractor that is generated through sonication with $f_r$. When $f=f_r$, the phase portrait consists of two big bell shape loops with one generating and enclosing the other. When $f=2f_r$, the phase portrait of the P2 oscillations looks like a rotated heart (Fig 8).\\
12-	The SH component of the $P_{sc}$ is higher when the bubble is sonicated with $2f_r$; however, maximum $P_{sc}$ amplitude, FU and UH component of the $P_{sc}$ are higher when $f=f_r$.\\
13-	Bigger bubbles scatter sound more strongly; however, for a given gas volume smaller microbubbles may produce stronger scattering due to their greater numbers compared to bigger bubbles.
\section{Acknowledgments}
The work is supported by the Natural Sciences and Engineering Research Council of Canada (Discovery Grant RGPIN-2017-06496), NSERC and the Canadian Institutes of Health Research ( Collaborative Health Research Projects ) and the Terry Fox New Frontiers Program Project Grant in Ultrasound and MRI for Cancer Therapy (project $\#$1034). A. J. Sojahrood is supported by a CIHR Vanier Scholarship.
\bibliography{apssamp}

\newpage

\appendix
\newcommand{\hbAppendixPrefix}{A}
\counterwithin{figure}{section}
\section{Bifurcation structure and the dynamical properties of the bubbles with $R_0=200nm-10 \mu m$}
Figure A.1(a), displays the bifurcation structure of the $\frac{R}{R_0}$ of the bubbles as a function of pressure for bubbles with $R_0= 200 nm-10 \mu m$ when $f=f_r$. We have omitted the chaotic range of oscillations as the main focus here is to compare the mechanism of PD of bubbles of different sizes. The bifurcation curves are plotted using the conventional bifurcation analysis method as here we are only interested in the period of the bubble oscillations. All bubbles undergo a period doubling from P1 to P2; the pressure threshold for PD increases as the bubble size decreases. Fig. A1(b), shows the corresponding maximum wall velocity. The wall velocity increases monotonically with pressure until the occurrence of PD. As soon as PD occurs, the maximum wall velocity starts decreasing with increasing pressure for bubbles with $R_0> 400 nm$. For bubbles with $R_0\leq 400 nm$ in diameter, the growth rate of wall velocity as a function of pressure elevation is reduced as soon as PD occurs. Smaller bubbles reach a higher P2 maximum wall velocities.\\
\begin{figure*}
	\begin{center}
		\scalebox{0.33}{\includegraphics{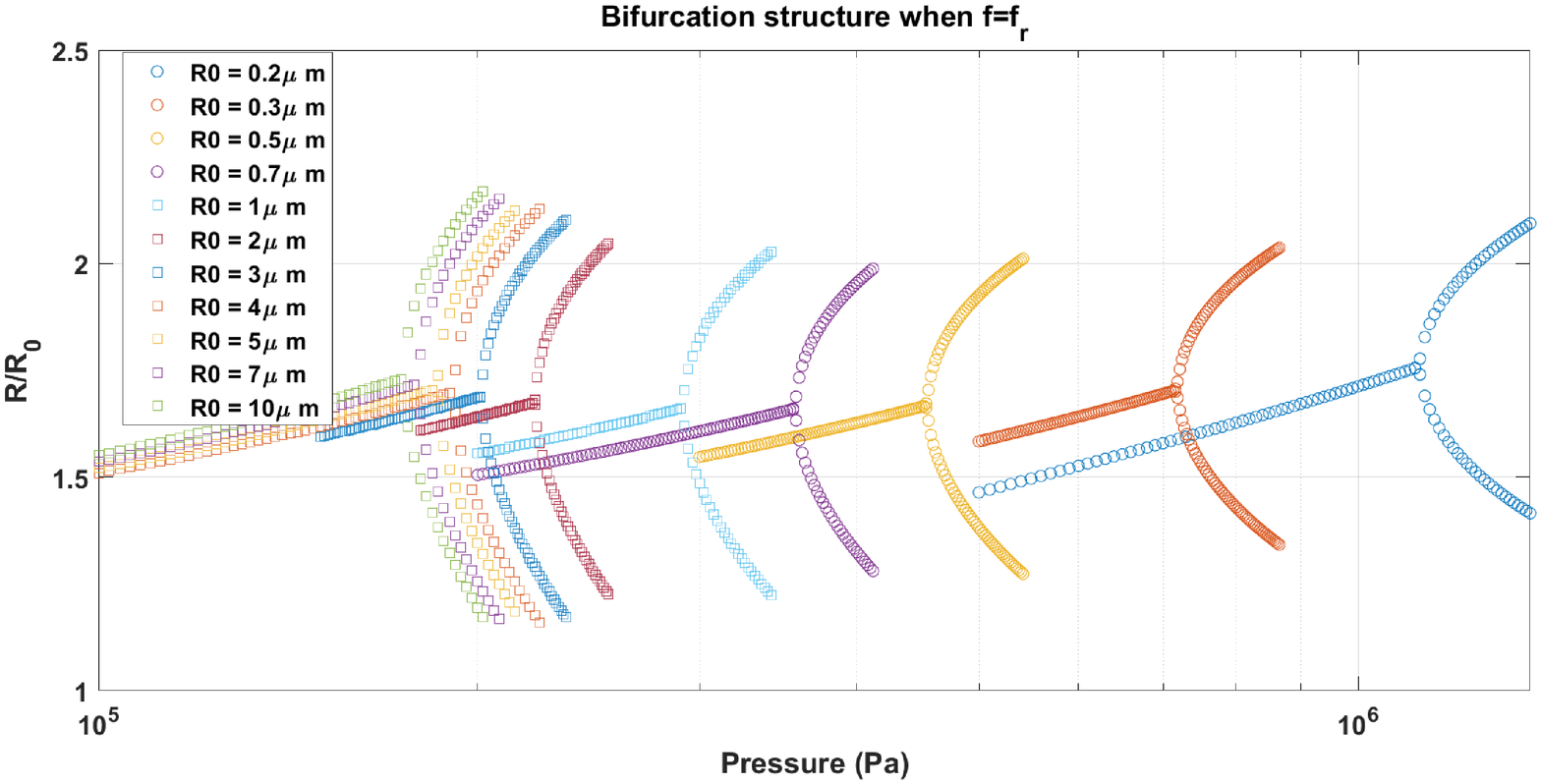}}\\
		(a) \\
		\scalebox{0.33}{\includegraphics{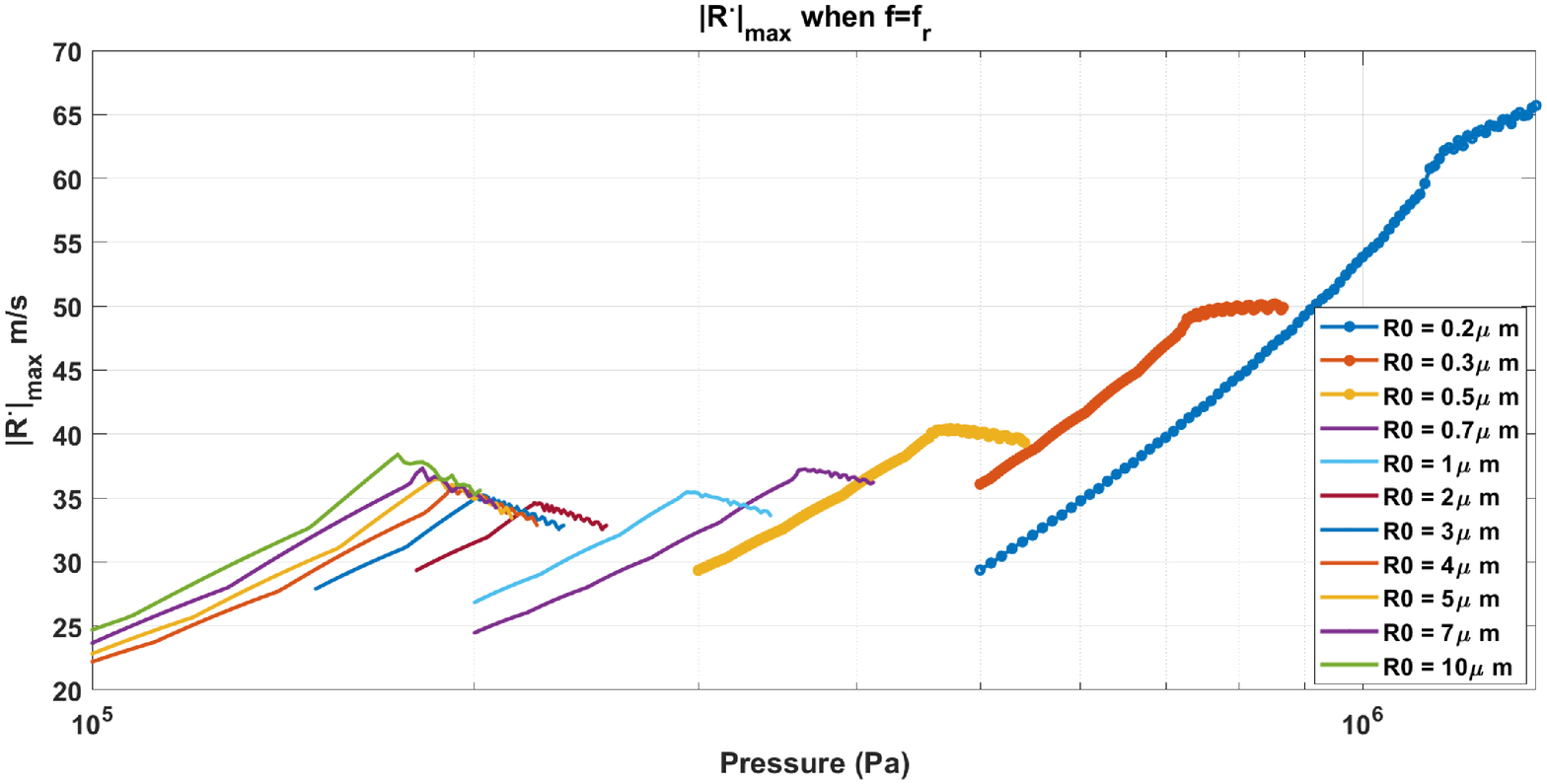}}\\
		(b) 
		\caption{Bifurcation structure of $\frac{R}{R_0}$ of the bubble oscillations as a function of pressure when $f=f_r$ ($R_0$ range 400nm-10 $\mu m$). b) The corresponding maximum wall velocity amplitude ($|\dot{R(t)}|_{max}$) .}
	\end{center}
\end{figure*}
Fig A.2(a) displays the bifurcation structure of the $\frac{R}{R_0}$ oscillations of the bubble a function of pressure presented for bubbles with $R_0= 200 nm-10 \mu m$ when $f=2f_r$.  Bigger bubbles undergo PD at lower pressures and lower $\frac{R}{R_0}$  amplitude because of less constrictions imposed by viscous forces. The mechanism of PD is through a PD bifurcation that evolves in a form of bow tie shape for bubbles with $R_0 > 370 nm$. The corresponding wall velocities in Fig. A.2b exhibit an exact opposite behavior when compare with the case of sonication with $f=f_r$ (fig. A.1b). The wall velocity grows very slowly as pressure increases until PD occurs. As soon as PD occurs, the growth rate of the wall velocity dramatically increases. Additionally, when PD coccus, wall velocities are 3-36 times smaller compared to when the bubbles are sonicated with $f=f_r$. Lower wall velocity is an important factor for the sustainable non-destructive oscillations of bubbles.\\
\begin{figure*}
	\begin{center}
		\scalebox{0.33}{\includegraphics{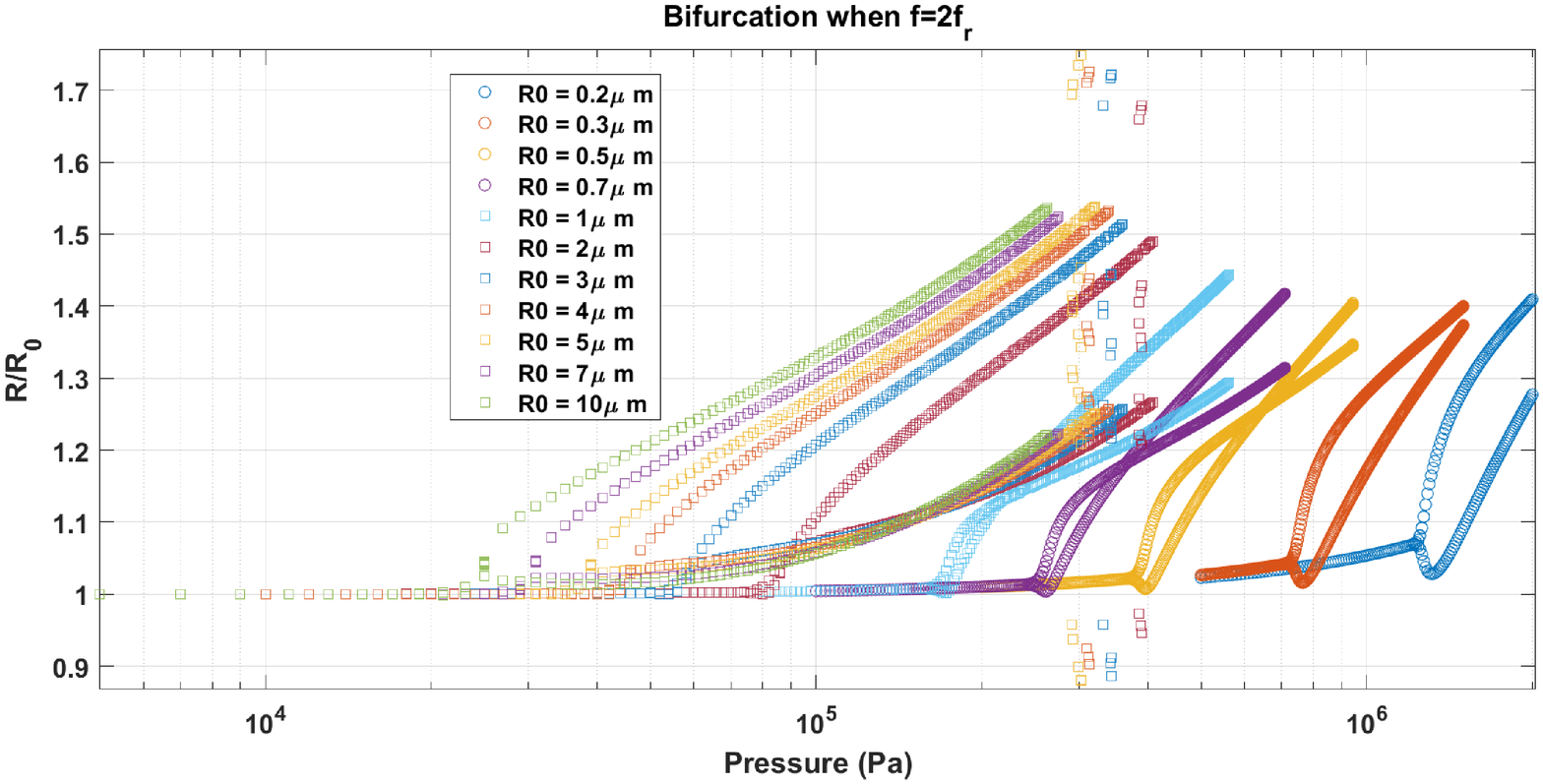}}\\
		(a) \\
		\scalebox{0.33}{\includegraphics{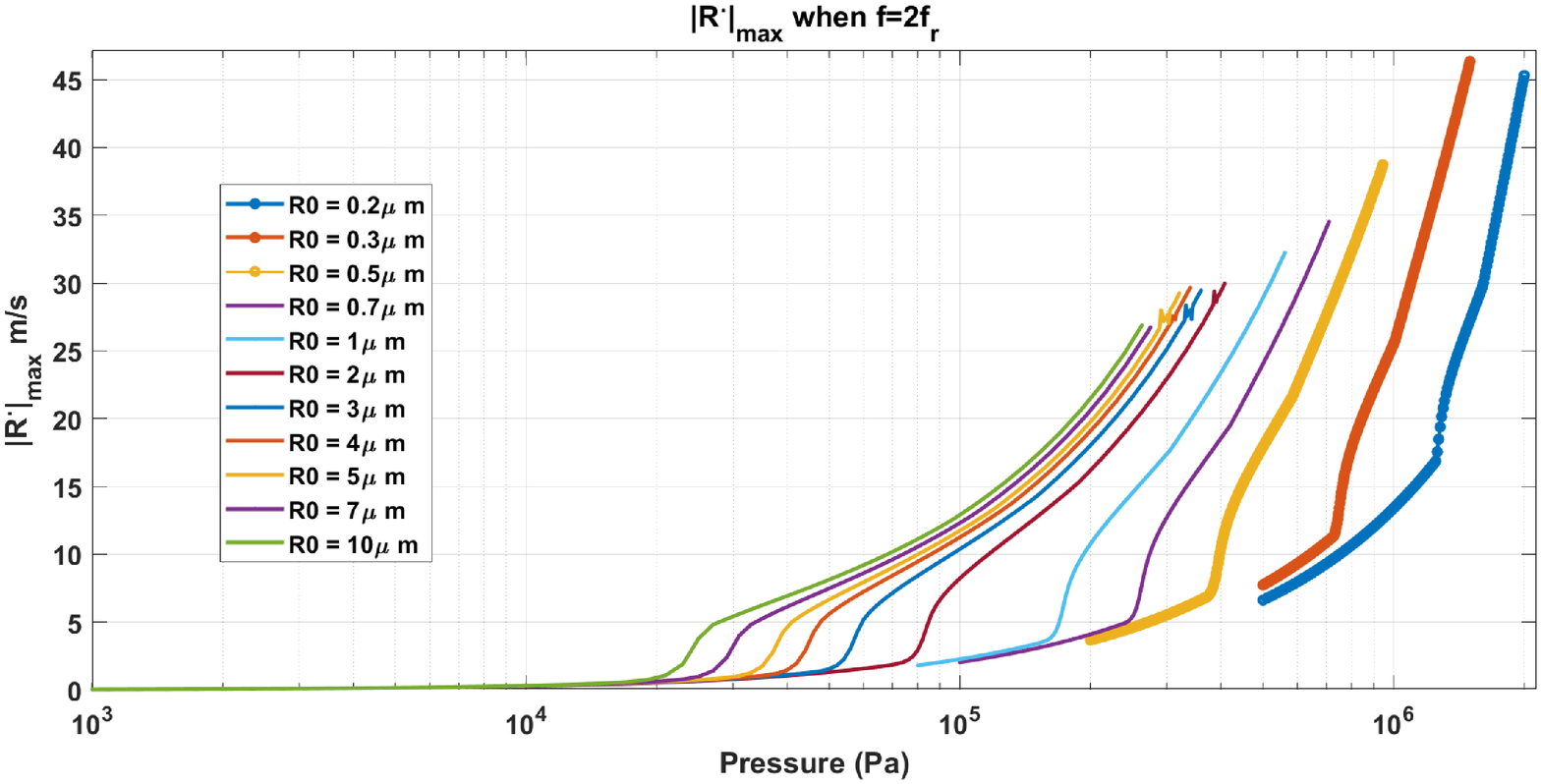}}\\
		(b) 
		\caption{Bifurcation structure of $\frac{R}{R_0}$ of the bubble oscillations as a function of pressure when $f=2f_r$ ($R_0$ range 400nm-10 $\mu m$). b) The corresponding maximum wall velocity amplitude ($|\dot{R(t)}|_{max}$) .}
	\end{center}
\end{figure*}
Figure A.3(a) shows the maximum value of $(P_{sc}^2)_{max}$ in the regime of non-destructive oscillations ($\frac{R}{R_0}\leq2$). The maximum value of  $(P_{sc}^2)_{max}$ increases monotonically with pressure. For bubbles that are able to exhibit PD while ($\frac{R}{R_0}\leq2$), $(P_{sc}^2)_{max}$ undergo a decrease as soon as PD occurs consistent with the predictions of [15] for coated bubbles. This phenomenon is discussed in full detail in [15].  Bigger bubbles achieve higher $(P_{sc}^2)_{max}$ and can be destroyed at lower pressures.\\
When $f=2f_r$ the maximum $(P_{sc}^2)_{max}$ increases monotonically with increasing pressure. As soon as PD occurs maximum $(P_{sc}^2)_{max}$ undergoes a rapid increase. For bubbles that exhibit the bow tie shape bifurcation ($R_0>370 nm$) the $(P_{sc}^2)_{max}$ continues increasing rapidly until a second pressure threshold. Above the second pressure threshold, the rate of change of $(P_{sc}^2)_{max}$ decreases. This pressure threshold is the same as the pressure where the bow tie point occurs in the conventional bifurcation diagram.\\ 
\begin{figure*}
	\begin{center}
		\scalebox{0.33}{\includegraphics{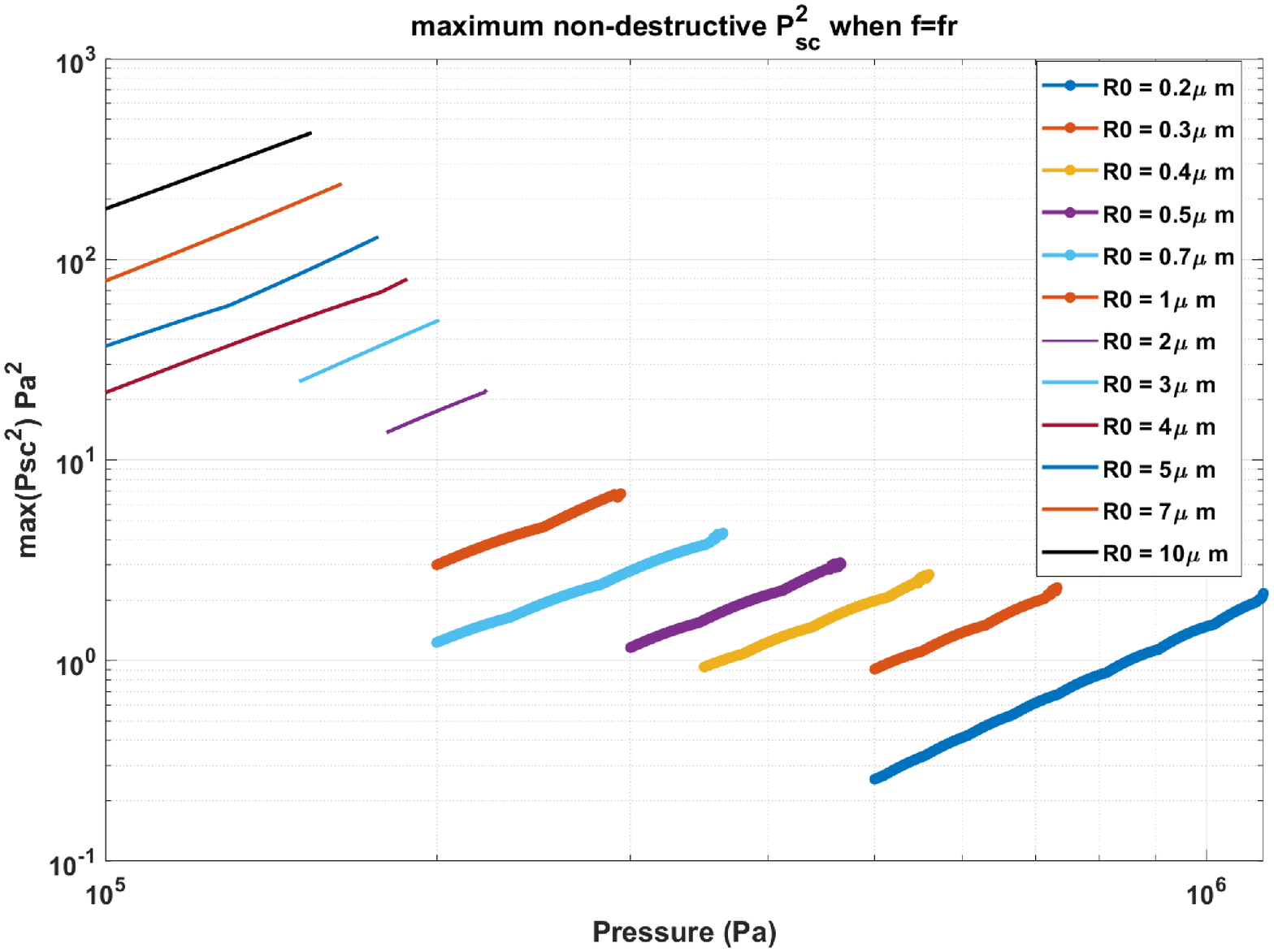}}	\\(a)\\
		\scalebox{0.33}{\includegraphics{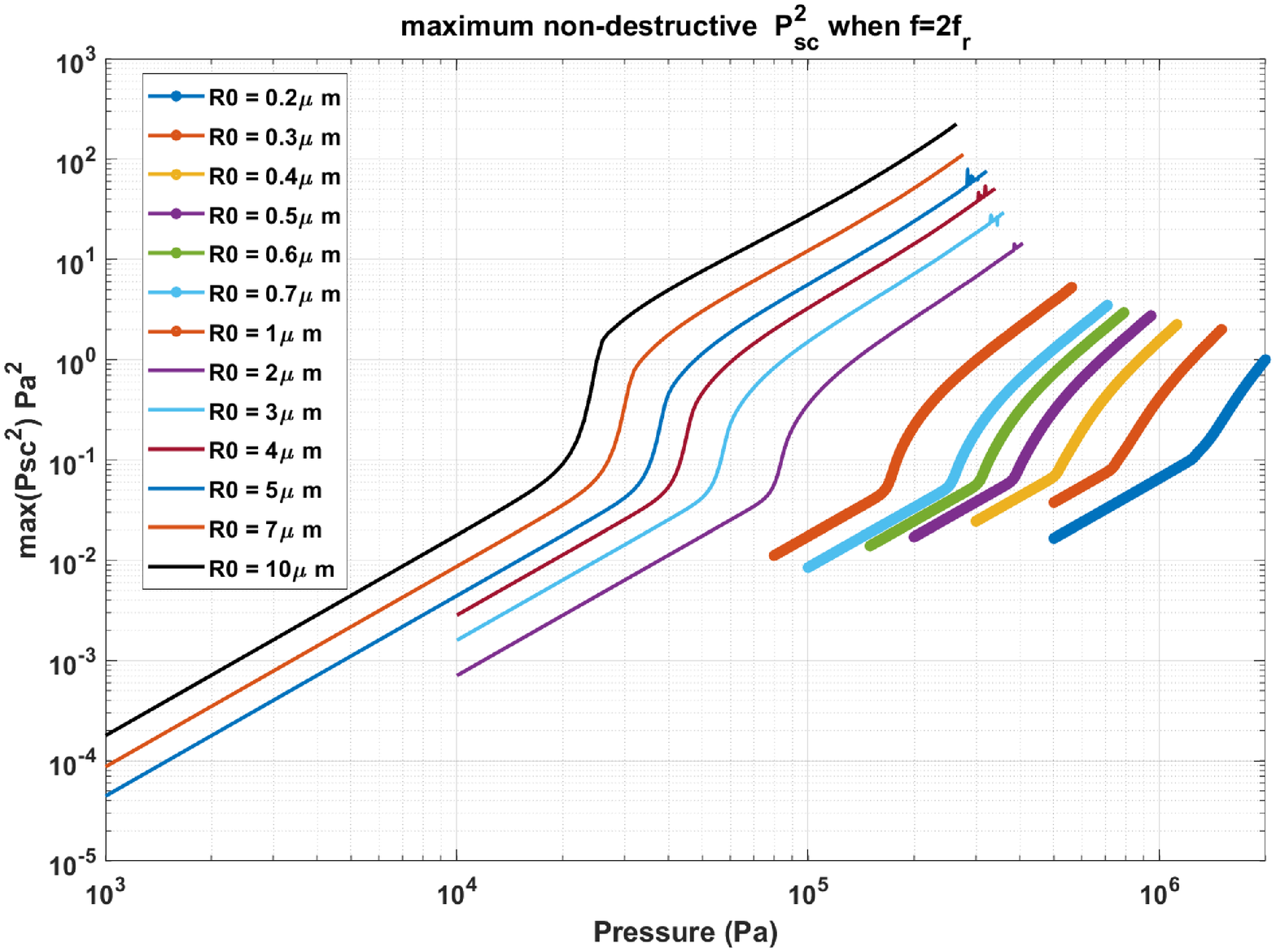}}\\
		(b) 
		\caption{Maximum value of non-destructive $P_{sc}^2$ ($\frac{R}{R_0}\leq2$) when : a) $f=f_r$, and b) $f=2f_r$.}
	\end{center}
\end{figure*}
Figure A.4a-b shows the fundamental component of the non-destructive $(P_{sc}^2)_{max}$ for $f=f_r$ and $f=2f_r$ respectively. The fundamental component of the $(P_{sc})$ exhibit the same behavior as $(P_{sc}^2)_{max}$ as a function of pressure. \\
\begin{figure*}
	\begin{center}
		\scalebox{0.6}{\includegraphics{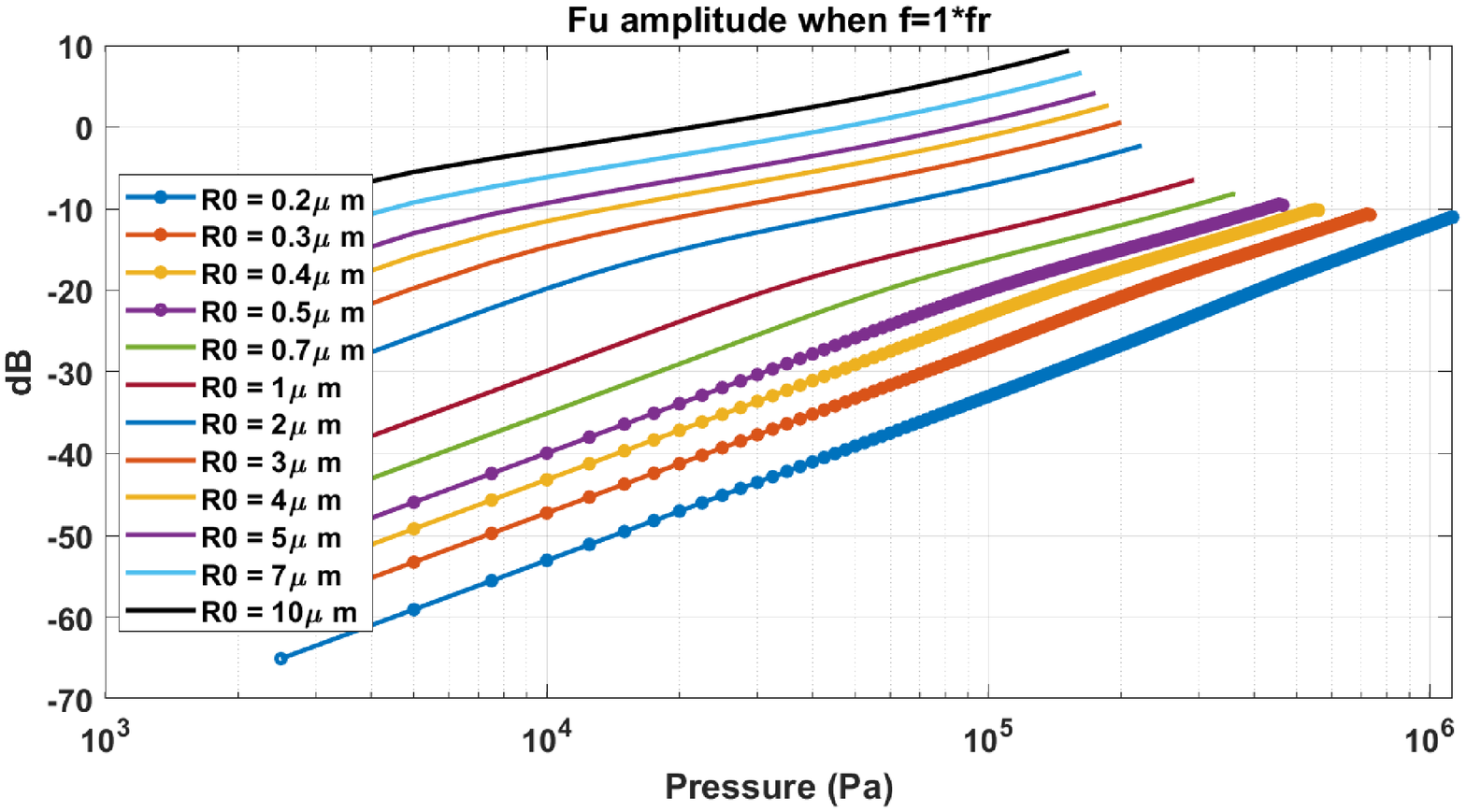}}\\(a)\\
		\scalebox{0.6}{\includegraphics{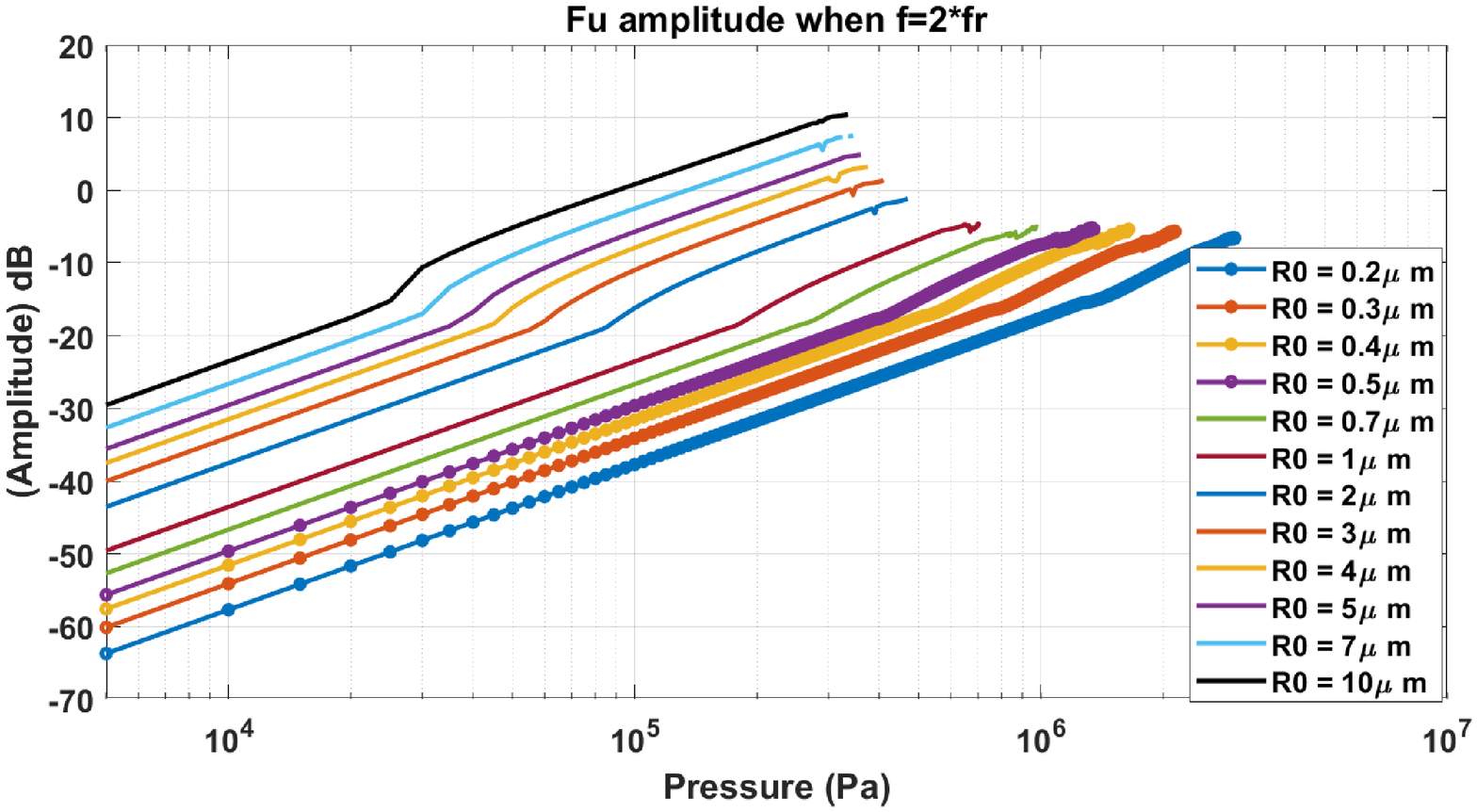}}\\
		(b) 
		\caption{Fundamental (FU) amplitude of the intensity of non-destructive  ($\frac{R}{R_0}\leq2$) $P_{sc}$ when : a) $f=f_r$, and b) $f=2f_r$.}
	\end{center}
\end{figure*}
Figure A.5 shows the amplitude of the SH component of the $(P_{sc})$ when $f=f_r$ and $f=2f_r$ respectively. When f=fr, only a fraction of the bubble sizes that are shown ($R_0<4 \mu m$) are able to undergo non-destructive PD ($\frac{R}{R_0}\leq2$); this is seen as a rapid increase in the SH component as the pressure increases above the PD threshold. However, the pressure range of P2 oscillations for non-destructive oscillations is very small. When $f=2f_r$ (Fig A5b), the SH component of the signal grows rapidly when PD occurs and the rate of increase decreases above a second pressure threshold. 
Fig A.6, illustrates the UH component of the $(P_{sc})$  when $f=f_r$ and $f=2f_r$ respectively. The UH component of the signal exhibits the same trend as the SH component shown in Fig A.5.
\begin{figure*}
	\begin{center}
		\scalebox{0.6}{\includegraphics{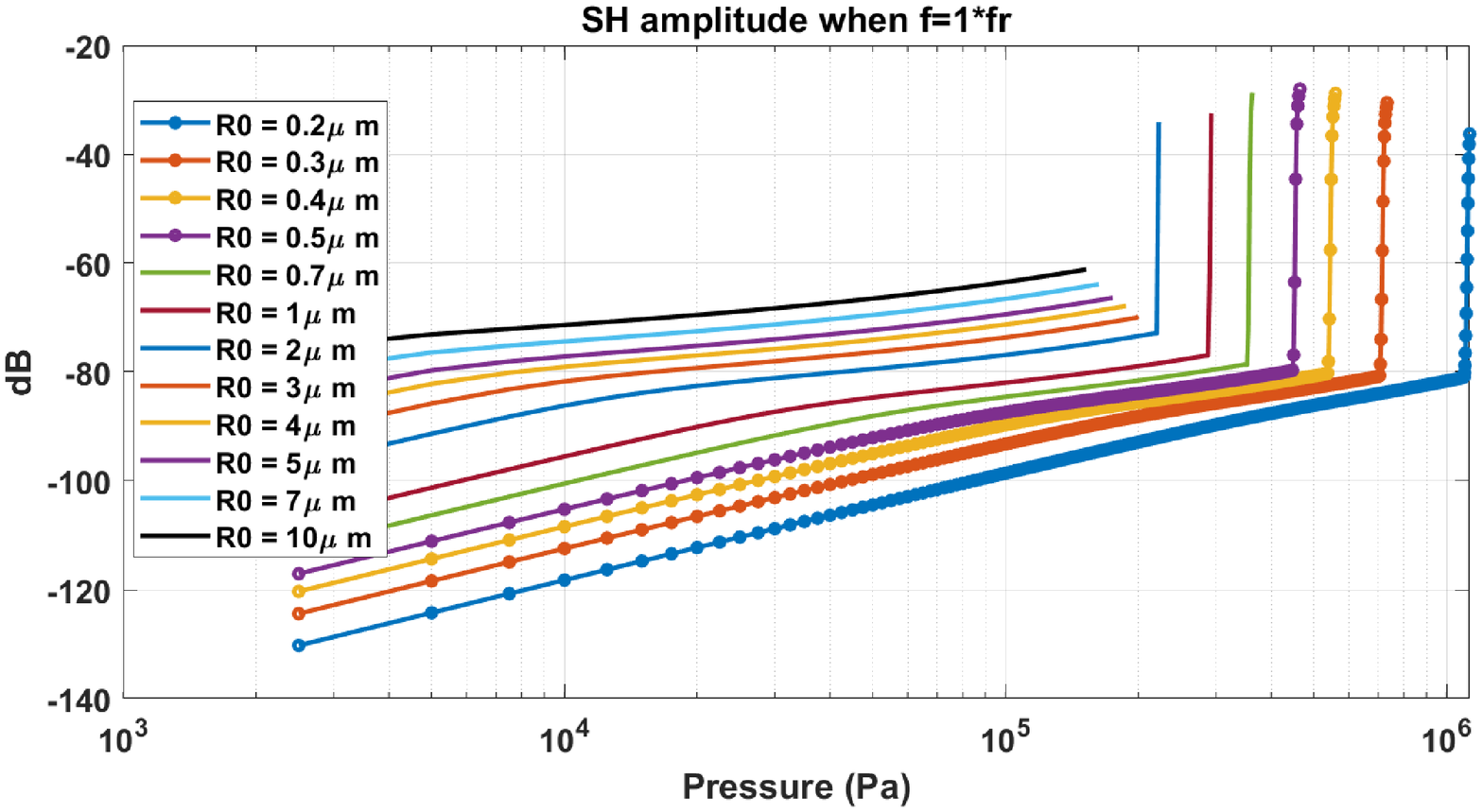}}\\ (a)\\
		\scalebox{0.6}{\includegraphics{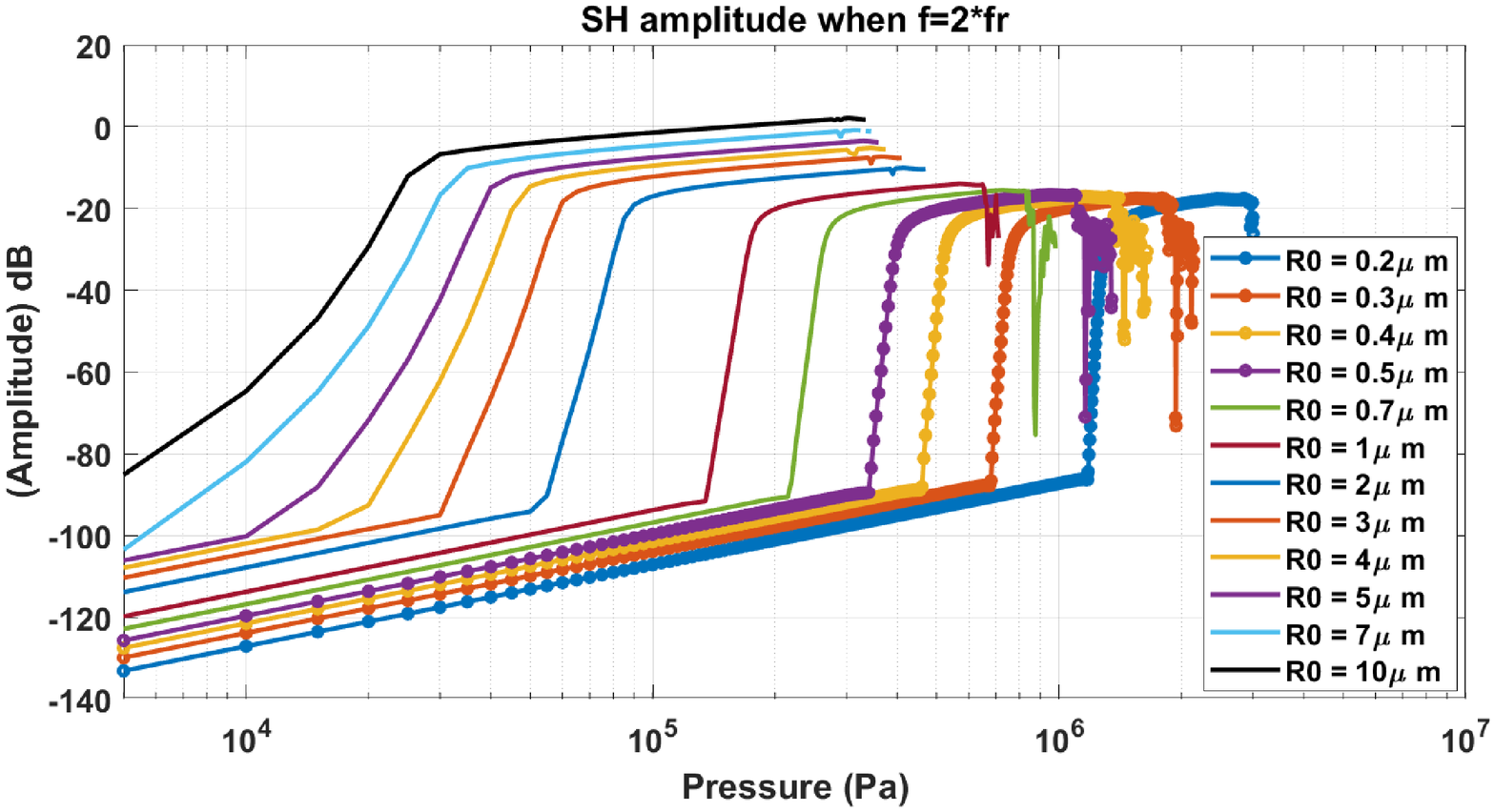}}\\
		(b) 
		\caption{Sub-harmonic (SH) amplitude of the intensity of non-destructive  ($\frac{R}{R_0}\leq2$) $P_{sc}$ when : a) $f=f_r$, and b) $f=2f_r$.}
	\end{center}
\end{figure*}
\begin{figure*}
	\begin{center}
		\scalebox{0.6}{\includegraphics{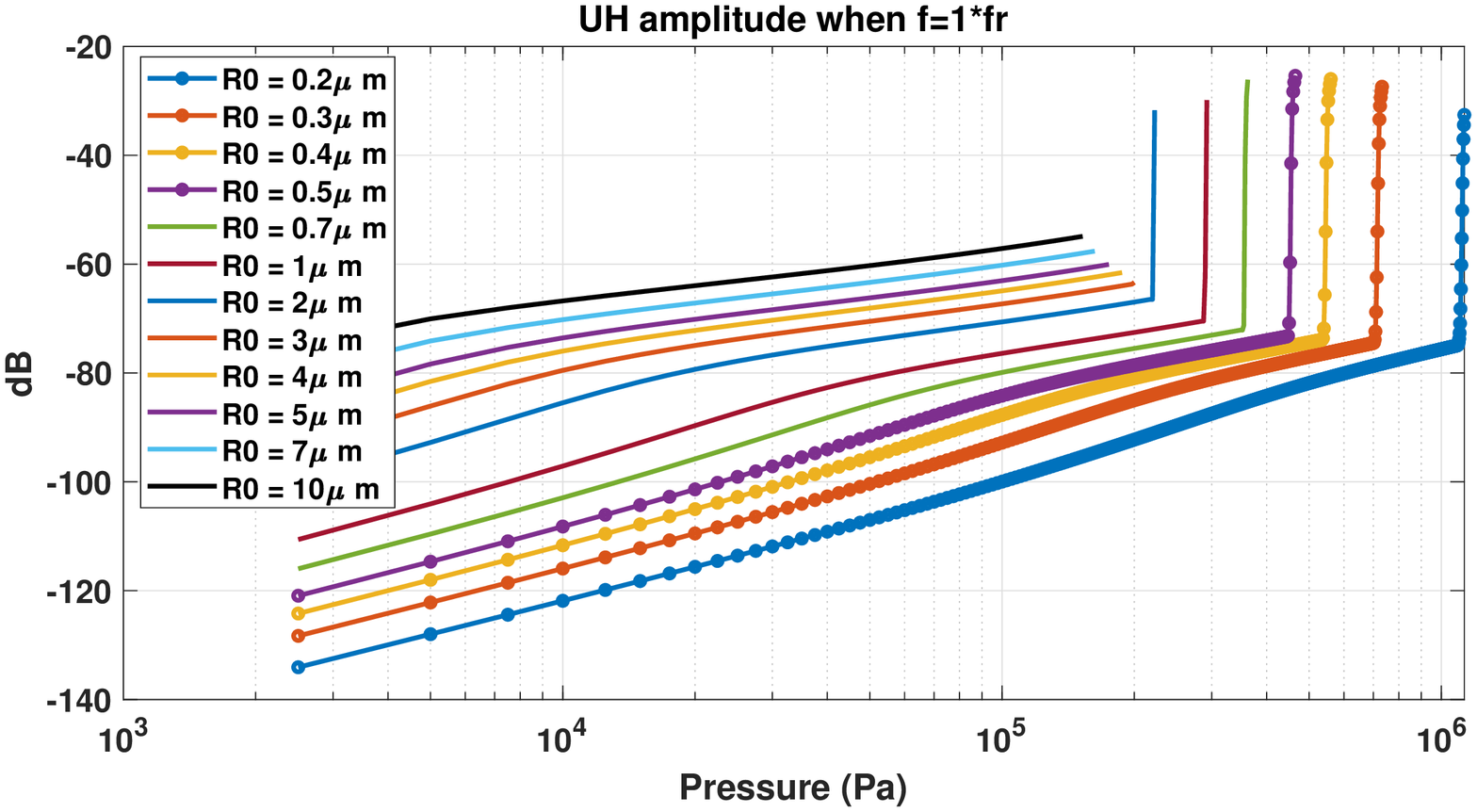}}\\(a)	\\
		\scalebox{0.6}{\includegraphics{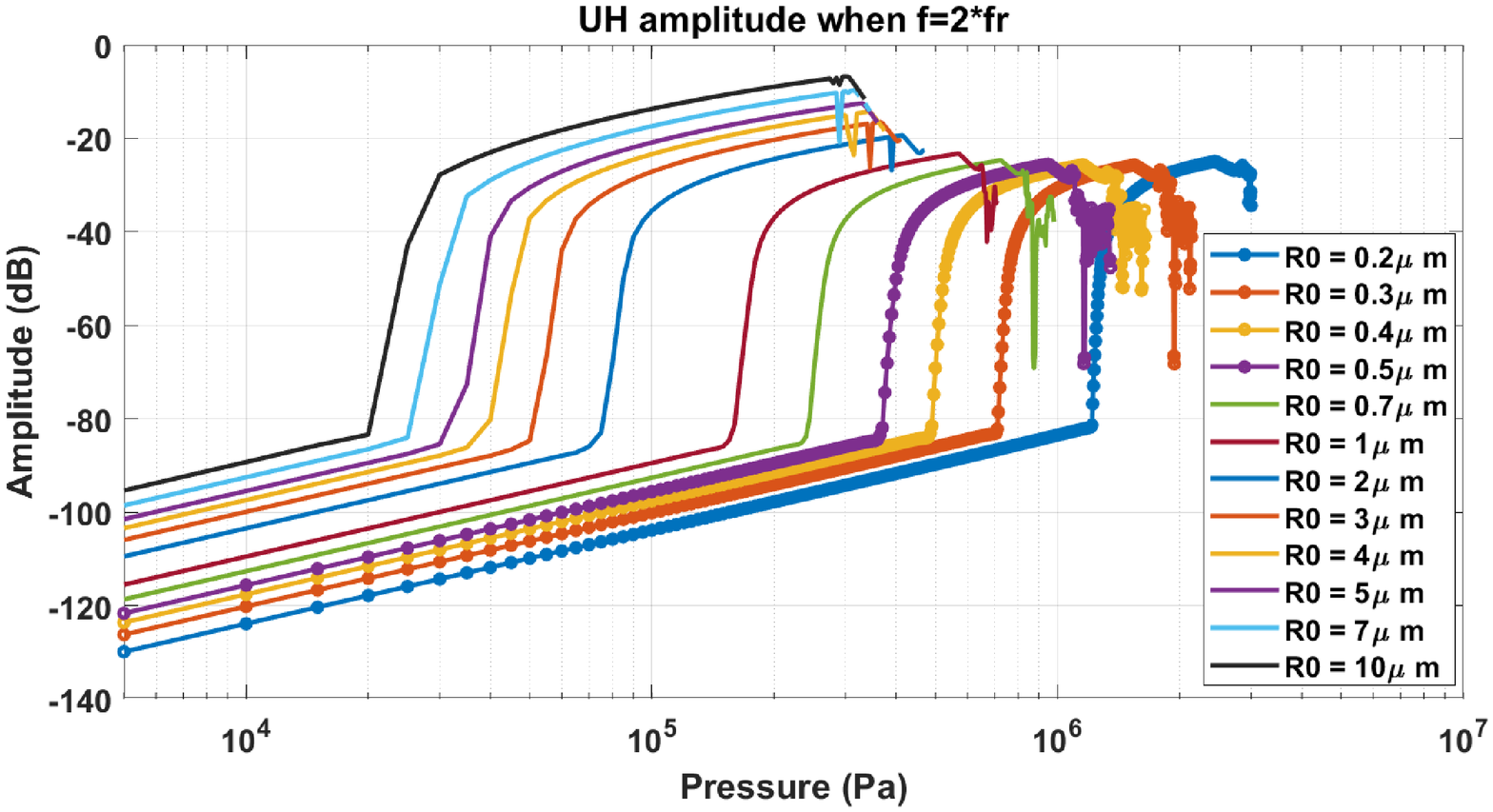}}\\
		(b) 
		\caption{Ultra-harmonic (UH) amplitude of the intensity of non-destructive  ($\frac{R}{R_0}\leq2$) $P_{sc}$ when : a) $f=f_r$, and b) $f=2f_r$.}
	\end{center}
\end{figure*}

\section{higher order attractors at $f_r$ and $2f_r$}
\begin{figure*}
	\begin{center}
		\scalebox{0.43}{\includegraphics{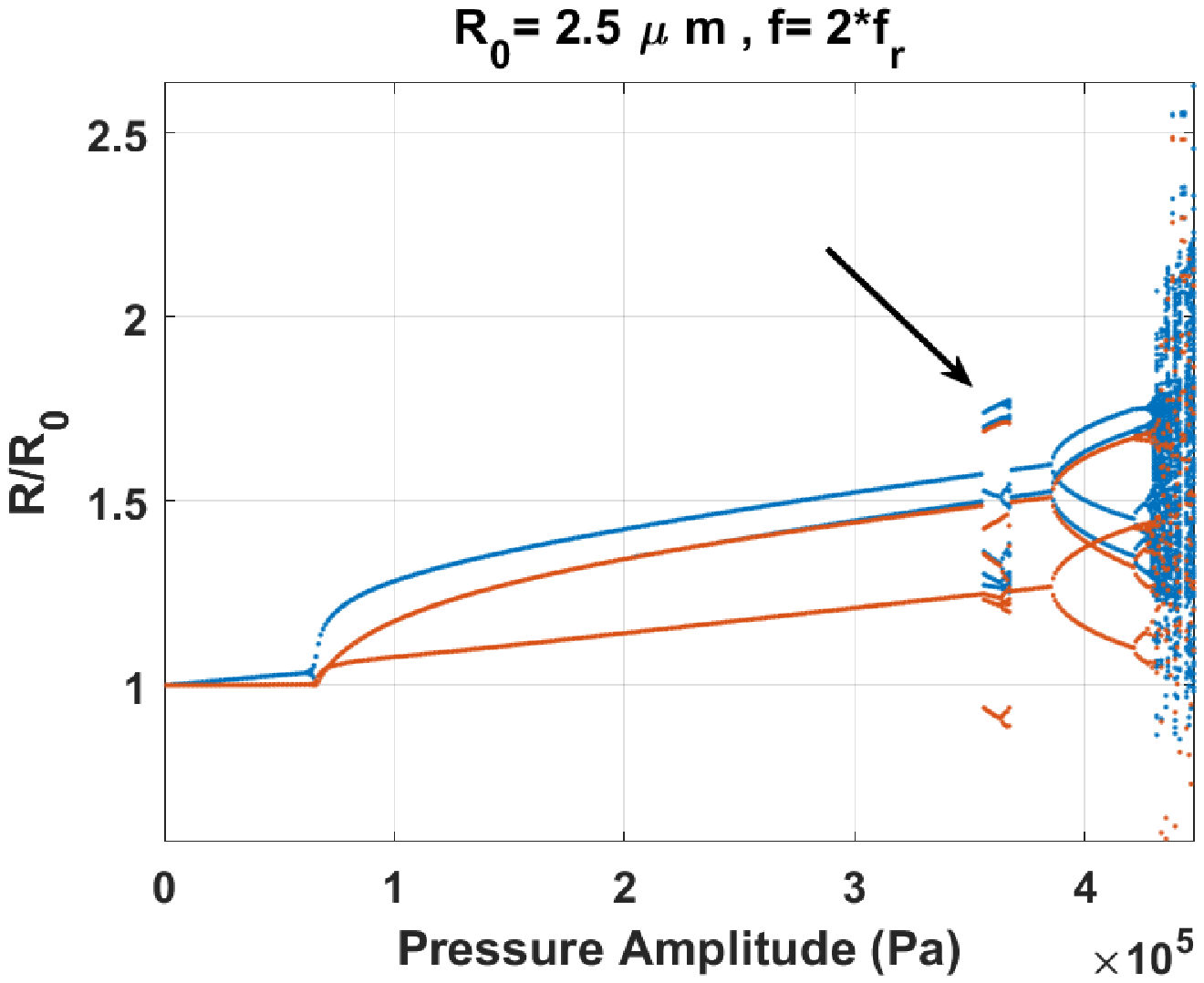}} \scalebox{0.43}{\includegraphics{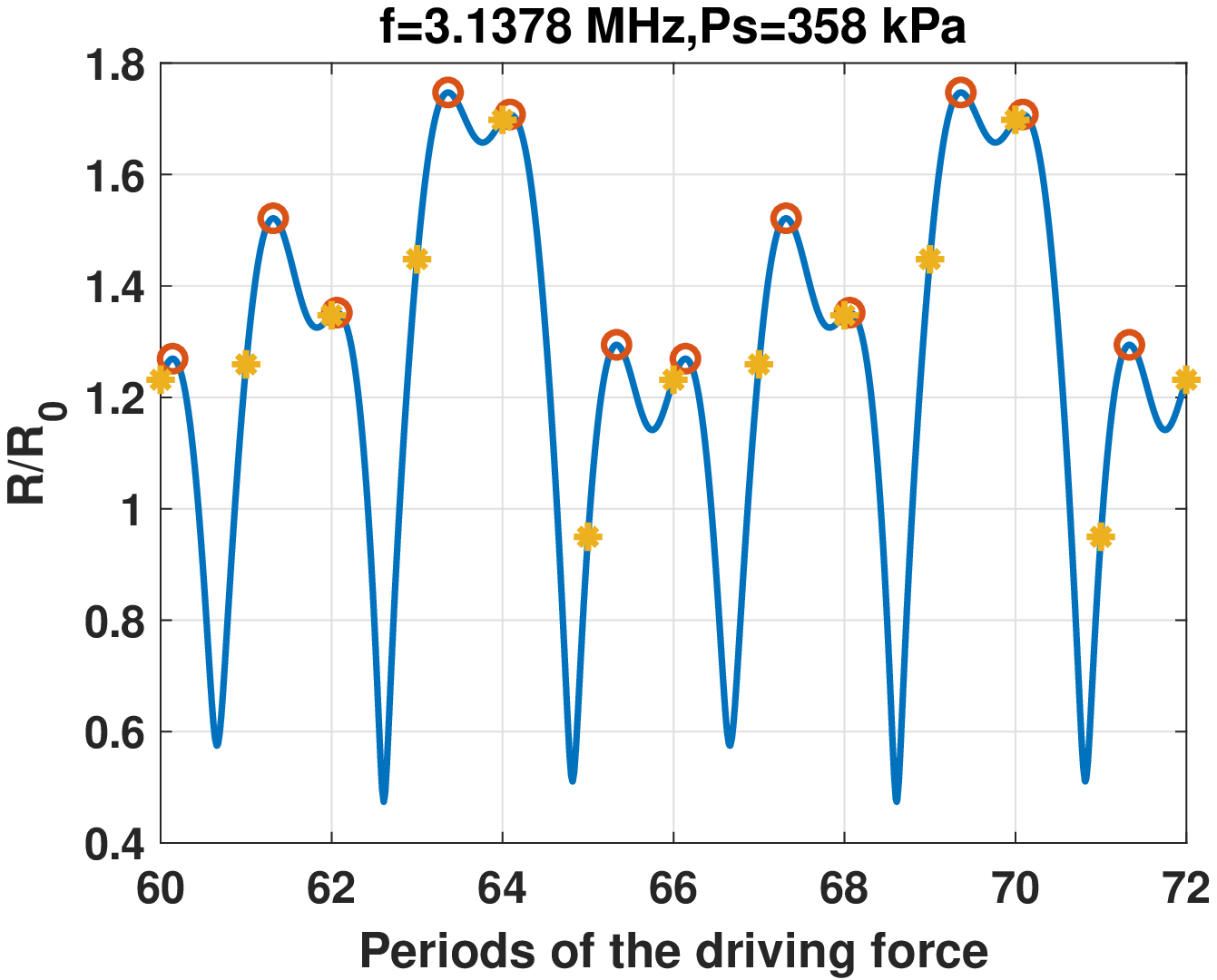}}\\
		\hspace{0.5cm} (a) \hspace{6cm} (b)\\
		\scalebox{0.43}{\includegraphics{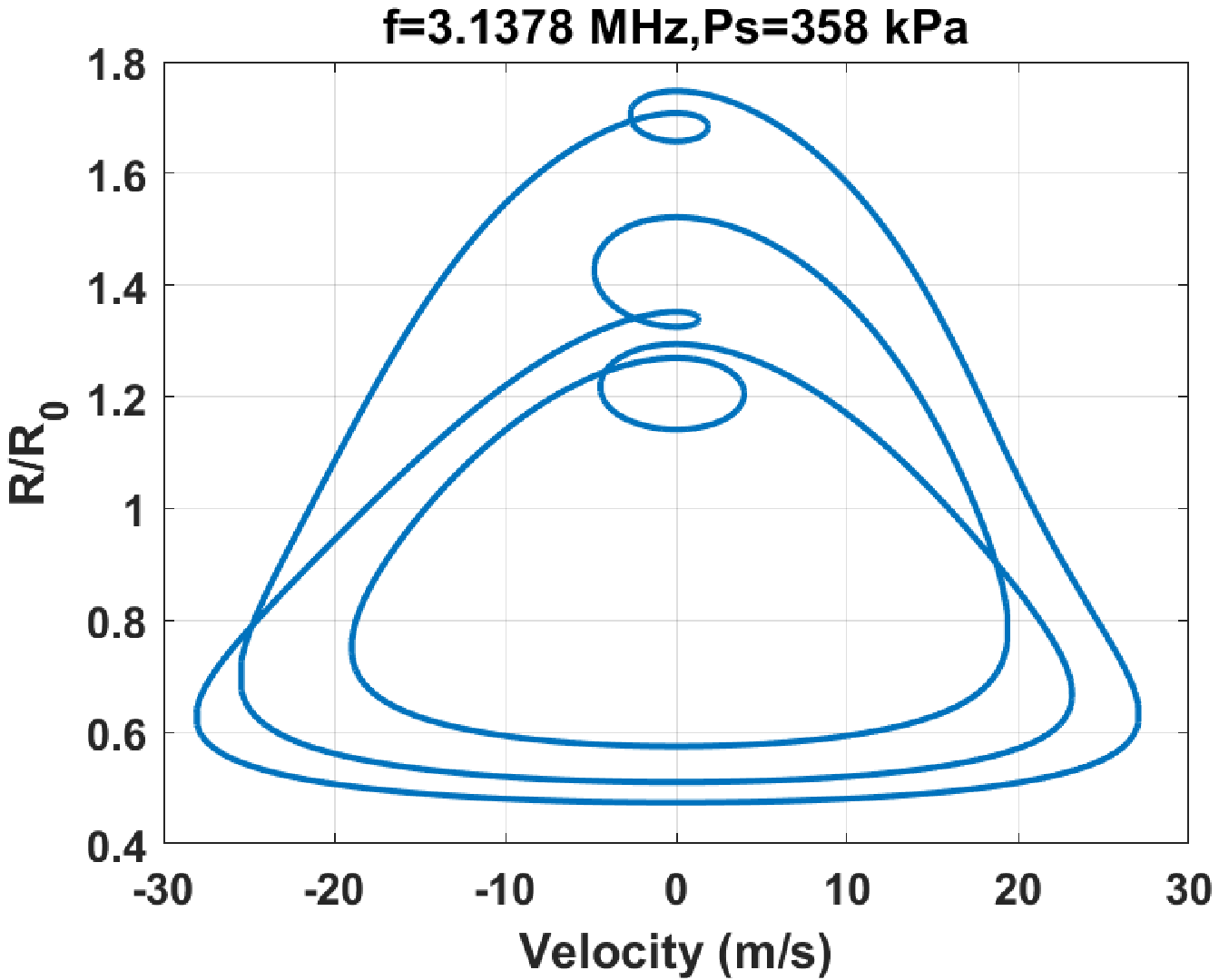}} \scalebox{0.43}{\includegraphics{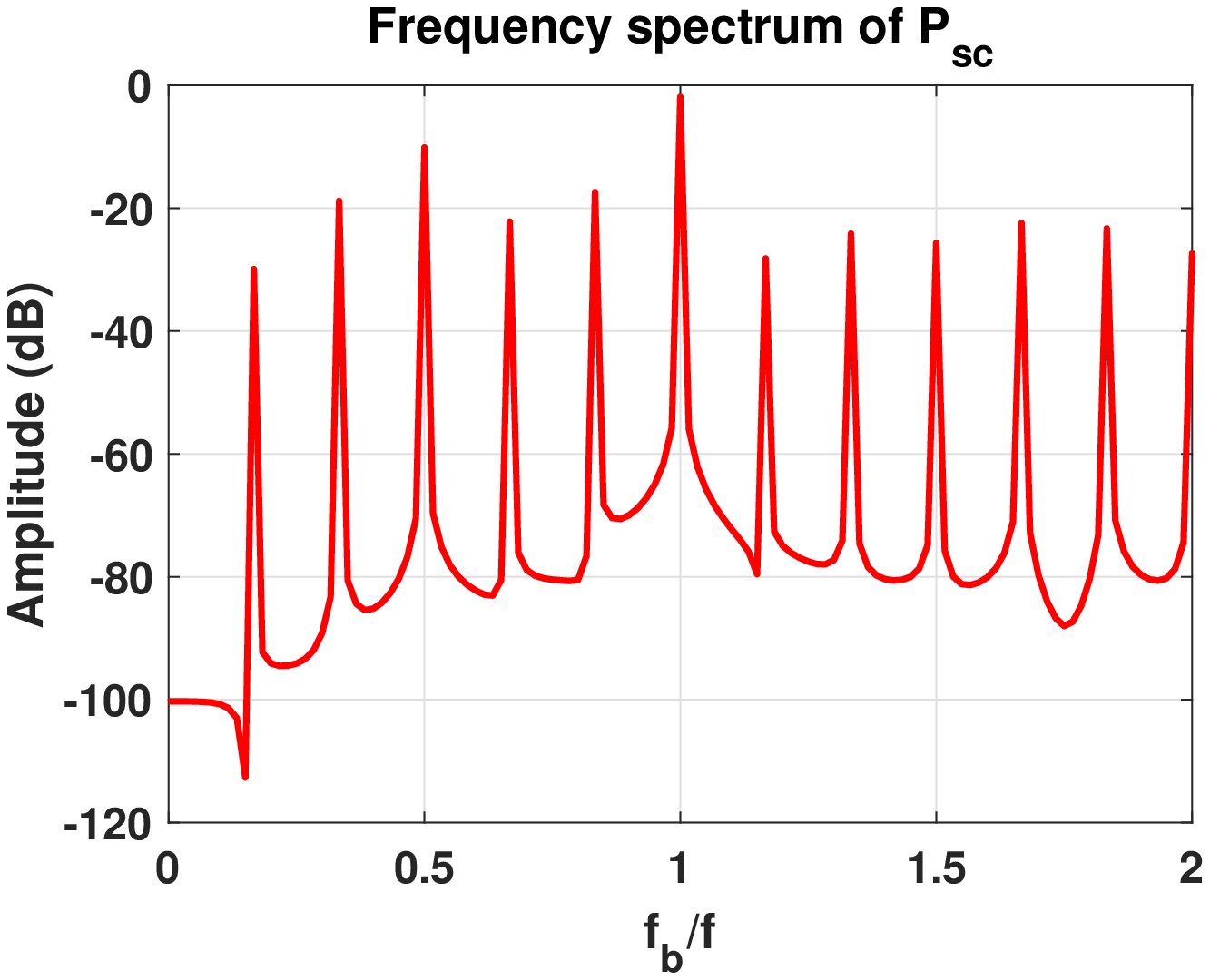}}\\
		\hspace{0.5cm} (c) \hspace{6cm} (d)\\
		\caption{Bifurcation structures of the bubble with $R_0=2.5 \mu m$ and $f=2f_r$ as a function of pressure highlighting a period 6 signal (black arrow). Time-series of the P6 $\frac{R}{R_0}$ oscillation as a function of the driving acoustic period when $f=2f_r$ and $P_A$=358 kPa. C) Phase portrait of the P6 attractor. d) the corresponding frequency spectrum of the backscattered pressure.}
	\end{center}
\end{figure*}
\begin{figure*}
	\begin{center}
		\scalebox{0.43}{\includegraphics{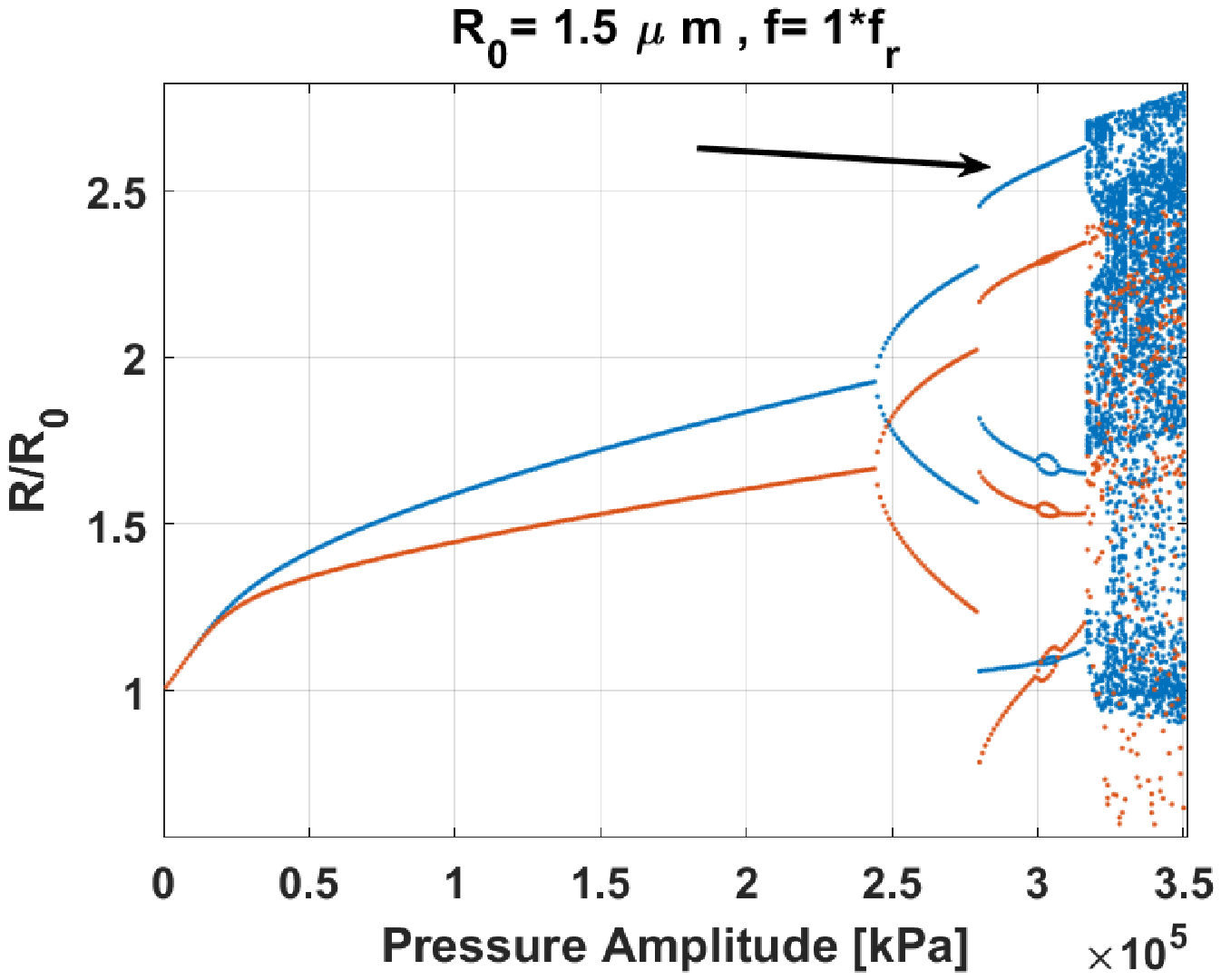}} \scalebox{0.43}{\includegraphics{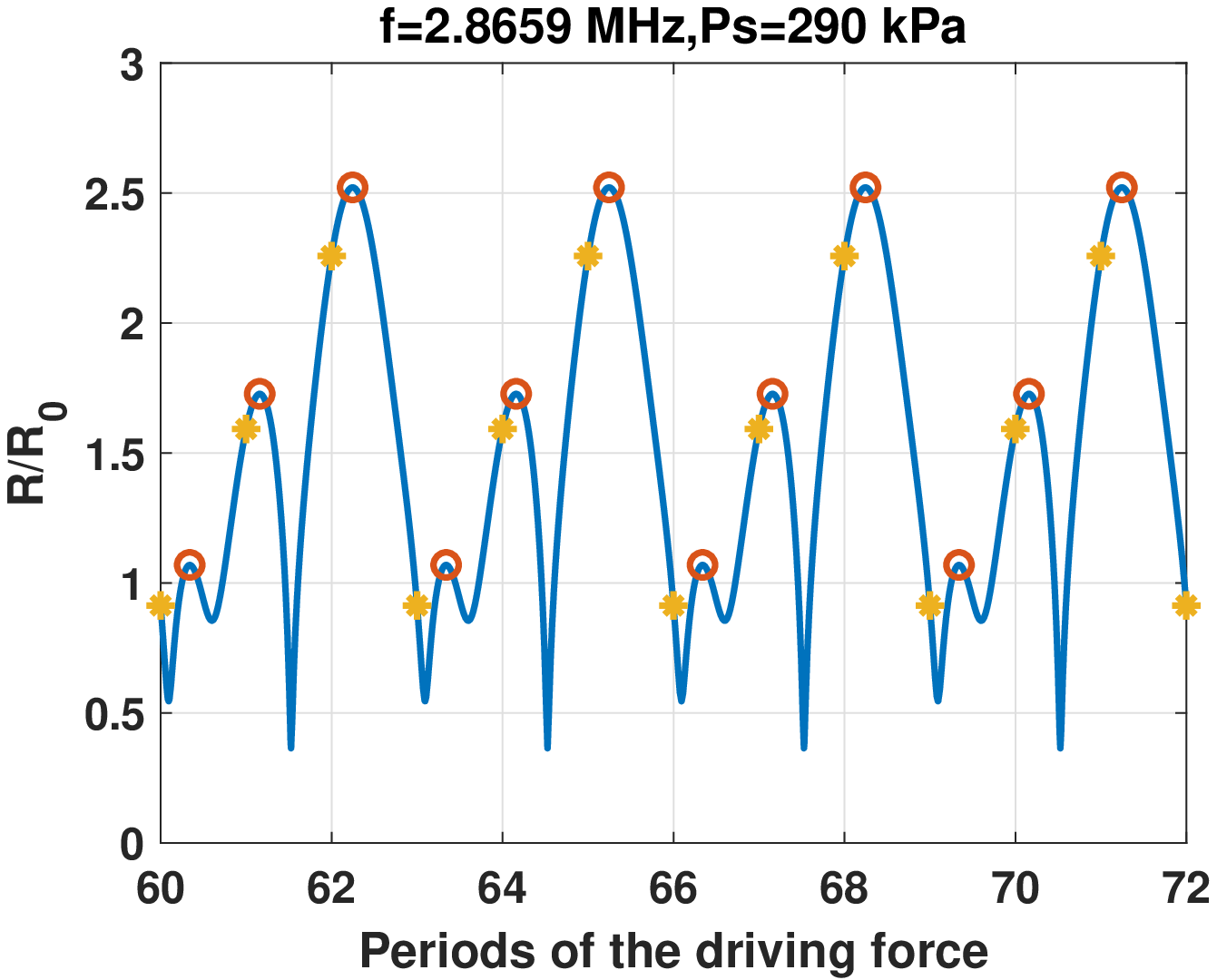}}\\
		\hspace{0.5cm} (a) \hspace{6cm} (b)\\
		\scalebox{0.43}{\includegraphics{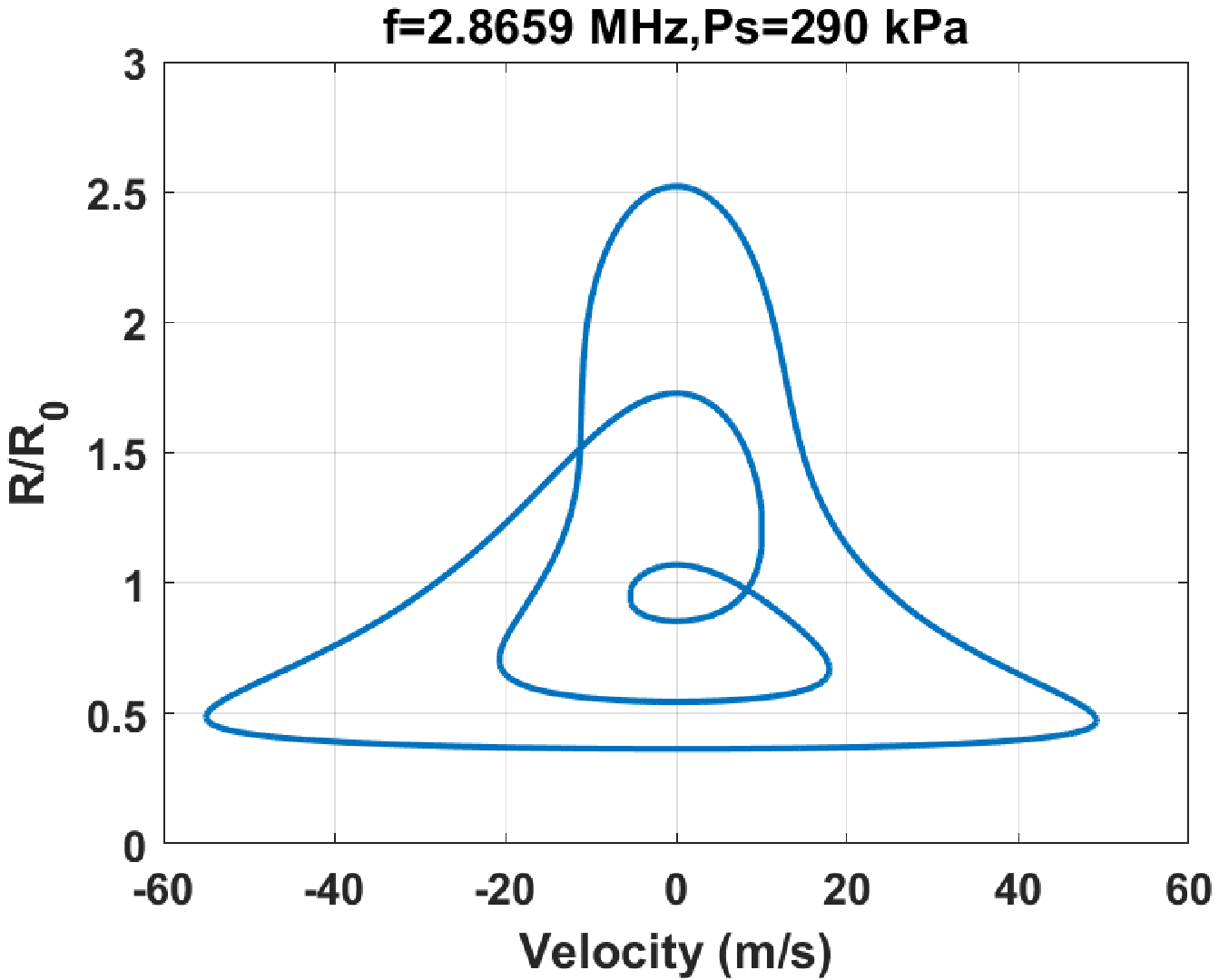}} \scalebox{0.43}{\includegraphics{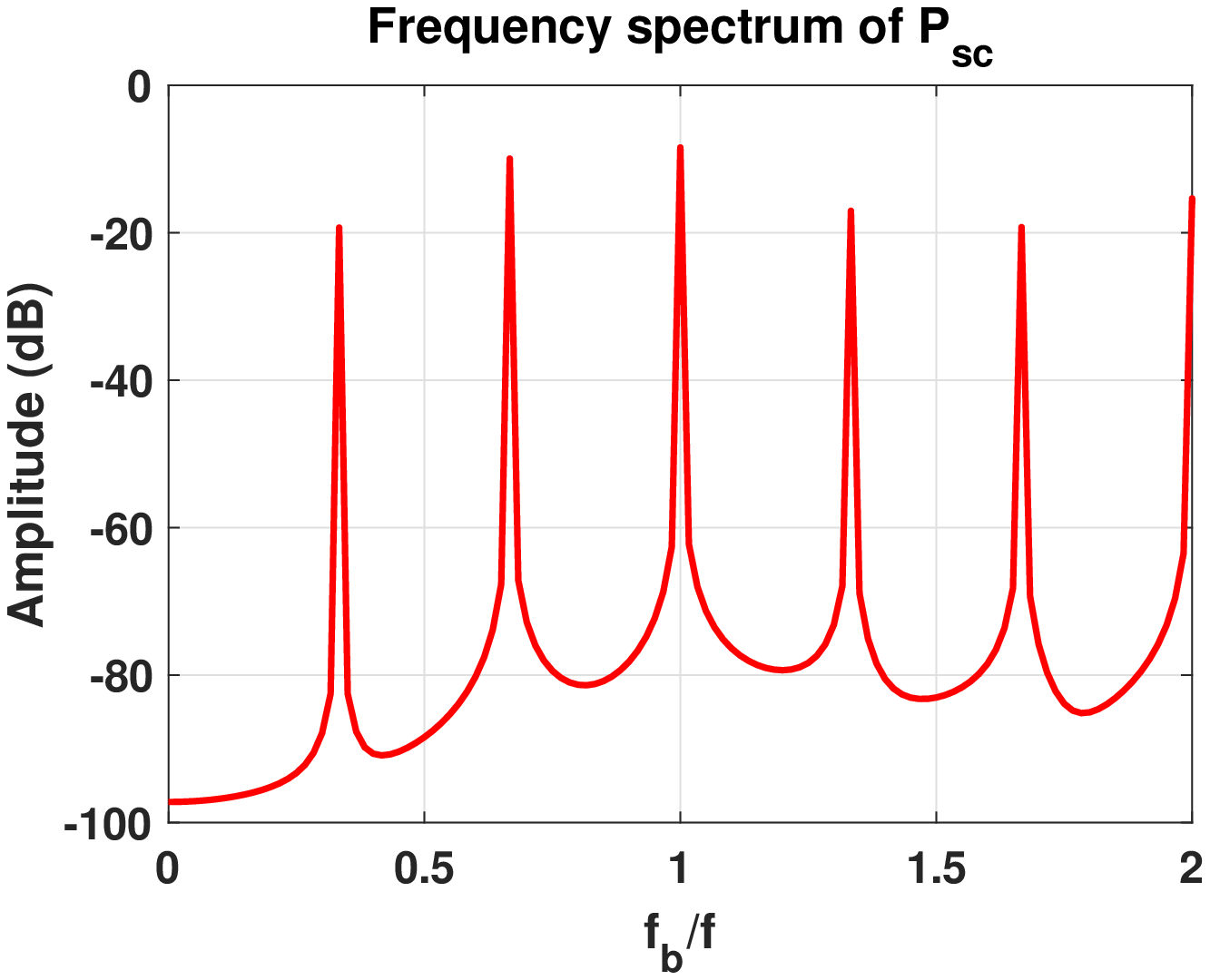}}\\
		\hspace{0.5cm} (c) \hspace{6cm} (d)\\
		\caption{Bifurcation structures of the bubble with $R_0=1.5 \mu m$ and $f=f_r$ as a function of pressure highlighting a period 3 signal (black arrow). Time-series of the P3 $\frac{R}{R_0}$ oscillation as a function of the driving acoustic period when $f=2f_r$ and $P_A$=290 kPa. C) Phase portrait of the P3 attractor. d) the corresponding frequency spectrum of the backscattered pressure..}
	\end{center}
\end{figure*}
Figure B.1 pays a closer attention on the P6 behavior that is seen for bubbles ($2.5<R_0<5$) when $f=2f_r$ and initial conditions are $R(0)=R_0$ and  $\dot{R(0)}=0$. This is a condition which is common in biomedical imaging applications [15]). Figure B.1a shows the generation of the P6 behavior (black arrow) through a saddle node bifurcation from P2 that only lasts for a small pressure window. The radial oscillations of $\frac{R}{R_0}\leq2$ as a function of period are shown in figure B.1b; the signal exhibits 6 maxima (red circles) that repeat themselves once every 6 acoustic cycles. The phase portrait of the signal has 3 loops; each are enclosing a smaller loop in Fig. B.1c. The back-scatter frequency spectrum is shown in Figure B.1d, depicting the existence of 5 SHs of f/6, f/3, f/2, 2f/3 and 5f/6 (with the f/2 component the strongest). The period 6 shown here (we name it P6-2) has distinct differences from the P6 that can be generated by sonicating a bubble with a frequency that is about 6 times the resonance frequency of the bubble [13]. The later is generated through a saddle node bifurcation [13] from a period 1 oscillations (we name it P6-1) while the former that is shown in figure B.1 is generated through a saddle node bifurcation from period 2 oscillations. Additionally, the P6-1 $\frac{R}{R_0}$ signal has one envelope with 6 or 5 maxima while a P6-2 $\frac{R}{R_0}$ signal has three envelopes each with 2 maxima. When the frequency spectrum of the $P_{sc}$ is considered, the f/6 is the strongest SH component of a P6-1 oscillation [13] while f/2 is the strongest SH component of a P6-2 oscillations.\\
Another interesting nonlinear oscillation that was observed in this paper is a P3 signal that is generated through a saddle node bifurcation from P2 oscillations when $f=f_r$. This behavior was observed for bubbles of size $1 \mu m$ $<R_0<$ $2\mu m$. Figure B.2a shows the P3 oscillation (black arrow) that is generated through a saddle node bifurcation from a P2 oscillation. The radial oscillations shown in fig. B.2b display a signal with three maxima with two repeating envelopes once every 3 acoustic cycles; one has two maxima and one has one maxima. The phase portrait in fig B.2c consists of two orbits sharing an internal bend. The frequency spectrum in fig B.2d, depicts three SHs with frequencies of f/3 and 2f/3 with 2f/3 stronger than the f/3 component. We name this a P3-2 oscillation and it has distinct differences from a P3-1 oscillation (a P3-1 occurs when a bubble is sonicated with a frequency that is approximately 3 times its resonance frequency and is generated via a saddle node bifurcation from a P1 oscillation [13]). The main difference is mechanism of generation as it is discussed above. The second difference is the shape of the radial oscillations; P3-1 has one envelope with 2 or 3 peaks that repeat itself once every three acoustic cycle. The phase portrait of a P3-1 oscillation consists of one orbit with two distinct internal bends and the f/3 component of the frequency spectrum is stronger than the 2f/3 component.

\end{document}